\documentclass[longauth]{aa}

\usepackage{graphicx}
\usepackage{txfonts}
\usepackage{comment}
\usepackage{caption}
\usepackage{subcaption}
\usepackage{longtable}
\usepackage{lscape}
\usepackage[]{hyperref}
\hypersetup{
   colorlinks=true,
   linkcolor= blue,
   filecolor= magenta,      
   urlcolor= blue,
   citecolor= blue,
}
\usepackage{float,amsmath,multirow}
\usepackage{color}
\usepackage[version=4]{mhchem}
\usepackage{booktabs}
\usepackage{soul} 

\usepackage{xcolor}
\usepackage{xifthen}
\usepackage[normalem]{ulem}
\newcommand{\stkout}[1]{\ifmmode\text{\sout{\ensuremath{#1}}}\else\sout{#1}\fi}
\newcommand{\edited}[2]{\ifthenelse{\isempty{#1}}{\textbf{#2}}{\ifthenelse{\isempty{#2}}{\textcolor{gray}{\stkout{#1}}}{\textcolor{gray}{\stkout{#1}} \textbf{#2}}}}

\usepackage{xstring,xifthen}
\newcommand{\replacedecimal}[2]{\ifthenelse{\isin{.}{#1}}{\text{\StrBefore{#1}{.}}\ensuremath{\overset{#2}{.}}\text{\StrBehind{#1}{.}}}{#1\ensuremath{^{#2}}}}

\begin{document}

   \title{
   CoCCoA: Complex Chemistry in hot Cores with ALMA}

   \subtitle{Selected oxygen-bearing species}

   \author{Y. Chen\inst{1}
          \and M. L. van Gelder\inst{1}
          \and P. Nazari\inst{1}
          \and C. L. Brogan\inst{2}
          \and E. F. van Dishoeck\inst{1,4} 
          \and H. Linnartz\inst{5} 
          \and J. K. J{\o}rgensen\inst{6}
          \and T. R. Hunter\inst{2,3}
          \and O. H. Wilkins\inst{7}
          \and G. A. Blake\inst{8}
          \and P. Caselli\inst{4}
          \and K.-J. Chuang\inst{5}
          \and C. Codella\inst{9,10}
          \and I. Cooke\inst{11}
          \and M. N. Drozdovskaya\inst{12}
          \and R. T. Garrod\inst{13,14}
          \and S. Ioppolo\inst{15}
          \and M. Jin\inst{16,17}
          \and B. M. Kulterer\inst{12}
          \and N. F. W. Ligterink\inst{18}
          \and A. Lipnicky\inst{2}
          \and R. Loomis\inst{2}
          \and M. G. Rachid\inst{5}
          \and S. Spezzano\inst{4}
          \and B. A. McGuire\inst{2,19}
          }

   \institute{Leiden Observatory, Leiden University, P.O. Box 9513, 2300 RA Leiden, the Netherlands\\
        \email{ychen@strw.leidenuniv.nl}
        \and National Radio Astronomy Observatory, Charlottesville, VA 22903, USA
        \and Harvard-Smithsonian Center for Astrophysics, Cambridge, MA 02138, USA
        \and Max Planck Institut für Extraterrestrische Physik (MPE), Giessenbachstrasse 1, 85748 Garching, Germany
        \and Laboratory for Astrophysics, Leiden Observatory, Leiden University, P.O. Box 9513, 2300 RA Leiden, The Netherlands
        \and Niels Bohr Institute, University of Copenhagen, Copenhagen, Denmark
        \and NASA Goddard Space Flight Center, Greenbelt, MD 20771, USA
        \and Division of Geological and Planetary Sciences, California Institute of Technology, Pasadena, CA 91125, USA
        \and INAF, Osservatorio Astrofisico di Arcetri, Largo E. Fermi 5, I-50125 Firenze, Italy
        \and Univ. Grenoble Alpes, CNRS, IPAG, F-38000 Grenoble, France
        \and Department of Chemistry, University of British Columbia, 2036 Main Mall, Vancouver BC V6T 1Z1, Canada
        \and Center for Space and Habitability, Universität Bern, Gesellschaftsstrasse 6, 3012 Bern, Switzerland
        \and Department of Chemistry, University of Virginia, Charlottesville, VA 22904, USA
        \and Department of Astronomy, University of Virginia, Charlottesville, VA 22904, USA
        \and Center for Interstellar Catalysis, Department of Physics and Astronomy, Aarhus University, Ny Munkegade 120, Aarhus C 8000, Denmark
        \and Astrochemistry Laboratory, Code 691, NASA Goddard Space Flight Center, Greenbelt, MD 20771, USA
        \and Department of Physics, Catholic University of America, Washington, DC 20064, USA
        \and Physics Institute, Space Research and Planetary Sciences, University of Bern, Sidlerstrasse 5, CH-3012 Bern, Switzerland
        \and Department of Chemistry, Massachusetts Institute of Technology, Cambridge, MA 02139, USA
             }

   \date{Received 2023 March 28; revised 2023 June 28; accepted 2023 August 1}

 
\abstract
{Complex organic molecules (COMs), especially oxygen-bearing species, have been observed to be abundant in the gas phase toward low-mass and high-mass protostars. Deep line surveys have been carried out only for a limited number of well-known high-mass star forming regions using the Atacama Large Millimeter/submillimeter Array (ALMA), which has unprecedented resolution and sensitivity. Statistical studies on oxygen-bearing COMs (O-COMs) in a large sample of high-mass protostars using ALMA are still lacking.}
{We aim to determine the column density ratios of six O-COMs with respect to methanol (\ce{CH3OH}) in a sample of 14 high-mass protostellar sources to investigate their origin through ice and/or gas-phase chemistry. The selected species are: acetaldehyde (\ce{CH3CHO}), ethanol (\ce{C2H5OH}), dimethyl ether (DME, \ce{CH3OCH3}), methyl formate (MF, \ce{CH3OCHO}), glycolaldehyde (GA, \ce{CH2OHCHO}), and ethylene glycol (EG, \ce{(CH2OH)2}).}
{We fit the spectra of 14 high-mass sources observed as part of the Complex Chemistry in hot Cores with ALMA (CoCCoA) survey and derive the column densities and excitation temperatures of the six selected O-COMs. The minor isotopologue of methanol CH$_3^{18}$OH is used to infer the column density of the main isotopologue \ce{CH3OH}, of which the lines are generally optically thick. We compare our O-COM ratios with those of 5 low-mass protostars available from the literature that are studied with ALMA, along with the results from experiments and simulations.}
{Although the CoCCoA sources have different morphologies and brightness in their continuum and methanol emission, the O-COM ratios with respect to methanol have very similar values in the high-mass and low-mass samples. DME and MF have the highest and most constant ratios within one order of magnitude, while the other four species have lower ratios and exhibit larger scatter by 1--2 orders of magnitude. The ratio between DME and MF is close to 1, which agrees well with previous observational findings. Current simulations and experiments can reproduce most observational trends with a few exceptions, for example, they tend to overestimate the abundance of ethanol and GA with respect to methanol.}
{The constant column density ratios of selected O-COMs among the low- and high-mass sources suggest that these species are formed in similar environments during star formation, probably on icy dust grains in the pre-stellar stages. Where deviations are found, hypotheses exist to explain the differences between observations and simulations/experiments, such as the involvement of gas-phase chemistry and different emitting areas of molecules.}

\maketitle
%

\section{Introduction}\label{sec_introduction}
Complex organic molecules (COMs), typically defined as carbon-bearing molecules with at least six atoms \citep{2009ARAA}, have been intensively studied over the past several decades due to their importance of linking atoms and simple molecules with prebiotic species \citep{Caselli2012, Jorgensen2020ARAA, Ceccarelli2022}. Up to now, more than 80 COMs have been detected in various environments \citep{McGuire2022}. Nearly or fully saturated COMs containing oxygen and nitrogen atoms have been widely observed in line surveys by radio telescopes toward protostars with different masses \citep[e.g.,][]{Belloche2016, Jorgensen2016, El-Abd2019, LeeCF2019, Csengeri2019, Belloche2020, Bianchi2020, MvG2020, Ligterink2020, Mininni2020, Colzi2021, Nazari2021, Nazari2022_NCOM, Hsu2022, Imai2022, Codella2022}. 
In particular, oxygen-bearing COMs (O-COMs) including methanol (\ce{CH3OH}),  acetaldehyde (\ce{CH3CHO}), dimethyl ether (DME, \ce{CH3OCH3}), and methyl formate (MF, \ce{CH3OCHO}) were first observed in massive star-forming regions more than two decades ago \citep{Cummins1986, Blake1987, Schilke1997}. 
They were also detected in subsequent observations toward low-mass protostars IRAS 16293–2422 \citep{Cazaux2003} and NGC 1333 IRAS 4A/2A \citep{Bottinelli2004, Jorgensen2005}. Larger O-COMs such as glycolaldehyde (GA, \ce{CH2OHCHO}) and ethylene glycol (EG, \ce{(CH2OH)2}) were first detected in the high-mass star-forming cluster Sgr B2(N) near the Galactic center \citep{Hollis2000, Holllis2002} and then in other protostellar sources \citep{Beltran2009, Maury2014}. Three abundant COMs, \ce{CH3OH}, CH$_3$CN, and DME were even detected in protoplanetary disks around more evolved young stellar objects \citep{Oberg2015Nature, Walsh2016, Brunken2022}.

However, the formation mechanisms of COMs are still under debate. The observed O-COMs were initially thought to exist exclusively in the gas phase in hot ($T\gtrsim100$ K) environments around protostars (i.e., hot cores/corinos), primarily from thermal desorption of ices \citep{2009ARAA, Jorgensen2020ARAA}. On the other hand, observations in the past decade started detecting them in cold ($T\sim$10 K) pre-stellar cores, albeit in low abundances \citep{Bacmann2012, Vastel2014, Jimnez-Serra2016, Soma2018, Scibelli2021}. These detections indicate that COMs may already be formed on the surfaces of dust grains in the early pre-stellar stages before any ice desorption or subsequent gas-phase chemistry occurs.
Simulations and experiments have found that solid-phase methanol can be formed in the ice mantles of dust grains. The formation begins with a series of CO hydrogenation under both energetic (e.g., UV radiation) and non-energetic (e.g., atom addition) conditions \citep{Hiraoka1998, Shalabiea1994, Watanabe2002, Fuchs2009, Cuppen2009, Simons2020}. Formation of bigger COMs on cold dust grains is also possible. Experiments by \cite{Fedoseev2015} show that GA and EG can be formed through surface hydrogenation of CO under cold dense cloud conditions ($T$ = 12 K). A follow-up experimental study by \cite{Chuang2016} mixed CO, H$_2$CO, and \ce{CH3OH} ices at $T$ = 15 K, and produced not only GA and EG, but also some MF in the solid phase. However, compared with observational results in the gas phase, MF was underproduced in their experiments while GA and EG were overproduced. The underproduction of MF was alleviated in \cite{Chuang2017} by introducing ultraviolet (UV) irradiation, which may photodissociate the initial ingredients in the mixed ice into reactive free radicals and therefore boost the formation of larger species. 
Based on these laboratory works, \cite{Simons2020} explored the dependence of the final grain mantle composition on the initial gas-phase composition and the dust temperature by chemical simulations. 
Their simulations were still able to produce MF, GA, and EG at temperatures as low as 8 K, but MF was again underestimated (without UV). A much higher abundance of MF is reproduced by \cite{Garrod2022}, who introduced a set of non-diffusive mechanisms to their simulations (see detailed discussions in Sect. \ref{sec_discussion}). These studies show that the abundances of some COMs like MF could depend on environmental conditions, and a debate still exists whether and to what extent gas-phase chemistry is involved in their formation \citep{Balucani2015, Ceccarelli2022}.

With the development of new observational techniques, it is now possible to make a more complete inventory of COMs in star-forming regions. The Protostellar Interferometric Line Survey \citep[PILS,][]{Jorgensen2016} using the Atacama Large Millimeter/submillimeter Array (ALMA) detected more than 20 COMs, including all the O-bearing species mentioned above plus ethanol (\ce{C2H5OH}), and some of their isotopologues, in the low-mass protostellar binary IRAS 16293–2422 A and B \citep{Jorgensen2018, Manigand2020}. Recent ALMA surveys of a larger sample of sources further confirm the ubiquity of COMs in both low- and high-mass protostars \citep{MvG2020, Yang2021, Nazari2022_NCOM}. The next step in COM studies is therefore to investigate similarities and differences in COM abundances and their ratios among different types of sources to constrain their formation routes. Indeed, surveys have revealed an interesting consistency of O-COM ratios with respect to methanol among different sources. \cite{Coletta2020} summarized single-dish observations of DME and MF in various objects, including pre-stellar cores, star-forming regions, a protostellar shock, and galactic center cores. They find that DME and MF abundances are strongly correlated with a ratio of about 1. \cite{MvG2020} report constant ratios of O-COMs with respect to methanol in five low-mass protostars. For N-bearing species, \cite{Nazari2022_NCOM} also find rather constant ratios with respect to methyl cyanide (CH$_3$CN) in more than 30 high-mass sources, though some species such as formamide (NH$_2$CHO) show larger scatter in their ratios. The constant abundance ratios revealed by observations are interesting since different objects have different physical environments, which are expected to influence the chemical evolution and alter the ratios among COMs. These similarities suggest that COMs are mainly formed under similar physical conditions, probably on the surfaces of dust grains during the cold pre-stellar phase.

Previous observational studies mostly focused on a small number of well-known sources, 
and many of them did not have enough spatial resolution and sensitivity to overcome beam dilution and optical depth issues. 
So far, a few ALMA surveys with enhanced sensitivities and resolutions on O-COMs in low-mass protostellar regions have been published \citep{MvG2020, Hsu2022}, while results for the high-mass counterparts, especially from ALMA surveys that cover minor isotopologues of methanol, are still lacking. We emphasize the importance of detecting optically thin lines of minor isotopologues of methanol, which requires higher sensitivity, in order to better constrain the column density of methanol and hence the ratios with respect to other O-COMs. There are several ALMA studies on O-COMs in the high-mass counterparts \citep[e.g.,][]{Csengeri2019, El-Abd2019, Mininni2020}, but all of them are case studies without using minor isotopologues of methanol in the analyses, and/or only focus a small set of O-COMs such as the \ce{C2H4O2} isomers. High-mass protostellar regions are not only important to the investigation of COM chemistry under different physical conditions, but given the higher temperatures and the possibly enhanced UV radiation along outflow cavity walls, it is also expected that the gas-phase data will give a better representation of the solid-phase abundances, as icy COMs are expected to be mostly thermally desorbed from dust grains. 
It is also timely to present the results of ALMA observations on high-mass sources after the latest development of a comprehensive model on COM chemistry in hot cores by \cite{Garrod2022}.

In this paper, we present the analysis of O-COMs observed in 14 high-mass star-forming regions in the Complex Chemistry in hot Cores with ALMA (CoCCoA) survey. 
The analysis is focused on six O-bearing COMs: \ce{C2H5OH}, \ce{CH3CHO}, DME, MF, GA, EG (including two conformers, anti and gauche, $a$-EG and $g$-EG), as well as two methanol isotopologues $^{13}$\ce{CH3OH} and CH$_3^{18}$OH. Covering minor isotopologues of methanol is essential because the lines of the main isotopologue, $^{12}$CH$_3^{16}$OH, are likely to suffer from a high optical depth, in which case its column density needs to be inferred from optically thin lines of a minor isotopologue. \ce{C2H5OH} and \ce{CH3CHO} are added to this sample for their potential detections in ices \citep{Schutte1999, Yang2022, McClure2023}. MF, GA, and EG are selected to study their relative abundances, based on laboratory and modeling findings in \cite{Chuang2017} and \cite{Simons2020}. DME is included because of its high abundance observed in star-forming regions.

This work will provide a base to compare to future observations of icy COMs by the \textit{James Webb} Space Telescope (JWST). To verify if COMs are formed in ices, a direct approach is to observe their vibrational absorption features in the infrared. However, this requires very high sensitivity and spectral resolution in the fingerprint wavelength range at $\sim$2--15 $\mu$m, which has only become feasible with the successful operation of JWST. So far, methanol is the only COM that has confirmed detections in interstellar ices, whereas several other solid-phase COMs such as \ce{C2H5OH} and \ce{CH3CHO} were only tentatively identified \citep[see review by][and references therein]{Boogert2015ARAA}. Recently, several JWST teams have started hunting for solid-phase COM features \citep{Yang2022, McClure2023}. Important laboratory measurements of COM infrared spectra are now available for JWST data analyses \citep{TvS2018, TvS2021, Rachid2020, Rachid2021, Rachid2022, Hudson2019, Hudson2020, Hudson2021, Hudson2022, Gerakines2022, Rocha2022}. The ultimate goal is to bridge the gas and grain chemistries and directly relate gas-phase spectra as presented here to solid-phase infrared features, from which we can gain a better understanding about the formation history of COMs. 


\begin{figure*}
   \centering
   \includegraphics[width=0.89\textwidth]{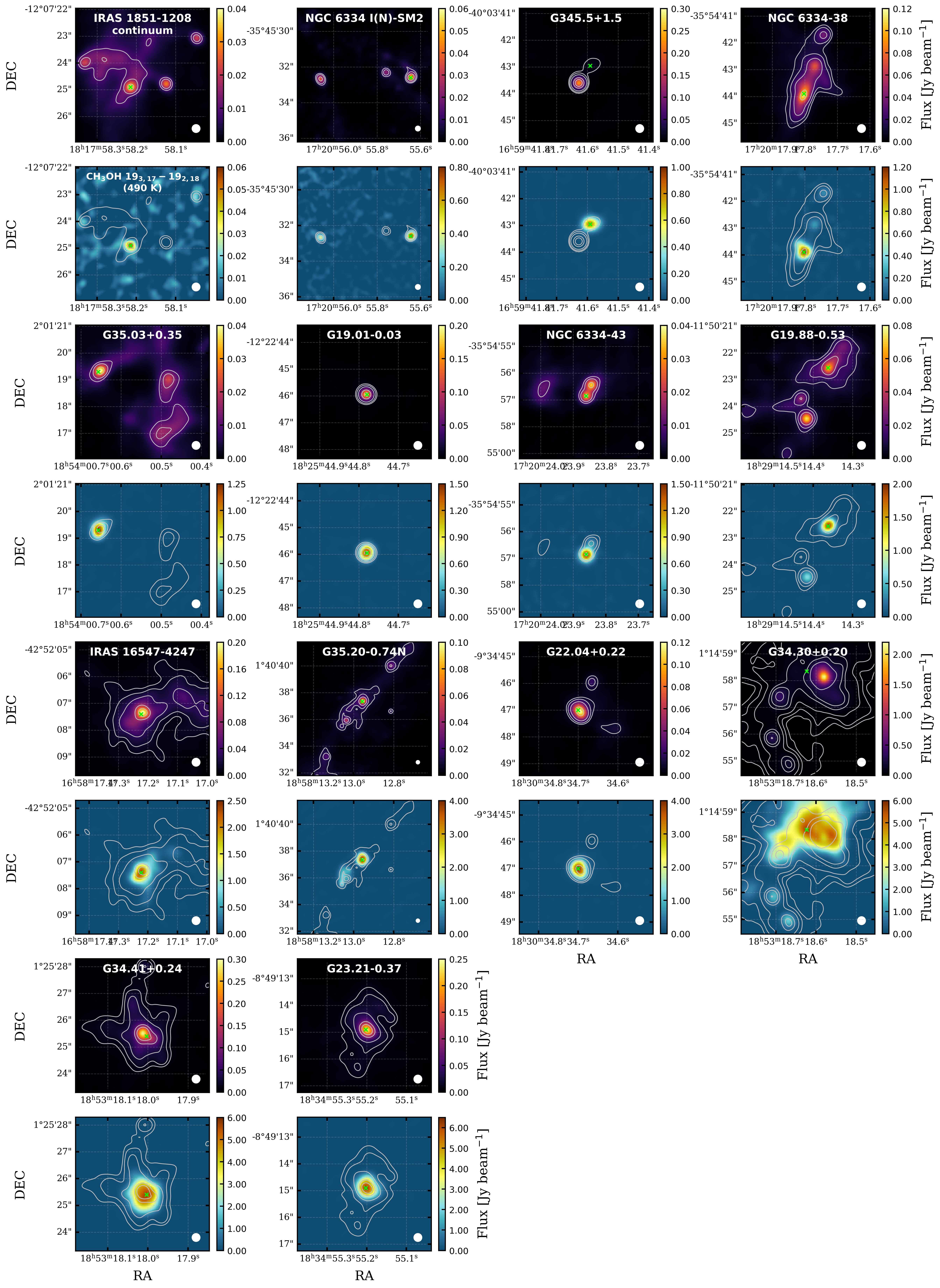}
   \caption{Continuum maps (odd rows, black-yellow) and moment 0 maps of the methanol line at 258.78 GHz (19$_{3,17}$--19$_{2,18}$) with $E_\mathrm{up}$= 490 K (even rows, blue-red) for the 14 high-mass CoCCoA sources. The white contours indicate the continuum emission at 3, 5, 10, 30, 50, and 100$\sigma$ level ($\sigma$ = 0.03 mJy beam$^{-1}$). The peak pixels from where the spectra were extracted are marked by crosses in light green. The sources are sorted by peak flux of methanol emission. The beam size (0.33$''$) is denoted by white ellipses in the lower right of each panel. The field of view for NGC 6334 I(N)-SM2 and G35.20-0.74 N is set to be twice as large as that of others in order to include more emission peaks.}
   \label{fig:maps}
\end{figure*}

\section{Observations} 
\subsection{Overview}
This paper presents early results from a first look at data from the Complex Chemistry in hot Cores with ALMA (CoCCoA) survey (PI: B. McGuire).  A complete overview of the survey will be provided in a separate publication once all observations are completed.  This will be followed by the public release of fully-reduced spectral-line image cubes. When complete, CoCCoA will comprise a dataset targeting 23 high-mass star-forming regions at a spatial resolution of $\sim$0.3$^{\prime\prime}$ using multiple configurations of the ALMA 12-m array as well as the ALMA 7-m array to capture both extended and compact emission. CoCCoA will ultimately include data from ALMA projects 2019.1.00246.S, 2019.2.00112.S, and 2022.1.00499.S.   

The sources were chosen to be within $\sim$1--4\,kpc, corresponding to a linear-scaled spatial resolution of $\sim$300 to 1200 au at the nominal 0.3$^{\prime\prime}$ angular resolution). Other than this distance requirement, the only other criterion used for selection was that the sources have a prior literature report of ``hot core chemistry,'' either described as such or evidenced by the presence of a rich array of emission lines of methanol and other COMs. The goal is to provide a diverse dataset to better sample the phase space of chemical complexity, and avoid biases toward the chemistry seen in the commonly observed extraordinary sources such as Sgr B2 \citep[see, e.g.,][]{El-Abd2019}.

The observations cover two spectral tunings per source in Band 6. The lower tuning covers 238.0--241.7\,GHz while the upper tuning covers 258.0--261.7\,GHz.  All observations are taken with 0.488\,MHz spectral resolution (twice the channel width due to online Hanning smoothing), corresponding to $\sim$0.6\,km\,s$^{-1}$.  These frequency ranges were chosen to maximize both the number of transitions and the range of upper-state rotational energy levels of key target molecules including those highlighted here. These first-look results use data from the upper tuning (258.0--261.7\,GHz) for which observations have been completed for 14 total sources listed in Table~\ref{table:sources}.

\begin{table*}
    \setlength{\tabcolsep}{0.12cm}
    \centering
    \footnotesize
    \caption{Names, phase center coordinates, coordinates where spectra were extracted, and properties of sources analyzed.}    
    \begin{tabular}{l r r r r c c c c c}
     \toprule
                    &   \multicolumn{2}{c}{Phase Center$^{\dagger}$}    &   \multicolumn{2}{c}{Extraction Location$^{\dagger}$} & $L$ & $D$ & $D_\text{GC}$ &  \ce{^16O}/\ce{^18O} & Refs.\\
     Source Name$^{\ddagger}$ & \multicolumn{1}{c}{R.A.} &  \multicolumn{1}{c}{Dec.} & \multicolumn{1}{c}{R.A.} & \multicolumn{1}{c}{Dec.} & (10$^4$ $L_{\odot}$) & (kpc) & (kpc) &  &  \\
     \midrule
     G19.01-0.03    &   18:25:44.80 &   --12:22:45.8 &   18:25:44.78    &   --12:22:45.95   &   1           &   4.0(3)   & 4.4 &  298$\pm$52  &   $L$: 1, $D$: 2\\
     G19.88-0.53    &   18:29:14.57 &   --11:50:23.0 &   18:29:14.36    &   --11:50:22.50   &   0.47        &   3.31     & 5.1 &  335$\pm$60  &   $L$: 3, $D$: 4\\
     G22.04+0.22    &   18:30:34.70 &   --09:34:47.0 &   18:30:34.70    &   --09:34:47.00   &   0.497(26)   &   3.4(5)   & 5.0 &  333$\pm$60  &   $L$: 5, $D$: 5\\
     G23.21-0.37    &   18:34:55.26 &   --08:49:15.3 &   18:34:55.20    &   --08:49:14.70   &   1.3         &   4.6      & 4.2 &  286$\pm$50  &   $L$: 6, $D$: 6\\
     G34.30+0.20    &   18:53:18.54 &   +01:14:57.9 &   18:53:18.62    &   +01:14:58.35   &   4.6         &   1.6      & 6.8 &  436$\pm$80  &   $L$: 7, $D$: 7\\
     G34.41+0.24    &   18:53:17.90 &   +01:25:25.0 &   18:53:18.02    &   +01:25:25.15   &   0.48        &   1.6      & 6.8 &  436$\pm$80  &   $L$: 8, $D$: 8\\
     G35.03+0.35    &   18:54:00.50 &   +02:01:18.0 &   18:54:00.66    &   +02:01:19.30   &   0.63        &   2.32     & 6.3 &  407$\pm$74  &   $L$: 9, $D$: 10\\ 
     G35.20-0.74N   &   18:58:13.00 &   +01:40:36.5 &   18:58:12.96    &   +01:40:37.35   &   3           &   2.2(2)   & 6.4 &  411$\pm$75  &   $L$: 11, $D$: 12\\
     G345.5+1.5     &   16:59:41.63 &   --40:03:43.6 &   16:59:41.59    &   --40:03:42.90   &   4.8         &   1.5      & 6.6 &  426$\pm$78  &   $L$: 13, $D$: 13\\
     IRAS 1851-1208 &   18:17:58.00 &   --12:07:27.0 &   18:17:58.22    &   --12:07:24.90   &   2.2         &   2.9      & 5.4 &  353$\pm$63  &   $L$: 14, $D$: 14\\
     IRAS 16547-4247 &  16:58:17.20 &   --42:52:07.0 &   16:58:17.22    &   --42:52:07.35   &   6.3         &   2.9      & 5.3 &  353$\pm$63  &   $L$: 13, $D$: 13\\
     NGC 6334-38    &   17:20:18.00 &   --35:54:55.0 &   17:20:17.80    &   --35:54:43.85   &   $<$20         &   1.7      & 6.4 &  412$\pm$75  &   $L$: 15, $D$: 15\\
     NGC 6334-43    &   17:20:23.00 &   --35:54:55.0 &   17:20:23.86    &   --35:54:56.90   &   $<$20         &   1.7      & 6.4 &  412$\pm$75  &   $L$: 15, $D$: 15\\
     NGC 6334 I(N)-SM2  &   17:20:55.00 &   --35:45:40.0 &   17:20:55.64    &   --35:45:32.60   &   0.07        &   1.3(1)   & 6.8 &  434$\pm$79  &   $L$: 16, $D$: 17\\
    \bottomrule
    \end{tabular}
    \begin{minipage}{\textwidth}
    $^\dagger$Coordinates given in J2000, Right Ascension (R.A) in units of $^\circ$:$^\prime$:$^{\prime\prime}$, Declination (Dec.) in units of hh:mm:ss.\\
    $^\ddagger$Other names exist for many of these, including different designations for the larger star-forming complex and for individual sub-sources within the complex.\\
    \textbf{References --} 
    [1] \citealt{Cyganowski2011},
    [2] \citealt{Williams2022},
    [3] \citealt{Issac2020},
    [4] \citealt{Ge2014},
    [5] \citealt{Towner2021},
    [6] \citealt{Tang2018},
    [7] \citealt{Csengeri2022},
    [8] \citealt{Konig2017},
    [9] \citealt{Beltran2014},
    [10] \citealt{Wu2014},
    [11] \citealt{Sanchez-Monge2013},
    [12] \citealt{Zhang2009},
    [13] \citealt{Faundez2004},
    [14] \citealt{Maud2015},
    [15] \citealt{WidicusWeaver2017},
    [16] \citealt{Sandell2000},
    [17] \citealt{Chibueze2014}.
    \end{minipage}
    \label{table:sources}
\end{table*}


\subsection{Data reduction}

The 12m-array CoCCoA data presented here from project 2019.1.00246.S, were observed between March 24, 2021 and April 2, 2021 using configuration C-5. The data were calibrated using the Cycle 8 version of the ALMA Pipeline (CASA version 6.2), including corrections for renormalization issues larger than 2\%. Due to the copious line emission emanating from the massive protostars within each targeted cluster, the default parameters of the {\tt findContinuum.py} procedure used by the pipeline to identify line-free channels did not yield optimal results (i.e., significant line contamination was present). Therefore, {\tt findContinuum.py} (available in the {\tt extern} directory of the pipeline distribution) was run manually with a few key parameters (primarily {\tt sigmaFindContinuum}\footnote{The {\tt sigmaFindContinuum} parameter sets how far above the corrected median of the baseline channels to place the initial threshold (the baseline channels ({\tt nBaselineChannels}) were drawn from the lowest $10\%$ of channels after excluding outliers). The {\tt narrow} parameter was also set to 2, instead of the default of 4, in order to preserve as many narrow windows as possible.}) adjusted in order to minimize the line contamination. 

After optimization of the channels used for continuum subtraction, the line-free channels were used to image the continuum and perform an iterative self-calibration \citep{Brogan2018b}. The self-calibration solutions were also applied to the continuum-subtracted line datasets. The imaging of both the continuum and data cubes employed the {\tt multiscale} functionality of CASA's {\tt tclean} with scales of 0, 5, and 15 (the {\tt cellsize} was chosen to oversample the beam on the smallest axis by a factor of $\sim$5). The {\tt robust} parameter employed is 0.5, which yields an angular resolution near $0.3\arcsec$; both the continuum and line images were subsequently convolved to exactly $0.3\arcsec$. The maximum recoverable scale of these data is ${\sim}2.5\arcsec$.

Considering that the data in the lower tuning are still under reduction, only the upper tuning was used for further analyses. The quality of the ALMA pipeline data of the lower tuning is good enough to see how many COM lines are included in each tuning. Among the 14 sources, G19.88-0.53 is taken as an example for line identification, since its spectrum shows less line blending despite the large line intensities. This ensures a sufficiently large number of detected lines without suffering too much from blending issues. The upper tuning covers about 70\% of the COM lines found in both tunings (not including \ce{CH3OH}), while the lower one is more abundant in strong lines of \ce{CH3OH} and CH$_3$CN, which are not directly relevant to our project. 
Therefore, we consider our results to be robust using only the data in the upper tuning.

\begin{sidewaysfigure*}[p]
   \centering
   \includegraphics[width=0.93\textwidth]{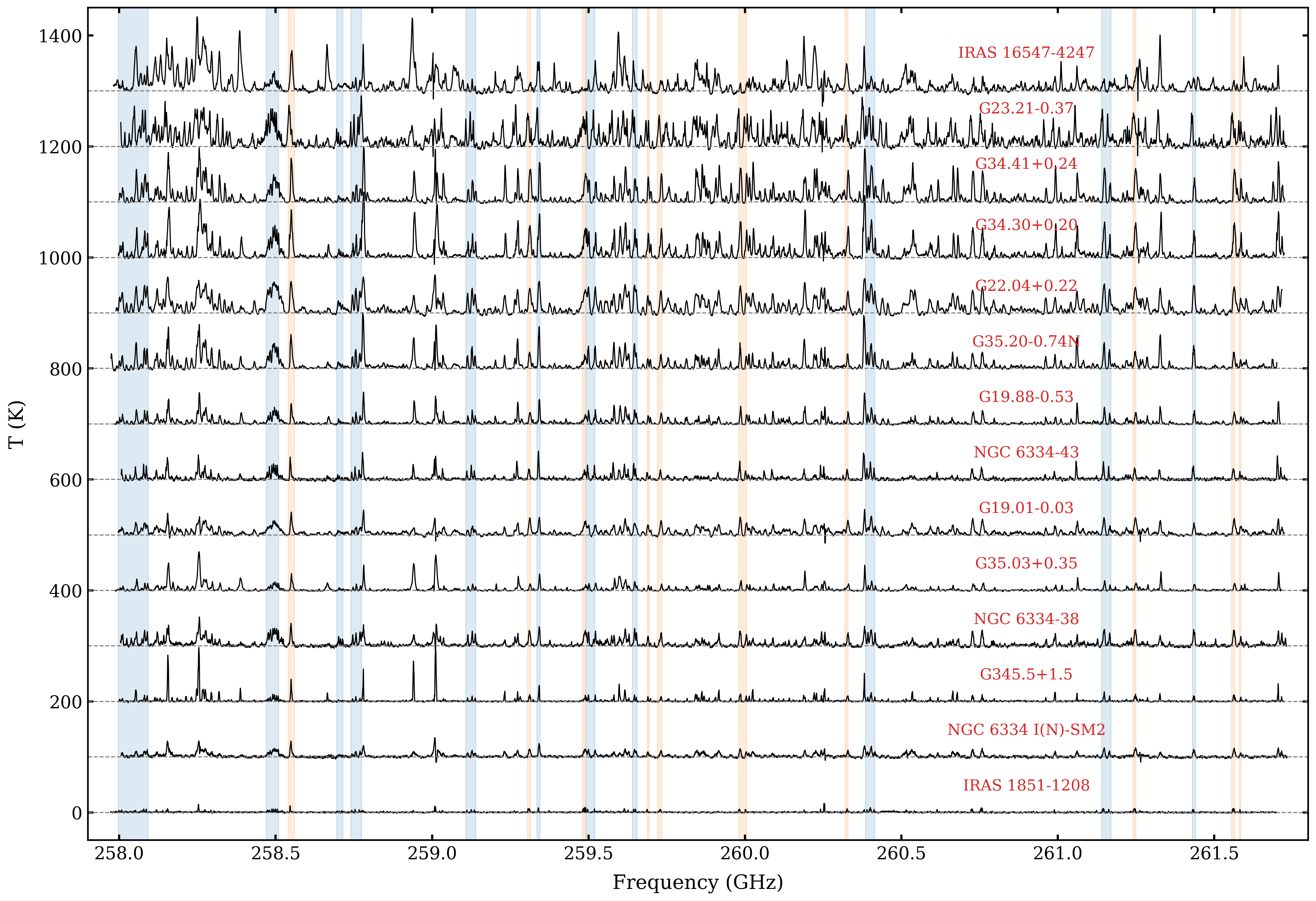}
      \caption{An overview of the spectra in the upper tuning (258.0--261.7 GHz) of the 14 high-mass sources observed in the CoCCoA survey. The line intensity has been converted from flux in Jy beam $^{-1}$ to brightness temperature in K using the Rayleigh-Jean approximation. The spectra are sorted by the average line intensities from bottom to top and each one is offset by 100 K. The shaded regions in blue and orange indicate the emission lines of MF and DME, respectively.}
         \label{fig:overview}
\end{sidewaysfigure*}

\begin{sidewaystable*}
    \setlength{\tabcolsep}{0.1cm}
    \centering
    \caption{Fitting results of CH$_3^{18}$OH, \ce{CH3CHO}, \ce{C2H5OH}, \ce{CH3OCH3} (DME), \ce{CH3OCHO} (MF), and \ce{CH2OHCHO} (GA).}
    \begin{tabular}{l c c c c c c c c c c c c c}
     \toprule
      & \ce{CH_3OH$^a$} & \multicolumn{4}{c}{CH$_3^{18}$OH (CDMS)$^b$} & \multicolumn{4}{c}{\ce{CH3CHO} (JPL)} & \multicolumn{4}{c}{\ce{C2H5OH} (CDMS)} \\
     \cmidrule(lr{0.5em}){2-2}\cmidrule(lr{0.5em}){3-6}\cmidrule(lr{0.5em}){7-10}\cmidrule(lr{0.5em}){11-14}
      & $N$ & \multicolumn{1}{c}{$N$} & \multicolumn{1}{c}{$T_\text{ex}$} & \multicolumn{1}{c}{FWHM} & \multicolumn{1}{c}{$v_\text{lsr}$} & \multicolumn{1}{c}{$N$} & \multicolumn{1}{c}{$T_\text{ex}$} & \multicolumn{1}{c}{FWHM} & \multicolumn{1}{c}{$v_\text{lsr}$} & \multicolumn{1}{c}{$N$} & \multicolumn{1}{c}{$T_\text{ex}$} & \multicolumn{1}{c}{FWHM} & \multicolumn{1}{c}{$v_\text{lsr}$}\\
     Source Name & cm$^{-2}$ & \multicolumn{1}{c}{(cm$^{-2}$)} & \multicolumn{1}{c}{(K)} & \multicolumn{1}{c}{(km s$^{-1}$)} & \multicolumn{1}{c}{(km s$^{-1}$)} & \multicolumn{1}{c}{(cm$^{-2}$)} & \multicolumn{1}{c}{(K)} & \multicolumn{1}{c}{(km s$^{-1}$)} & \multicolumn{1}{c}{(km s$^{-1}$)} & \multicolumn{1}{c}{(cm$^{-2}$)} & \multicolumn{1}{c}{(K)} & \multicolumn{1}{c}{(km s$^{-1}$)} & \multicolumn{1}{c}{(km s$^{-1}$)}\\
     \midrule
     G19.01-0.03 (c1)$^c$ & (3.6$\pm$1.1)$\times10^{18}$ & (1.2$\pm$0.3)$\times10^{16}$ & [180]$^d$ & 3.0 & 63.0 & & & & & (3.1$\pm$0.8)$\times10^{16}$ & 160$\pm$40 & 3.2 & 62.3 \\
     G19.01-0.03 (c2)$^c$ & & & & & & (8.0$\pm$3.0)$\times10^{15}$ & [150] & 3.2 & 58.5 & (3.3$\pm$0.6)$\times10^{16}$ & 160$\pm$40 & 3.2 & 58.5 \\
     G19.88-0.53 & (4.4$\pm$1.3)$\times10^{18}$ & (1.3$\pm$0.3)$\times10^{16}$ & [180] & 2.7 & 46.2 & (4.0$\pm$1.5)$\times10^{15}$ & 100$\pm$30 & 3.0 & 46.4 & (5.5$\pm$1.5)$\times10^{16}$ & 170$\pm$30 & 3.5 & 46.4 \\
     G22.04+0.22 & (1.3$\pm$0.4)$\times10^{19}$ & (4.0$\pm$1.0)$\times10^{16}$ & [130] & 4.5 & 53.5 & (4.5$\pm$1.0)$\times10^{16}$ & 180$\pm$20 & 4.5 & 52.5 & (1.7$\pm$0.4)$\times10^{17}$ & [160] & 5.5 & 53.0 \\
     G23.21-0.37 & (2.6$\pm$0.7)$\times10^{19}$ & (9.0$\pm$2.0)$\times10^{16}$ & [170] & 3.0 & 76.8 & (4.0$\pm$0.6)$\times10^{16}$ & 150$\pm$30 & 3.5 & 77.0 & (2.4$\pm$0.4)$\times10^{17}$ & 160$\pm$40 & 3.3 & 77.2 \\
     G34.30+0.20 (c1) & (9.6$\pm$2.8)$\times10^{18}$ & (2.2$\pm$0.5)$\times10^{16}$ & [140] & 3.5 & 56.5 & (1.3$\pm$0.3)$\times10^{16}$ & 130$\pm$20 & 3.5 & 56.3 & (1.2$\pm$0.3)$\times10^{17}$ & 170$\pm$30 & 3.3 & 56.5 \\
     G34.30+0.20 (c2) & & & & & & & & & & (5.0$\pm$1.0)$\times10^{16}$ & 170$\pm$30 & 3.0 & 60.0 \\
     G34.41+0.24 (c1) & (7.9$\pm$2.3)$\times10^{18}$ & (1.8$\pm$0.4)$\times10^{16}$ & [160] & 3.5 & 60.0 & (1.2$\pm$0.3)$\times10^{16}$ & 140$\pm$20 & 3.0 & 60.0 & (8.0$\pm$2.0)$\times10^{16}$ & 180$\pm$20& 3.0 & 60.0\\
     G34.41+0.24 (c2) & & & & & & (9.0$\pm$2.0)$\times10^{15}$ & [160] & 3.0 & 56.8 & (8.0$\pm$2.0)$\times10^{15}$ & 180$\pm$20 & 3.0 & 56.8 \\
     G35.03+0.35 & (1.8$\pm$0.7)$\times10^{18}$ & (4.5$\pm$1.5)$\times10^{15}$ & [150] & 2.5 & 44.8 & (2.5$\pm$1.0)$\times10^{15}$ & 110$\pm$30 & 4.0 & 44.8 & (2.8$\pm$0.7)$\times10^{16}$ & 120$\pm$40 & 3.5 & 44.6 \\
     G35.20-0.74N (c1) & (8.2$\pm$2.5)$\times10^{18}$ & (2.0$\pm$0.5)$\times10^{16}$ & [140] & 3.5 & 31.0 & (5.0$\pm$1.0)$\times10^{15}$ & 100$\pm$20 & 3.0 & 31.0 & (8.5$\pm$1.5)$\times10^{16}$ & 160$\pm$20 & 4.0 & 31.2 \\
     G35.20-0.74N (c2) & & & & & & (1.8$\pm$0.5)$\times10^{15}$ & 100$\pm$20 & 1.7 & 33.5 & (1.8$\pm$0.4)$\times10^{16}$ & 160$\pm$20 & 1.5 & 33.7 \\
     G345.5+1.5 & (1.4$\pm$0.3)$\times10^{18}$ & (3.2$\pm$0.5)$\times10^{15}$ & 120$\pm$30 & 2.0 & -15.8 & (6.0$\pm$2.0)$\times10^{14}$ & [120] & 2.0 & -16.3 & (1.4$\pm$0.3)$\times10^{16}$ & 140$\pm$40 & 2.0 & -16.3\\
     IRAS 1851-1208 & (5.3$\pm$2.0)$\times10^{17}$ & (1.5$\pm$0.5)$\times10^{15}$ & [180] & 2.0 & 35.0 & (9.0$\pm$2.0)$\times10^{14}$ & 150$\pm$30 & 2.0 & 35.0 & (2.0$\pm$0.5)$\times10^{15}$ & [180] & 1.9 & 35.3 \\
     IRAS 16547-4247 & (2.1$\pm$0.8)$\times10^{18}$ & (6.0$\pm$2.0)$\times10^{15}$ & [200] & 2.2 & -35.5 & (2.0$\pm$0.4)$\times10^{16}$ & 150$\pm$30 & 3.0 & -35.5 & (6.0$\pm$1.5)$\times10^{16}$ & 140$\pm$40 & 3.0 & -35.2 \\
     NGC 6334-38 & (4.9$\pm$1.5)$\times10^{18}$ & (1.2$\pm$0.3)$\times10^{16}$ & [120] & 2.7 & -5.0 & (1.8$\pm$0.5)$\times10^{15}$ & 100$\pm$30 & 2.8 & -5.1 & (1.6$\pm$0.3)$\times10^{16}$ & 190$\pm$50 & 2.6 & -5.2\\
     NGC 6334-43 & (5.4$\pm$1.6)$\times10^{18}$ & (1.3$\pm$0.3)$\times10^{16}$ & [160] & 3.2 & 0.8 & (2.3$\pm$0.5)$\times10^{15}$ & 100$\pm$50 & 3.0 & 1.0 & (2.0$\pm$0.5)$\times10^{16}$ & 180$\pm$30 & 3.0 & 0.8 \\
     NGC 6334 I(N)-SM2 & (3.9$\pm$1.1)$\times10^{18}$ & (9.0$\pm$2.0)$\times10^{15}$ & [150] & 3.0 & -3.0 & (2.8$\pm$1.0)$\times10^{15}$ & 80$\pm$30 & 3.0 & -2.5 & (1.3$\pm$0.4)$\times10^{16}$ & 180$\pm$20& 3.0 & -3.0 \\
     \midrule
      & & \multicolumn{4}{c}{\ce{CH3OCH3} (CDMS)} & \multicolumn{4}{c}{\ce{CH3OCHO} (JPL)} & \multicolumn{4}{c}{\ce{CH2OHCHO} (CDMS)} \\
     \cmidrule(lr{0.5em}){3-6}\cmidrule(lr{0.5em}){7-10}\cmidrule(lr{0.5em}){11-14}
      & & \multicolumn{1}{c}{$N$} & \multicolumn{1}{c}{$T_\text{ex}$} & \multicolumn{1}{c}{FWHM} & \multicolumn{1}{c}{$v_\text{lsr}$} & \multicolumn{1}{c}{$N$} & \multicolumn{1}{c}{$T_\text{ex}$} & \multicolumn{1}{c}{FWHM} & \multicolumn{1}{c}{$v_\text{lsr}$} & \multicolumn{1}{c}{$N$} & \multicolumn{1}{c}{$T_\text{ex}$} & \multicolumn{1}{c}{FWHM} & \multicolumn{1}{c}{$v_\text{lsr}$} \\
     Source Name & & \multicolumn{1}{c}{(cm$^{-2}$)} & \multicolumn{1}{c}{(K)} & \multicolumn{1}{c}{(km s$^{-1}$)} & \multicolumn{1}{c}{(km s$^{-1}$)} & \multicolumn{1}{c}{(cm$^{-2}$)} & \multicolumn{1}{c}{(K)} & \multicolumn{1}{c}{(km s$^{-1}$)} & \multicolumn{1}{c}{(km s$^{-1}$)} & \multicolumn{1}{c}{(cm$^{-2}$)} & \multicolumn{1}{c}{(K)} & \multicolumn{1}{c}{(km s$^{-1}$)} & \multicolumn{1}{c}{(km $s^{-1}$)}\\
     \midrule
     G19.01-0.03 (c1) & & (1.1$\pm$0.2)$\times10^{17}$ & 130$\pm$20 & 3.8 & 61.8 & (5.0$\pm$1.0)$\times10^{16}$ & 180$\pm$40 & 3.5 & 62.5 & (7.0$\pm$2.3)$\times10^{15}$ & <120 & 3.5 & 61.5 \\
     G19.01-0.03 (c2) & & (1.0$\pm$0.2)$\times10^{17}$ & 130$\pm$20 & 3.5 & 57.8 & (6.0$\pm$1.0)$\times10^{16}$ & 180$\pm$40 & 3.5 & 58.5 & (7.0$\pm$2.3)$\times10^{15}$ & <120 & 3.5 & 58.5 \\
     G19.88-0.53 & & (1.1$\pm$0.3)$\times10^{17}$ & 130$\pm$20 & 3.5 & 46.0 & (1.1$\pm$0.3)$\times10^{17}$ & 180$\pm$40 & 4.0 & 46.0 & (6.0$\pm$1.3)$\times10^{15}$ & 160$\pm$40 & 2.5 & 46.4 \\
     G22.04+0.22 & & (3.5$\pm$0.5)$\times10^{17}$ & 120$\pm$20 & 5.5 & 52.3 & (7.0$\pm$2.0)$\times10^{17}$ & 130$\pm$30 & 5.5 & 52.5 & (3.9$\pm$0.8)$\times10^{16}$ & 150$\pm$50 & 5.0 & 52.5\\
     G23.21-0.37 & & (4.0$\pm$0.8)$\times10^{17}$ & 120$\pm$10 & 3.0 & 77.2 & (3.0$\pm$0.5)$\times10^{17}$ & 170$\pm$30 & 3.3 & 77.2 & (3.7$\pm$0.8)$\times10^{16}$ & 200$\pm$40 & 3.5 & 76.8 \\ 
     G34.30+0.20 (c1) & & (3.7$\pm$0.6)$\times10^{17}$ & 130$\pm$20 & 3.5 & 56.5 & (1.6$\pm$0.3)$\times10^{17}$ & 140$\pm$40 & 3.2 & 56.5 & (1.0$\pm$0.3)$\times10^{16}$ & 180$\pm$30 & 3.5 & 55.5 \\
     G34.30+0.20 (c2) & & (1.1$\pm$0.3)$\times10^{17}$ & 130$\pm$20 & 3.0 & 60.0 & (5.0$\pm$1.5)$\times10^{16}$ & 140$\pm$40 & 2.5 & 60.0 & & & & \\
     G34.41+0.24 (c1) & & (2.2$\pm$0.3)$\times10^{17}$ & 130$\pm$20 & 3.3 & 60.0 & (1.0$\pm$0.2)$\times10^{17}$ & 160$\pm$20 & 3.0 & 60.0 & (1.1$\pm$0.3)$\times10^{16}$ & 160$\pm$40 & 3.0 & 57.0 \\
     G34.41+0.24 (c2) & & (2.2$\pm$0.3)$\times10^{17}$ & 130$\pm$20 & 3.3 & 56.5 & (1.0$\pm$0.2)$\times10^{17}$ & 160$\pm$20 & 3.0 & 56.8 & & & & \\
     G35.03+0.35 & & (6.3$\pm$1.7)$\times10^{16}$ & 130$\pm$20 & 3.7 & 45.0 & (4.7$\pm$1.3)$\times10^{16}$ & 150$\pm$50 & 4.0 & 45.3 & (2.0$\pm$0.7$\times10^{15}$ & [150] & 3.5 & 44.5 \\
     G35.20-0.74N (c1) & & (1.3$\pm$0.3)$\times10^{17}$ & 130$\pm$20 & 4.0 & 31.0 & (1.5$\pm$0.3)$\times10^{17}$ & 140$\pm$30 & 4.0 & 31.2 & (7.4$\pm$2.5)$\times10^{15}$ & [140] & 3.5 & 31.2 \\
     G35.20-0.74N (c2) & & (3.5$\pm$1.0)$\times10^{16}$ & 130$\pm$20 & 1.7 & 33.7 & (2.0$\pm$0.5)$\times10^{16}$ & 140$\pm$30 & 1.2 & 33.7 & & & & \\
     G345.5+1.5 & & (4.3$\pm$0.7)$\times10^{16}$ & 120$\pm$20 & 2.0 & -16.1 & (2.1$\pm$0.4)$\times10^{16}$ & 120$\pm$20 & 2.2 & -15.8 & <1.4 $\times10^{15}$ & [120] & 2.2 & -15.8\\
     IRAS 1851-1208 & & (2.6$\pm$0.5)$\times10^{16}$ & 110$\pm$20 & 2.0 & 35.1 & (1.4$\pm$0.2)$\times10^{16}$ & 180$\pm$40 & 2.0 & 35.2 & <7.4$\times10^{14}$ & [180] & 1.7 & 35.0 \\
     IRAS 16547-4247 & & (6.0$\pm$1.5)$\times10^{16}$ & 180$\pm$20 & 2.7 & -35.5 & (8.0$\pm$2.0)$\times10^{16}$ & 200$\pm$20 & 3.3 & -35.5 & (1.5$\pm$0.3)$\times10^{16}$ & [200] & 3.0 & -35.5 \\
     NGC 6334-38 & & (1.5$\pm$0.3)$\times10^{17}$ & 110$\pm$20 & 3.0 & -5.2 & (3.0$\pm$0.5)$\times10^{17}$ & 120$\pm$30 & 3.0 & -5.2 & (5.0$\pm$1.2)$\times10^{15}$ & 140$\pm$70 & 2.8 & -5.2 \\
     NGC 6334-43 & & (9.5$\pm$2.5)$\times10^{16}$ & 120$\pm$20 & 3.0 & 0.7 & (8.5$\pm$2.0)$\times10^{16}$ & 160$\pm$40 & 3.0 & 0.8 & (2.7$\pm$0.7)$\times10^{15}$ & [160] & 3.0 & 0.8 \\
     NGC 6334 I(N)-SM2 & & (8.0$\pm$2.0)$\times10^{16}$ & 100$\pm$20 & 4.5 & -3.0 & (6.0$\pm$1.0)$\times10^{16}$ & 150$\pm$30 & 5.0 & -2.5 & (4.4$\pm$1.5)$\times10^{15}$ & 180$\pm$40 & 3.0 & -2.5 \\
     \bottomrule
    \end{tabular}
    \begin{minipage}{\textwidth}
    $^a$The column density of the major isotopologue of methanol is inferred from the minor isotopologue CH$_3^{18}$OH and the isotopic ratio between \ce{^16O} and \ce{^18O}.\ 
    $^b$The spectroscopic databases used in the fitting.\ 
    $^c$Component 1 and 2 (some sources exhibit two components in their spectra).\ 
    $^d$$T_\mathrm{ex}$ in square brackets means that it was fixed to that value during the fitting. In this case, the uncertainties of $N$ were determined only for this particular $T_\mathrm{ex}$.\ 
    \end{minipage}
    \label{table:results_1}
\end{sidewaystable*}

\begin{table*}[t]
    \setlength{\tabcolsep}{0.12cm}
    \centering
    \caption{Fitting results of $a$-\ce{(CH2OH)2} ($a$-EG) and $g$-\ce{(CH2OH)2} ($g$-EG).}    
    \begin{tabular}{l c c c c c c c c}
     \toprule
      & \multicolumn{4}{c}{\ce{a-(CH_2OH)_2} (CDMS)} & \multicolumn{4}{c}{\ce{g-(CH_2OH)_2} (CDMS)}\\
     \cmidrule(lr{0.5em}){2-5}\cmidrule(lr{0.5em}){6-9}
      & \multicolumn{1}{c}{$N$} & \multicolumn{1}{c}{$T_\text{ex}$} & \multicolumn{1}{c}{FWHM} & \multicolumn{1}{c}{$v_\text{lsr}$} & \multicolumn{1}{c}{$N$} & \multicolumn{1}{c}{$T_\text{ex}$} & \multicolumn{1}{c}{FWHM} & \multicolumn{1}{c}{$v_\text{lsr}$} \\
     Source Name & \multicolumn{1}{c}{(cm$^{-2}$)} & \multicolumn{1}{c}{(K)} & \multicolumn{1}{c}{(km s$^{-1}$)} & \multicolumn{1}{c}{(km s$^{-1}$)} & \multicolumn{1}{c}{(cm$^{-2}$)} & \multicolumn{1}{c}{(K)} & \multicolumn{1}{c}{(km s$^{-1}$)} & \multicolumn{1}{c}{(km s$^{-1}$)} \\
     \midrule
     G19.01-0.03 & (4.0$\pm$0.8)$\times10^{16}$ & [180] & 4.0 & 62.0 & (2.0$\pm$0.4)$\times10^{16}$ & [180] & 3.5 & 61.0 \\ 
     G19.88-0.53 & (5.7$\pm$1.2)$\times10^{16}$ & 220$\pm$20 & 3.0 & 46.7 & (2.9$\pm$0.7)$\times10^{16}$ & [220] & 3.0 & 46.7\\
     G22.04+0.22 & (5.9$\pm$1.5)$\times10^{16}$ & 160$\pm$40 & 4.5& 52.0 & (4.6$\pm$0.9)$\times10^{16}$ & [160] & 4.5 & 52.0 \\
     G23.21-0.37 & (1.4$\pm$0.4)$\times10^{17}$ & 180$\pm$40 & 3.8 & 77.0 &  (1.0$\pm$0.4)$\times10^{17}$ & [180] & 3.5 & 76.8 \\
     G34.30+0.20 & (3.1$\pm$0.7)$\times10^{16}$ & [140] & 3.5 & 55.5 & (2.2$\pm$0.5)$\times10^{16}$ & [140] & 3.5 & 55.5 \\
     G34.41+0.24 & (1.8$\pm$0.5)$\times10^{16}$ & 160$\pm$40 & 3.0 & 60.8 & (8.2$\pm$1.8)$\times10^{15}$ & [160] & 3.0 & 60.0 \\
     G35.03+0.35 & (4.3$\pm$1.0)$\times10^{16}$ & 220$\pm$40 & 3.5 & 43.5 & (1.9$\pm$0.7)$\times10^{16}$ & [220] & 3.5 & 43.5 \\
     G35.20-0.74N & (5.3$\pm$1.6)$\times10^{16}$ & 180$\pm$40 & 4.0 & 30.6 & (2.6$\pm$0.8)$\times10^{16}$ & [180] & 3.5 & 30.6 \\
     G345.5+1.5 & <3.3$\times10^{15}$ & [120] & 2.0 & -15.0 & <1.6$\times10^{15}$ & [120] & 2.0 & -15.0 \\
     IRAS 1851-1208 & <1.0$\times10^{15}$ & [180] & 2.0 & 35.0 & <2.0$\times10^{15}$ & [180] & 2.0 & 35.0 \\
     IRAS 16547-4247 & (9.0$\pm$2.0)$\times10^{16}$ & 180$\pm$40 & 3.2 & -35.5 & (5.0$\pm$1.0)$\times10^{16}$ & [180] & 3.0 & -35.7 \\
     NGC 6334-38 & (6.5$\pm$1.6)$\times10^{15}$ & [120] & 2.8 & -5.2 & (6.5$\pm$2.4)$\times10^{15}$ & [120] & 2.8 & -5.2 \\
     NGC 6334-43 & (2.1$\pm$0.6)$\times10^{16}$ & 200$\pm$40 & 3.0 & 0.8 & (1.5$\pm$0.4)$\times10^{16}$ & [200] & 3.0 & 0.7\\
     NGC 6334 I(N)-SM2 & (9.0$\pm$2.0)$\times10^{15}$ & 180$\pm$50 & 4.0 & -3.0 & <5.0 $\times10^{15}$ & [180] & 3.0 & -3.0\\
     \bottomrule
    \end{tabular}
    \label{table:results_2}
\end{table*}

\section{Methods}\label{sec_methods}
\subsection{Spectral analyses}\label{sec_spectral_analyses}
Figure~\ref{fig:maps} shows the continuum maps and the integrated intensity (moment 0) maps of the \ce{CH3OH} 19$_{3,17}$--19$_{2,18}$ transition with upper energy $E_\mathrm{up}$ = 490 K and Einstein $A$ coefficient $A_\mathrm{ij} = 9.27\times10^{-5}$ s$^{-1}$. This particular methanol line was chosen because it is unblended and has the lowest $E_\mathrm{up}$ and the highest $A_\mathrm{ij}$ in the upper tuning, which is expected to show the most extended methanol emission. Figure~\ref{fig:overview} presents an overview of the spectra in the upper tuning, with several representative line features of MF and DME indicated by shaded areas. The spectra are extracted from the peak pixel of the methanol intensity maps. 
The moment 0 maps of other selected O-COMs of one example source G19.88-0.53 are shown in Fig. \ref{fig:OCOM_mom0_G19.88}, which confirmed that their emission peaks at similar regions to the methanol emission. In most sources, the methanol emission peaks at the same location as the continuum. However, in some bright sources such as G34.30+0.20 (Fig.~\ref{fig:maps}), the methanol emission shows a ring shape around the continuum peak, which is likely due to the high optical depth of dust \citep{DeSiomone2020, MvG2022}. In this case, we picked the brightest pixel on the ring to extract the spectra. For the two bright sources G34.41+0.24 and G23.21-0.37, the methanol emission peaks at the same location as the continuum, but the spectrum extracted from the central pixel has too much line blending, so we deliberately chose a pixel offset from the actual peak ($\sim$9 pixels offset for G34.41 and $\sim$3 pixels offset for G23.21). 

After extracting all the spectra, we performed line identification and spectral fitting using the spectral analysis software CASSIS\footnote{\url{http://cassis.irap.omp.eu/}} \citep{Vastel2015CASSIS}. The spectroscopic data in CASSIS are taken from two databases: the Jet Propulsion Laboratory database \citep[JPL;][]{Pickett1998} and the Cologne Database for Molecular Spectroscopy \citep[CDMS;][]{Muller2001, Muller2005, Endres2016}. Detailed references for each species can be found in the Appendix A of \cite{MvG2020}. Vibrational corrections were applied to GA and EG \citep[more information can be found in Sect. 5 of][]{Jorgensen2016}. Some species only have data available in one database, while others are included in both databases. In the latter case, we used the database with smaller uncertainties in central frequency or that has been used more frequently in previous work. We first went through the detection inventory of the PILS survey \citep{Jorgensen2016} and checked for each species whether all the transitions in the databases have corresponding line features in the observed spectra. 
The detection results are presented in Sect. \ref{sec_spectra_detection}.

In the next step, we chose the six O-COMs and the two methanol isotopologues ($^{13}$\ce{CH3OH} and CH$_3^{18}$OH) mentioned in Sect. \ref{sec_introduction} for detailed spectral fitting using CASSIS. The two conformers of EG ($a$-EG and $g$-EG) were fitted separately due to their different sets of transitions. The column density ($N$), excitation temperature ($T_\mathrm{ex}$), and full width half maximum (FWHM) of each molecule were fitted for each source. Here we assume one $T_\mathrm{ex}$ for each species, that is, the populations of all levels can be characterized by a single $T_\mathrm{ex}$, which is often called “local thermodynamic equilibrium (LTE)” in radio astronomy. However, the LTE here does not necessarily refer to its strict definition that $T_\mathrm{ex}$ approaches the kinetic temperature $T_\mathrm{kin}$ under high-density conditions. In fact, COMs are likely to be subthermally excited ($T_\mathrm{ex}$ < $T_\mathrm{kin}$) in hot cores, while the observed lines can still be well characterized by one $T_\mathrm{ex}$. Examples provided in Fig. 6 of \cite{Johnstone2003} and Fig. 8 of \cite{Jorgensen2016} show that for the case of methanol, densities of 10$^9$--10$^{10}$ cm$^{-3}$ are needed for thermalization. Correspondingly, physical models of the envelopes of high-mass protostars \citep[e.g.,][]{vanderTak2013} indicate densities of 10$^7$--10$^9$ cm$^{-3}$ at temperatures of 100--300 K on scales of a few hundred au. At such densities, the level populations of COMs may not yet be fully in LTE, while the fitting results in Sect. \ref{sec_results} show that the single-$T_\mathrm{ex}$ assumption is reasonable and works well. 

We adopted two methods for fitting the spectra: $\chi^2$ minimization or visual inspection when the former is not applicable. For each source, the radial velocity $v\mathrm{_{lsr}}$ and FWHM are determined for each species based on strong unblended lines. The uncertainties of $v\mathrm{_{lsr}}$ and FWHM are smaller than 0.5 km s$^{-1}$. The difference of $v\mathrm{_{lsr}}$ among the O-COMs is within 1 km s$^{-1}$ in most cases. 
For sources where the lines are narrow and unblended, grid-fitting was used to determine the best-fit values as well as the uncertainties by calculating $\chi^2$ in the parameter space and making the contour plot on the $N$--$T_\mathrm{ex}$ plane \citep[also see Sect. 3.1 in][]{MvG2020}. For each species, $\chi^2$ was calculated from the difference between the LTE model and the observed spectrum around unblended lines. 
We started with a sparse grid with broad ranges of $N$, $T_\mathrm{ex}$, and FWHM and large intervals between the grid points. We then gradually narrowed down to smaller ranges and smaller intervals. Finally, we ended up with a fine grid around the best-fit grid point, from which we could make the contour plot and estimate the 2$\sigma$ uncertainties. This grid-fitting method works well with weaker sources where most of the lines are unblended. However, for bright sources where the lines are very broad and blended, grid-fitting does not converge to a solution and the results need to be visually inspected. 

When it came to fitting by visual inspection, we started with an initial guess of the parameters and adjusted them to a better fit until no improvement could be made. This is more efficient and reliable for complex spectra with blended lines, since we can monitor the change intuitively and interactively. This method was also adopted by \cite{MvG2020} and \cite{Nazari2021,Nazari2022_NCOM}. Uncertainties were estimated by comparisons shown in Fig.~\ref{fig:uncertainty}, in which the three panels correspond to the upper/lower limits and the best-fit. We can see in panel (a) that with an underestimated $T_\mathrm{ex}$, transitions with low $E_\mathrm{up}$ tend to overestimate the observation while the high-$E_\mathrm{up}$ lines are normally fit, and vice versa in panel (c).

In some cases, grid-fitting did not work well for $T_\mathrm{ex}$ even when there are enough unblended lines available for a certain species. This is because a robust estimation on $T_\mathrm{ex}$ requires the unblended lines to cover a wide range of $E_\mathrm{up}$ so that the models can be sensitive to temperature changes. For species such as CH$_3^{18}$OH, GA, and $g$-EG, there are few unblended lines covering a wide range of $E_\mathrm{up}$. These lines are either too weak to be detected in faint sources, or severely blended by strong lines in bright sources. Under these circumstances, the results of grid-fitting will not be able to constrain the $T_\mathrm{ex}$, often accompanied with huge uncertainties. As a solution, $T_\mathrm{ex}$ was fixed to that of MF, which has the most identified transitions and therefore the best constraint on $T_\mathrm{ex}$, and only $N$ was fitted and estimated for uncertainties. Besides, the $T_\mathrm{ex}$ of $g$-EG is always set as the same as that of $a$-EG. We consider the fitting results of $N$ still representative since the difference would be within a factor of two if we change the fixed value of $T_\mathrm{ex}$ by 20--50 K in the range of 100--250 K, which is a typical temperature range for hot cores \citep[e.g., Fig.~2 in][]{Ligterink2015}. Similarly, if the FWHM cannot be constrained to better than 0.5 km s$^{-1}$ uncertainty due to the blending of lines, it was also fixed to a value (e.g., if the possible range is 3.5--4.0 km s$^{-1}$, we use 3.8 km s$^{-1}$).

Additionally, there are two special cases encountered during the fitting: one is that the spectra of several sources (i.e., G19.01-0.03, G34.30+0.20, G34.41+0.24, and G35.20-0.74N) show double-peaked line profiles in some O-COMs, which can not be fitted when only one component. These double-peaked features appear in nearly all the transitions, and thus are unlikely to be due to self-absorption. An example of two-component fitting is given by Fig. \ref{fig:fitting_example_2components}, which shows that the spectra can be well fitted by two components with different $v\mathrm{_{lsr}}$ and $N$, while $T_\mathrm{ex}$ remains the same (see Table \ref{table:results_2} for relevant sources and species). It is likely that these sources are not spatially resolved (e.g., maps of G19.01-0.03 show a perfect beam shape in Fig. \ref{fig:maps}), and there is more than one physical component contained in the beam. The second case is that in the two sources G23.21-0.37 and NGC 6334-38, the MF lines are found to be highly optically thick. All strong lines with $A_\mathrm{ij} \gtrsim 10^{-4}$ s$^{-1}$ are saturated, and therefore do not show correct intensity ratios against weaker lines. In this case, the fitting was only based on weak lines with $A_\mathrm{ij} < 10^{-4}$ s$^{-1}$.

\subsection{Isotope ratio calibration for methanol}\label{sec_isotope_ratio}
As mentioned in Sect. \ref{sec_spectral_analyses}, the column density of \ce{CH3OH} needs to be inferred from its minor isotopologues since the main isotopologue itself is likely to be optically thick. The isotope ratios of $\mathrm{^{16}O/^{18}O}$ and $\mathrm{^{12}C/^{13}C}$ can be calculated from the distance to the Galactic center ($D_\mathrm{GC}$) using the equations in \cite{Wilson1994ARAA} and \cite{Milam2005}:
\begin{eqnarray}
    (\mathrm{^{16}O/^{18}O}) & = & (58.8\pm11.8)D_\text{GC} + (37.1\pm82.6) \label{eqn:18O_ratio}\\
    (\mathrm{^{12}C/^{13}C}) & = & (6.21\pm1.00)D_\text{GC} + (18.71\pm7.37)
    \label{eqn:13C_ratio}
\end{eqnarray}
where $D_\mathrm{GC}$ can be derived from the coordinates of the sources and their distances to Earth. The two values before and after $D_\mathrm{GC}$ (with uncertainties) are the slope and intercept of these linear relationships, respectively. In the vicinity of the solar system, $D_\mathrm{GC}$ is 8.05 kpc, which gives $\mathrm{^{16}O/^{18}O}\sim 510$ and $\mathrm{^{12}C/^{13}C}\sim 69$. In the literature, 560 and 70 are commonly used for nearby low-mass sources. For the high-mass sources that are farther away from the solar system, isotope ratios can be calculated from $D_\mathrm{GC}$ before applying them to infer the column density of $^{12}$CH$_3^{16}$OH (Table~\ref{table:sources}). If both $^{13}$\ce{CH3OH} and CH$_3^{18}$OH are optically thin, we can expect their column density ratio to be 7--8. However, our fitting results show a ratio of 2--5, which indicates that $^{13}$\ce{CH3OH} is also (marginally) optically thick. Therefore, only CH$_3^{18}$OH is used to calculate the column density of methanol.

If the errors of both the slope and the intercept are considered in error propagation, the uncertainties of $\mathrm{^{16}O/^{18}O}$ would be around 30\% of the ratios themselves (e.g., if $\mathrm{^{16}O/^{18}O}$ = 300, the error will be $\sim$90). However, a large portion of the total uncertainty comes from the intercept error (82.6), which is even two times larger than the intercept itself (37.1). Considering that only the slope in Eq. \eqref{eqn:18O_ratio} contains the information of the trend between the $\mathrm{^{16}O/^{18}O}$ ratio and $D_\mathrm{GC}$, and the intercept error only represents the scatter of the sources from which the equation was originally fitted, we did not include the intercept error in the error propagation. This yields a decrease in the uncertainties of $\mathrm{^{16}O/^{18}O}$ from $\sim$30\% to $\sim$18\%. Results that include the intercept error are shown in Fig. \ref{fig:O-COM ratio full error}.

\section{Results}\label{sec_results}
\subsection{Morphology}
The continuum maps of the 14 CoCCoA sources show different morphologies (see odd rows in Fig.~\ref{fig:maps}). Six out of 14 show a single peak in both continuum and methanol emission. Others have multiple peaks in continuum but only one or two peaks in methanol emission, usually corresponding to the brightest continuum peak(s). G345.5+1.5 is an exception that its methanol peak is off-set from the continuum peak, and the corresponding region in continuum does not show a peak feature. G35.20-0.74N is an interesting source where the four continuum peaks are located along a line and the methanol emission tends to follow the same alignment. Except for several sources that are not well resolved, most sources show extended weak continuum emission aside the flux peaks, implying the existence of dusty envelopes or parent cores. More detailed studies on the morphology of CoCCoA sources will be presented in a future paper.

\subsection{Spectra and detection}\label{sec_spectra_detection}
As expected from the selection of sources and frequency range, the extracted spectra are rich in COM lines for all the 14 sources in the upper tuning. An overview of the full spectra of each source is presented in Fig.~\ref{fig:overview}. We can see that the spectral appearance of our sources is diverse in intensities and line widths. Bright sources such as G34.41+0.24 and G23.21-0.37 have very strong and broad lines ($\gtrsim$5 km s$^{-1}$) that are severely blended and can only be fitted by visual inspection. There are also faint sources like IRAS 1851-1208 and NGC 6334 I(N)-SM2 where the lines are much weaker and narrower ($<$2 km s$^{-1}$). Their spectra are clean enough for $\chi^2$ fitting, but some less abundant species and some weak lines (with lower $A_\mathrm{ij}$) may remain undetected. The spectra of the other sources have intermediate intensities and line widths (3--4 km s$^{-1}$), which are easiest to fit. 

\begin{figure*}[!ht]
   \centering
   \includegraphics[width=\textwidth]{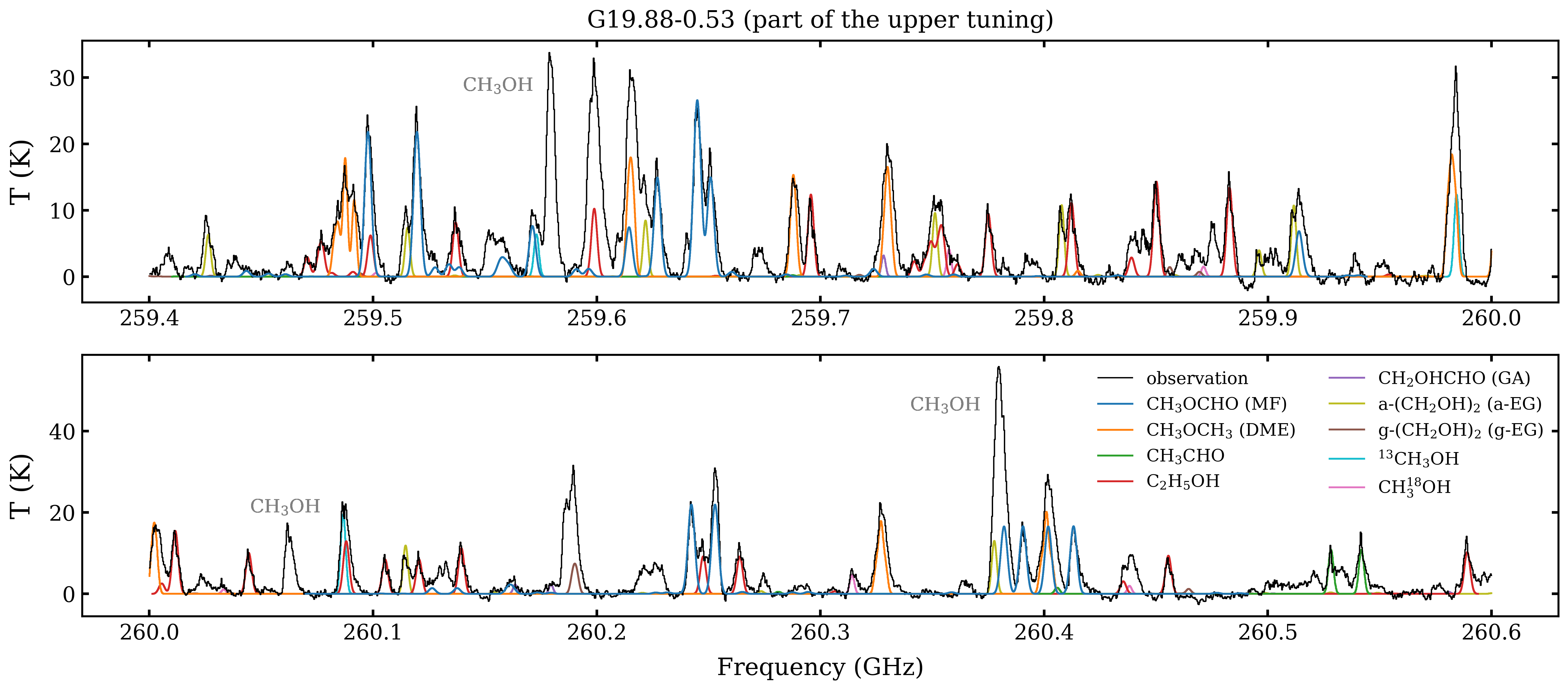}
      \caption{Best fit of the observed spectrum of G19.88-0.53 in a line-rich frequency range (259.4--260.6 GHz). The black and colorful spectra correspond to the observed spectrum and the modeled spectra of each species, respectively. Only the O-bearing COMs that are targeted in this paper (excluding \ce{CH3OH}) are shown here. Two strong lines of methanol are labeled by gray text in the panels. The same figure but for the full upper tuning is given in Fig.\ \ref{fig:fitting_example_full}.}
         \label{fig:fitting_example}
\end{figure*}

The 14 sources provide a rich inventory of detections. More than ten O-bearing COMs and about five N-bearing species are detected in the upper tuning. These detections include the originally targeted species as well as other O- and N-bearing species, covering alcohols, aldehydes, esters, ethers, ketones, acids, and the simplest sugar (GA). We do not detect methoxymethanol (CH$_3$OCH$_2$OH) since our frequency range only covers its weak lines with $A_\mathrm{ij} < 10^{-5}$ s$^{-1}$ and no corresponding line features are spotted in the spectra. Besides the COMs mentioned above, two simple O-bearing molecules, ketene (H$_2$CCO) and formic acid (t-HCOOH), have 1 and 3 transitions covered in the upper tuning, respectively. 
The N-bearing species include HNCO, CH$_3$CN, C$_2$H$_5$CN, and NH$_2$CHO. 
Some abundant species such as HNCO and CH$_3$CN also have their $^{13}$C isotopogues detected. A hydrocarbon molecule, propyne (CH$_3$CCH), is also detected but with all its transitions covered in the lower tuning. Simpler molecules such as SiO, SO, and SO$_2$ only have very few strong lines covered in the upper tuning and do not affect our analyses on O-COMs. Species that are detected in the CoCCoA sources but are not the focus of this paper will be studied in future works.

Figure~\ref{fig:fitting_example} shows the best-fit model of G19.88-0.53 in a line-rich frequency range (259.4-260.6 GHz). A version for the full upper tuning is given by Fig. \ref{fig:fitting_example_full}, and Figs. \ref{fig:fitting_per_species_1}-\ref{fig:fitting_per_species_3} show zoom-in panels for selected unblended lines of each species. The model contains the minor isotopologues of methanol and the six O-COMs that we focus on in this paper. More than 70\% of the line features can be fitted quite well with uncertainties of < 30\% \text{assuming a single excitation temperature}. Apart from several strong features that are attributed to the main isotopologue of methanol and some simple molecules (e.g., SO and \ce{SO2}), weaker features that are not well-fitted by the model likely originate from other COMs that are not included in the model, such as acetone (CH$_3$COCH$_3$), N-bearing COMs, and some minor isotopologues of the detected species. The identified transitions of the selected O-COMs are listed in Table \ref{table:transition}, where transitions that are above 3$\sigma$ and not fully blended with other strong lines are considered as ``identified.'' We can see that most of the identified transitions have an upper energy level of 100--300 K.

\subsection{Column density and excitation temperature}\label{sec_N_Tex}
The column densities and excitation temperatures of all sources are presented in Tables~\ref{table:results_1} and \ref{table:results_2}. With the uniform beam size of 0.33$''$, most sources have a methanol column density of 10$^{18}$--10$^{19}$ cm$^{-2}$, except the weakest source IRAS 1851-1208, which has $N=7\times10^{17}$cm$^{-2}$. Among the six O-COMs that we focus on (excluding methanol isotopologues), DME and MF are the most abundant, with column densities about one order of magnitude higher (10$^{16}$--10$^{17}$ cm$^{-2}$) than those of the other four species. These two species always have a considerable number of strong lines available for fitting in the spectra of all sources (as shown by the shaded areas in Fig. \ref{fig:overview}). \ce{C2H5OH} has somewhat fewer distinct transitions (unblended and large $A_\mathrm{ij}$), and it is not always abundant enough to produce strong lines. In some sources, the column density of \ce{C2H5OH} is of the same order of magnitude as DME and MF, while in some other sources the difference can reach up to two orders of magnitude. This big variation is also present for $a$-EG, which has many strong transitions covered in our data. In some sources, the number of detected lines of $a$-EG are as high as those of DME, while in others there are only a few obvious line features. The fitting of GA and $g$-EG is more difficult, since they only have two to three distinct transitions, which are often subject to blending issues. In summary, the average column densities of the six O-COMs rank as DME $\sim$ MF $>$ \ce{C2H5OH} $>$ ($a$+$g$)EG $>$ \ce{CH3CHO} $>$ GA. Since the absolute values of $N$ are related to the physical environments of the parent sources, it is more useful to look at the relative abundances of O-COMs, that is, their column density ratios with respect to methanol (see Sect. \ref{sec_discussion} for details).

As for the excitation temperature, \ce{C2H5OH}, MF, and $a$-EG tend to have a warm $T_\mathrm{ex}$ of $\gtrsim$150 K, while \ce{CH3CHO} and DME have a relatively lower $T_\mathrm{ex}$ of 100--130 K. This may because different species has a different emitting area of the hot core, for example, some species are emitting from a slightly colder and more extended region, which is not well resolved in our sample according to Fig. \ref{fig:OCOM_mom0_G19.88}). Hot cores are known to have temperature gradients \citep[e.g.,][]{vdTak1999, vdTak2000, Beltran2018, Gieser2019} but only probed on scales larger than the observing beams. Since our analyses are based on the spectra at the peak pixels, the temperature structure on larger scales is not expected to affect our results. The lower $T_\mathrm{ex}$ of \ce{CH3CHO} and DME is also consistent with earlier single-dish findings of other high-mass sources \citep{Bisschop2007, Isokoski2013} and the results of low-mass sources in \cite{MvG2020}. As mentioned in Sect. \ref{sec_spectral_analyses}, CH$_3^{18}$OH, GA, and $g$-EG usually have fixed $T_\mathrm{ex}$ due to a lack of unblended lines covering a wide range of $E_\mathrm{up}$. 

\begin{figure*}[!ht]
    \centering
    \includegraphics[width=\textwidth]{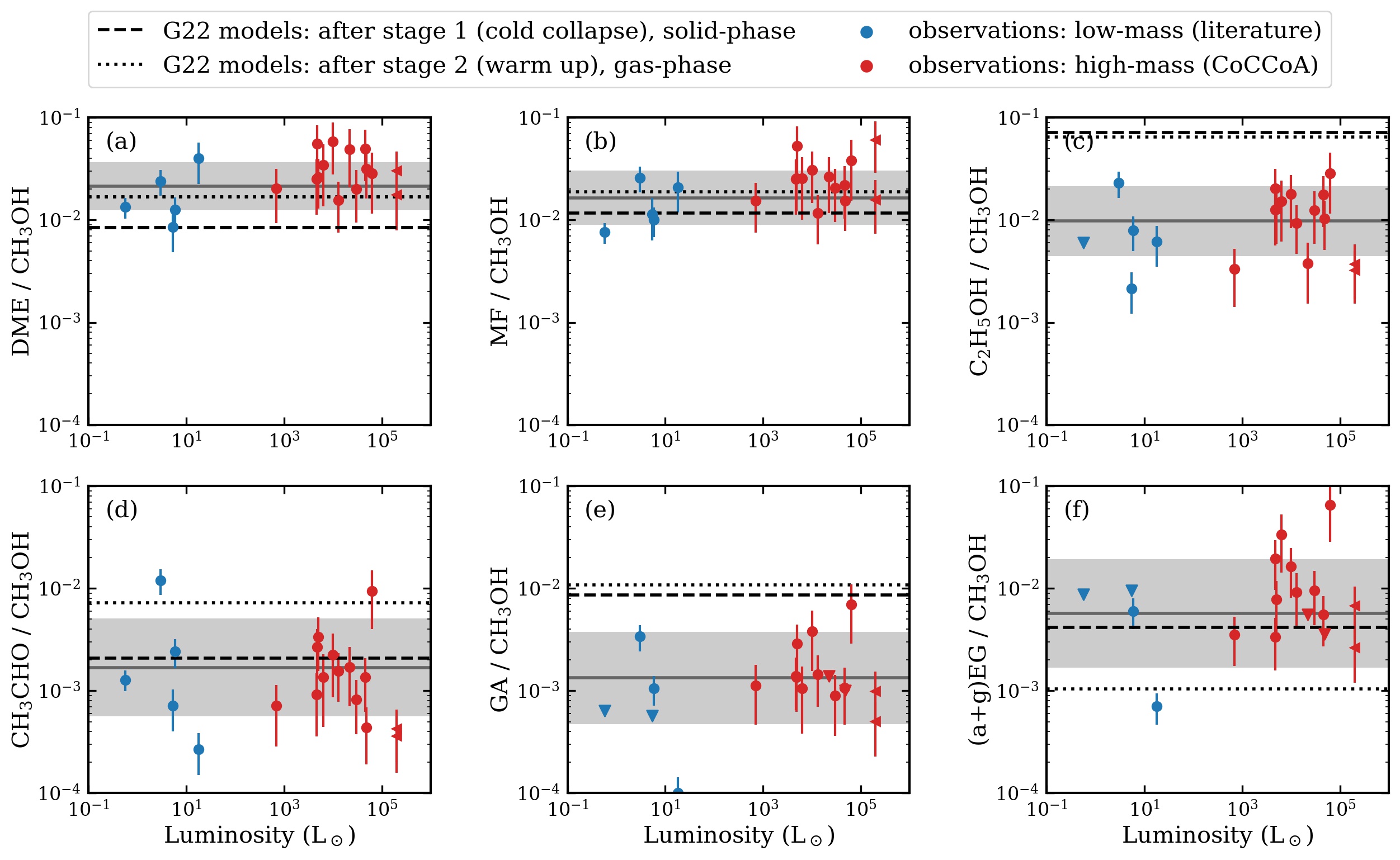}
    \caption{Column density ratios of six O-COMs with respect to methanol versus bolometric luminosity. From left to right, top to bottom is dimethyl ether (DME, \ce{CH3OCH3}), methyl formate (MF, \ce{CH3OCHO}), ethanol (\ce{C2H5OH}), acetaldehyde (\ce{CH3CHO}), glycolaldehyde (GA, \ce{CH2OHCHO}), and ethylene glycol (EG, \ce{(CH2OH)2}). The red points show the CoCCoA data of 14 high-mass sources derived in this work, and the blue ones show the literature data of five low-mass sources: IRAS 16293-2242 A \& B \citep{Manigand2020, Jorgensen2018}, S68N, B1c, and B1-bS \citep{MvG2020}. Upper limits are denoted by downward triangles instead of circles. The data points of NGC 6334-38 and NGC 6334-43 are denoted by leftward triangles due to their upper limits on luminosity. The solid line in gray corresponds to the mean column density ratio of each species, and the shaded area in gray shows the standard deviation weighted by the uncertainty on log scales. The mean and standard deviation are calculated from both the high-mass sources and the low-mass ones. The dashed and dotted lines in black correspond to the modeled COM ratios at the end of the cold collapse stage (stage 1) and the warm-up stage (stage 2) in G22, respectively.} 
    \label{fig:O-COM ratios}
\end{figure*}

\section{Discussion}\label{sec_discussion}
\subsection{Column density ratios}
To further investigate the COM chemistry in our sample, we calculated the column density ratios of O-COMs with respect to methanol and summarize these column density ratios in Fig.~\ref{fig:O-COM ratios}. The column densities of the two conformers of EG, $a$-EG and $g$-EG, are summed up in the last subplot, although they were fitted separately. The uncertainty of the column density ratios was propagated from the uncertainties of the O-COM column densities and the $^{16}$O/$^{18}$O ratio (calculated in Sect. \ref{sec_isotope_ratio}) of each source. We compare our observational results of high-mass protostellar sources with the results of low-mass sources in literature \citep{Jorgensen2018, Manigand2020, MvG2020} and the simulation results of COM chemistry in \cite{Garrod2022}. For consistency, this figure does not include results of other COM observations \citep[e.g.,][]{Csengeri2019, Yang2021} considering that they did not use minor isotopologues to derive the column density of methanol as we did.  

From Fig.~\ref{fig:O-COM ratios} we can see that there is no obvious difference in the O-COM ratios between low-mass and high-mass sources. However, molecules in the two groups should have experienced different physical conditions, such as temperature and fluence of energetic particles (UV photons, X-rays, cosmic rays) before the hot cores/corinos formed. This implies that these species are likely formed under similar conditions, which points to a common and early pre-stellar stage before the star formation processes begin to differentiate.

The six selected O-COMs have different column density ratios with respect to methanol, varying by over two orders of magnitude between $0.01$\% and $10$\%. Among the six species, DME and MF have the highest and similar ratios with respect to methanol (2--3\%). 
\ce{C2H5OH} and EG have the intermediate ratios but with a much larger scatter. \ce{CH3CHO} and GA have the lowest ratios at around 0.1\%.

\begin{figure}[]
   \centering
   \includegraphics[width=\hsize]{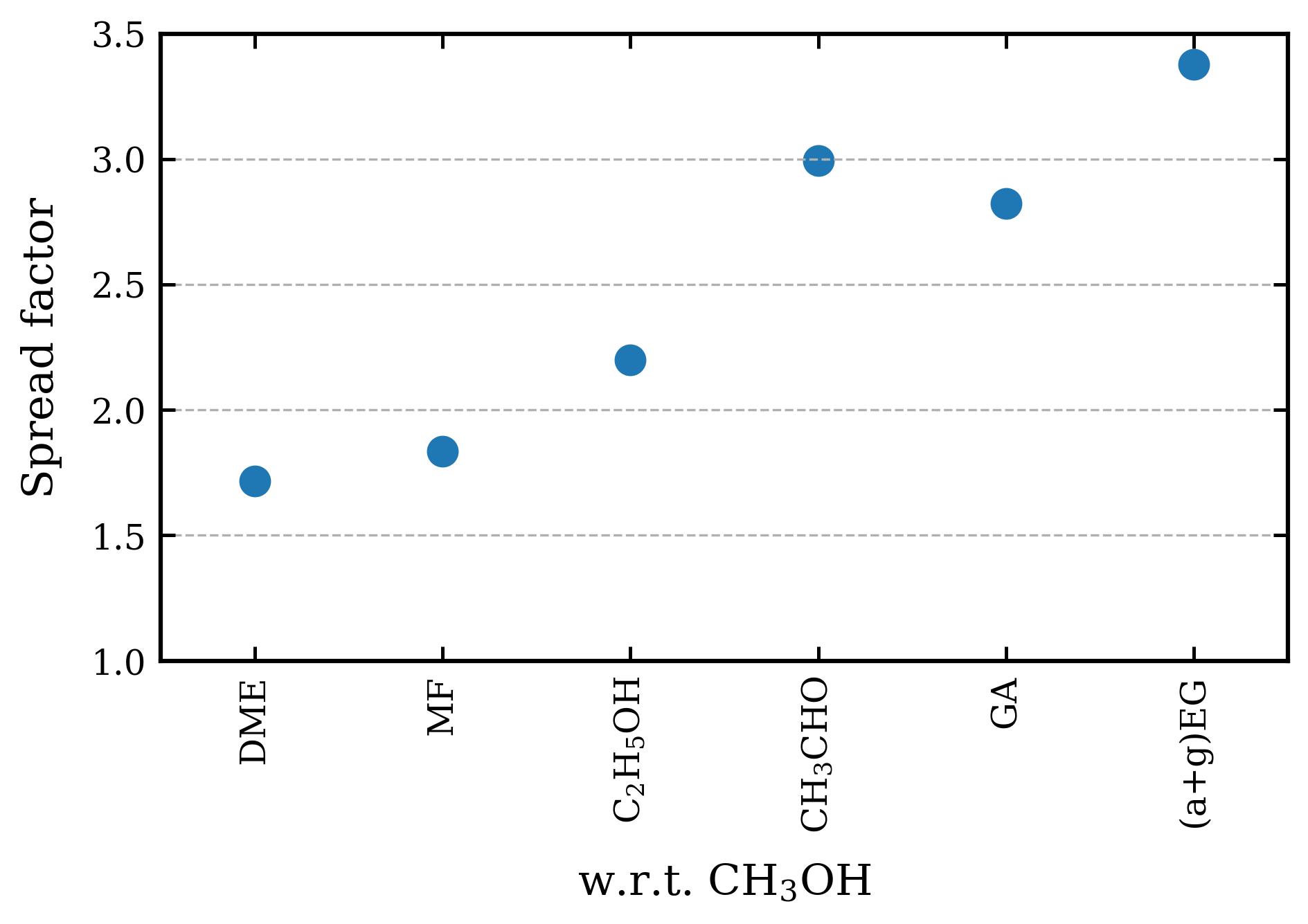}
      \caption{Spread factor (log$_{10}$ of the 1$\sigma$ scatter) for the ratios plotted in Fig.~\ref{fig:O-COM ratios} of different O-COMs with respect to methanol.}
      \label{fig:spread_factor}
\end{figure}

Figure~\ref{fig:spread_factor} shows the spread factor, that is, $\log_{10}$ of the standard deviation of O-COM ratios in log scales of our combined sample of low-mass and high-mass protostars. It is obvious that DME and MF have smaller scatter than the other four species. The scatter may result from an observational effect, that different species have different gas-phase emitting areas depending on their sublimation temperatures from dust grains \citep{Nazari2021, Nazari2022_NCOM}. The column densities that we derive from spectral fitting represent the abundances averaged over the observational beam. If the actual emitting area of a species is smaller than the area that we can resolve, then we are suffering from a beam dilution issue. This is likely the case in the CoCCoA observations since the moment 0 maps of selected O-COMs given in Fig. \ref{fig:OCOM_mom0_G19.88} show that the emission of these molecules are barely or not spatially resolved. The actual resolved area associated with a specific beam size depends on not only the angular resolution (which is constant across the CoCCoA sample), but also the distance to a source (which varies by a factor of 3--4 across the sample). Moreover, the size of the hot core, as defined by the radius where $T=100$ K, depends on the square root of the luminosity of the source \citep{Bisschop2007}:
\begin{equation}
    R_{T=100\ \mathrm{K}} \approx 15.4 \sqrt{\frac{L}{L_\odot}}\ \mathrm{au}.
\end{equation}
Therefore, there is a beam dilution factor between the observationally inferred column density and the actual one. This factor can differ from source to source and from species to species, and hence can lead to the scatter in COM ratios. These effects are discussed
 and quantified in more details in \cite{Nazari2022, Nazari2023}. 

More generally, it should be noted that surveys of large samples of low- and high-mass protostars have found that some fraction of sources do not show any methanol or COM emission \citep[e.g.,][paper I]{Yang2021, MvG2022}. The reasons for this absence of COM emission are varied but include the possible presence of a disk which lowers the overall temperature structure \citep[][paper II]{Nazari2022}, as well as different evolutionary stages, such as the presence of an H II region \citep[][modeling paper]{Nazari2023}.

\subsection{Observations versus simulations}\label{sec_obs_vs_sim}
The comparison with simulations is mainly based on the state-of-the-art models in \cite{Garrod2022}, hereafter G22. The G22 models simulate the chemistry coupled in three phases: gas-phase, grain/ice-surface, and bulk-ice mantle. The last two phases are collectively known as the solid phase. The evolution of hot cores is treated as two stages, a cold collapse stage followed by a static (fixed density) warm-up stage once the central protostar has formed. There are three warm-up timescales used in the models: $5\times10^4$ yr (fast), $2\times10^5$ yr (medium), and $1\times10^6$ yr (slow). A major update in the G22 models is to include non-diffusive chemistry on surfaces and in bulk ices, which is proposed to be important in interstellar ices based on laboratory work \citep{Fedoseev2015, Linnartz2015}. G22 test the effect of different non-diffusive mechanisms along with other parameters in about 20 models (see Table 1 in G22). The ``final'' model includes all the discussed non-diffusive mechanisms and we use it as the fiducial model for further discussion. In Fig.~\ref{fig:O-COM ratios}, the horizontal dashed lines correspond to the COM ratios (w.r.t. methanol) in the solid phase at the end of the collapse stage, and the horizontal dotted lines correspond to the gas-phase ratios after the warm-up stage with the medium warm-up speed. In addition to Fig. \ref{fig:O-COM ratios} that shows the O-COM ratios w.r.t. methanol, we also present a number of ratios between two O-COMs in Fig. \ref{fig:2COM_ratio}. In the following subsections, we will compare our observational results of each species to the ``final'' model with the ``medium'' warm-up speed in G22.

\subsubsection{\ce{CH3OCH3} (DME) and \ce{CH3OCHO} (MF)}\label{sec_DME_MF}
DME and MF have the most stable column density ratios in observations and the best match with the G22 simulations (Fig.~\ref{fig:O-COM ratios}a,b). Their high abundances are often underproduced by experiments and simulations \citep{Fedoseev2015, Chuang2016, Simons2020, Jin2020}. However, the inclusion of the new formation routes in G22 is able to reproduce our observational results on the gas-phase ratios of DME and MF with respect to methanol.

In the G22 models, more than 60\% of the DME is formed in ices, through the reaction between methylene (CH$_2$) and methanol via
\begin{align}
    &\mathrm{CH_2 + CH_3OH \rightarrow CH_3OCH_2}\label{DME_ice_1},\ \mathrm{and}\\
    &\mathrm{H + CH_3OCH_2 \rightarrow CH_3OCH_3}\label{DME_ice_2}.
\end{align}
Nearly 90\% of CH$_3$OCH$_2$ radicals are formed by the combination of CH$_2$ and CH$_3$O, in which CH$_3$O comes from the hydrogenation of formaldehyde (H$_2$CO) or H abstraction from methanol. As the reactant of two reactions, CH$_2$ directly affects the formation of DME ices. G22 set the activation energy barrier of the grain-surface reaction
\begin{equation}\label{CH2_formation}
    \mathrm{C + H_2 \rightarrow CH_2}
\end{equation}
to zero \citep{Krasnokutski2016, Henning2019NatAs}, and added about 20 grain-surface CH$_2$-related reactions to the network (see Table~4 in G22). This greatly enhances the efficiency of forming CH$_2$ and its subsequent contribution to COMs formation. Nevertheless, this assumption is not fully supported by the combined experimental/theoretical work of \cite{Lamberts2022}, who argued that the reaction between C and H$_2$ is unlikely to be fully barrierless on water ices. However, the effects to the full kinetic model may not be significant with a modest non-zero barrier of reaction~(\ref{CH2_formation}). G22 also introduced a set of methylidyne (CH) reactions (Table~5 in G22) which can form CH$_2$ and larger hydrocarbons barrierlessly. These reactions can make up the CH$_2$ formation when reaction~(\ref{CH2_formation}) has a barrier. According to G22, the inclusion of CH and CH$_2$ chemistry in the solid phase enhances the abundance of solid-phase DME by more than a factor of 2. This emphasizes the importance of including the carbon hydrogenation to the chemical network \citep{Qasim2020}.

Besides the bottom-up formation of DME from CH$_2$, the photodissociation of CH$_4$ may also provide important ingredients. Several experimental studies were able to produce DME from UV-irradiated ices of \ce{CH3OH} and CH$_4$ \citep{Oberg2009, Paardekooper2016, Yocum2021}. However, \cite{Fedoseev2015} and \cite{Chuang2016, Chuang2017} did not observe DME formation in their experiments with CH$_4$ or H$_2$CO not included in the deposition, even when UV was introduced in \cite{Chuang2017}.

According to G22, the remaining 40\% DME is formed in the gas phase 
through two reactions:
\begin{align}\label{DME_gas}
    &\mathrm{CH_3OH + CH_3OH_2^+ \rightarrow CH_3OCH_4^+ + H_2O},\ \mathrm{and}\\
    &\mathrm{CH_3OCH_4^+ + NH_3 \rightarrow CH_3OCH_3 + NH_4^+}.
\end{align}
In the first step, protonated DME is formed via the reaction between methanol and protonated methanol. The second step is the proton transfer to ammonia \citep{Charnley1995, Rodgers2001, Taquet2016}, where ammonia comes from ice sublimation and therefore the reaction sequence would not be efficient under cold conditions \citep{Skouteris2019}.

MF in the solid phase forms on grain surfaces mainly through the non-diffusive reaction \citep{Chuang2016}
\begin{equation}\label{MF_ice_1}
    \mathrm{HCO + CH_3O \rightarrow CH_3OCHO}.
\end{equation}
A small contribution is made by the newly introduced 3-body excited-formation (3-BEF) reactions in the bulk ice
\begin{equation}\label{MF_ice_2}
    \mathrm{CH_3O + CO \rightarrow CH_3OCO},
\end{equation}
whereby the hydrogenation of H$_2$CO would produce excited CH$_3$O that can overcome the reaction barrier \citep{Jin2020}. The CH$_3$OCO radicals can then recombine with H atoms to form MF. Reactions (\ref{MF_ice_1}) and (\ref{MF_ice_2}) occur in the cold collapse stage and contribute about 70\% of the total MF. In the gas phase, the newly added reaction \citep{Balucani2015}
\begin{equation}\label{MF_gas_1}
    \mathrm{O + CH_3OCH_2 \rightarrow CH_3OCHO + H}
\end{equation}
becomes the main production route when $T>100$ K. The gaseous CH$_3$OCH$_2$ radicals can be released from ice mantles, or converted from DME by OH abstraction \citep{Shannon2014}:
\begin{equation}\label{MF_gas_2}
    \mathrm{CH_3OCH_3 + OH \rightarrow CH_3OCH_2 + H_2O}.
\end{equation}

The formation of both DME and MF is strongly related to methanol and its precursor CH$_3$O. Simulations with different input $n$(H)/$n$(CO) ratios by \cite{Simons2020} and experiments by \cite{Santos2022} proposed that the final step of methanol formation is dominated by
\begin{equation}\label{methanol_formation}
    \mathrm{CH_3O + H_2CO \rightarrow CH_3OH + HCO}.
\end{equation}
\cite{Chuang2016} noticed that H$_2$CO is a prerequisite for the formation of MF, that is, without the input of H$_2$CO there would be no MF detected in the outcome, which was also the case in \cite{Fedoseev2015} where only H and CO were used. Since CH$_3$O is necessary to form MF, it is inferred that the H abstraction from \ce{CH3OH} yields primarily CH$_2$OH, while CH$_3$O mainly comes from the hydrogenation of H$_2$CO. This may explain why DME and MF can retain relatively stable abundance ratios with respect to methanol among a large sample of sources, since the formation of all the three species tends to be strongly related with the same precursor CH$_3$O. However, it is not clear whether the fact that DME/MF ratio $\sim$1 (Fig.~\ref{fig:2COM_ratio}a) is a pure coincidence or there is some chemical balance between the two COMs.

\subsubsection{\ce{C2H5OH}}\label{sec_ethanol}
The observed abundance ratio of ethanol is on average one order of magnitude lower than the simulation results in G22. The scatter is also relatively large compared to DME and MF. The G22 models give nearly the same abundance after the collapse stage and the warm-up stage. This is because in their models ethanol is formed almost entirely in the early cold collapse stage on dust grains. The dominant reactions are
\begin{equation}\label{EtOH_ice_1}
    \mathrm{C_2H_5 + OH \rightarrow C_2H_5OH}
\end{equation}
in the bulk ice, and
\begin{equation}\label{EtOH_ice_2}
    \mathrm{CH_2 + CH_3OH \rightarrow C_2H_5OH}
\end{equation}
on the surface. The C$_2$H$_5$ radicals in reaction (\ref{EtOH_ice_1}) are formed by diffusive reactions of atomic H with ethane (C$_2$H$_6$) and ethylene (C$_2$H$_4$). These routes may be supported by the experiments in \cite{Chuang2020} showing that ethanol can be formed through non-energetic processing of C$_2$H$_2$ ices. However, the efficiency of these reactions is not well constrained. 

The OH radicals in reaction (\ref{EtOH_ice_1}) come from photodissociation of water ice by external UV radiation. The G22 models used the same initial visual extinction ($A_{V,\ \mathrm{init}}$ = 3) throughout the molecular cloud, while in reality, $A_V$ is higher in the inner part of the cloud. This difference may result in an overestimation of the UV intensity in the cold collapse stage, and hence the overproduction of ethanol on dust grains.

Besides the production routes, there is also a recently proposed gas-phase destruction mechanism, the ``ethanol tree'' network \citep{Skouteris2018}, that is not included in the G22 models. This new network starts with the H abstraction from ethanol by halogen atoms or OH radicals, leading to two reactive radicals: \ce{CH3CHOH} and \ce{CH2CH2OH}. These radicals are further converted into formic acid (HCOOH) and formaldehyde (\ce{H2CO}) by reacting with O atoms, along with other minor products (see branching ratios in Fig.~1 of Skouteris et al.). As a result, the abundance of gas-phase ethanol decreases over time. \cite{Skouteris2018} predicts the ethanol/GA ratio to fall from $\sim$200 to $\sim$10 in about 1000 years. In our observations, the \ce{C2H5OH}/GA ratios are around 10 (see Fig.~\ref{fig:2COM_ratio}d). This suggests that the gas-phase destruction of ethanol in a later stage may play a role in explaining its overproduction in chemical models.

\subsubsection{\ce{CH3CHO}}\label{sec_acetaldehyde}
In the G22 simulations, the abundance ratio of \ce{CH3CHO} with respect to methanol in the solid phase after the collapse stage agrees well with our observational data. However, there are discrepancies among the warm-up stages with different timescales. The longer the warm-up stage is, the more \ce{CH3CHO} is produced, indicating substantial gas-phase formation. In the G22 models, only 25\% of the total amount of \ce{CH3CHO} is formed in ices during the cold collapse stage ($T\sim10$ K) through the hydrogenation of ketene
\begin{equation}\label{CH3CHO_ice_1}
    \mathrm{CH_2CO + 2H \rightarrow CH_3CHO}.
\end{equation}
About 35\% is produced in the early warm-up stage ($T<100$ K) by
\begin{equation}\label{CH3CHO_ice_2}
    \mathrm{CH_2 + H_2CO \rightarrow CH_3CHO}
\end{equation}
on the grain surface, and
\begin{equation}\label{CH3CHO_ice_3}
    \mathrm{CH_3 + CO + H \rightarrow CH_3CHO}
\end{equation}
in the bulk ice.
Nearly 40\% is formed in the gas phase through the reaction
\begin{equation}\label{CH3CHO_gas_1}
    \mathrm{C_2H_5 + O \rightarrow CH_3CHO + H}
\end{equation}
after the desorption from dust grains at $T>200$ K in the warm-up stage. 
\cite{Vazart2020} give a brief overview on the recorded gas-phase formation routes of \ce{CH3CHO} and explore some new reactions by theoretical computations. They confirm that reaction (\ref{CH3CHO_gas_1}) is efficient in the temperature range of 7--300 K.

The observed ratios of \ce{CH3CHO} only match the modeled values after the cold collapse stage and the fast warm-up stage (the latter of which is not shown in Fig. \ref{fig:O-COM ratios}d), which implies that the formation of \ce{CH3CHO} in the early stage may be more dominant than suggested by the G22 models. The experimental studies by \cite{Fedoseev2022} proposed a similar formation route for \ce{CH3CHO}, that is, the hydrogenation of ketene on cold (10 K) surfaces of mixed C, H, CO, and H$_2$O ices. They proposed that H$_2$CCO can be formed through hydrogenation of CCO radicals, which are the product of the barrierless association reaction between CO molecules and C atoms. 
Unfortunately, the intermediate products, CCO and HCCO, are very difficult to observe in experiments due to their high reactivity, and the relevant reaction rates remain undetermined. \cite{Chuang2020, Chuang2021} also proposed formation routes from C$_2$H$_2$ ices to \ce{CH3CHO} at $T=10$ K under both non-energetic and energetic conditions. To better understand the difference between the chemical modeling results in G22 (especially the gas-phase ratios after the warm-up stage) and our observational results, we need to know more about the relative importance of \ce{CH3CHO} formation in the solid and gas phase.

\begin{figure*}[!ht]
   \centering
   \includegraphics[width=1\textwidth]{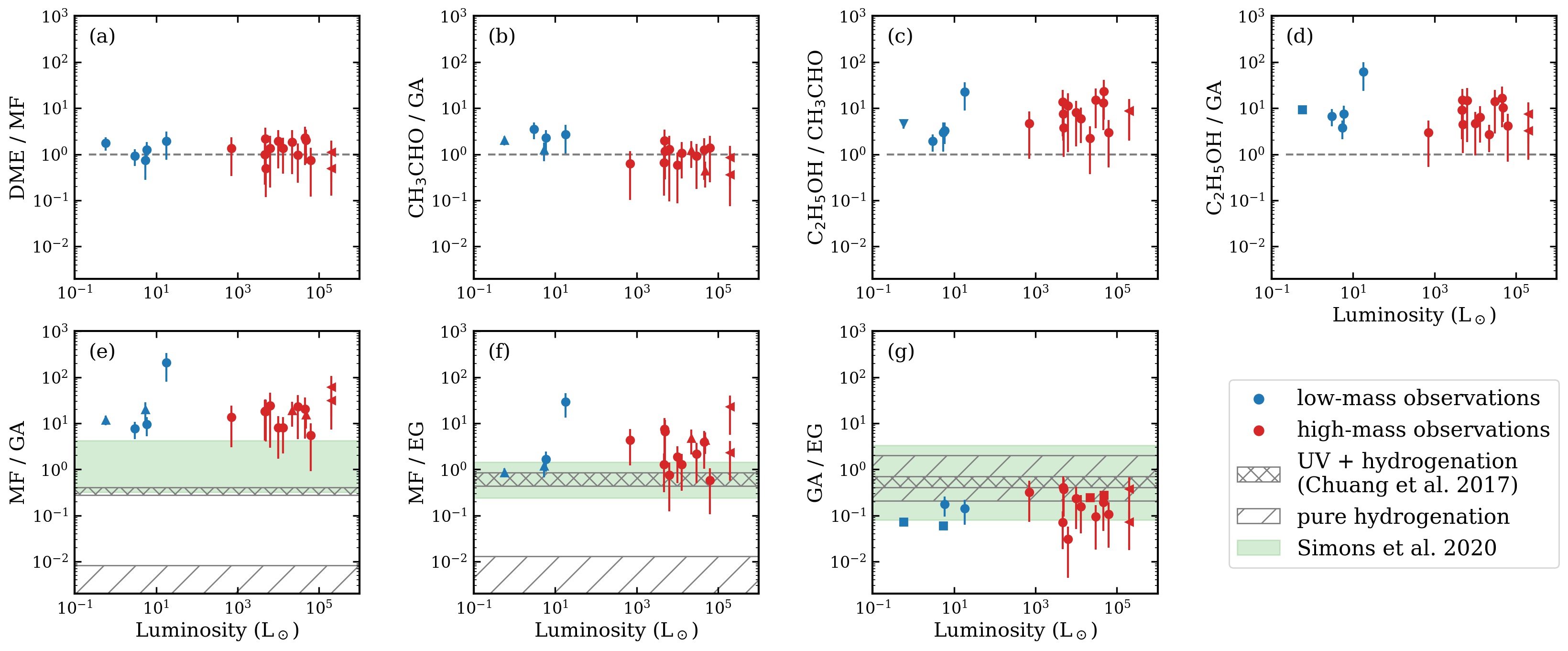}
      \caption{Observed abundance ratios between seven pairs of O-COMs. Data points in blue and red correspond to the 5 low-mass sources from literature and the 14 high-mass sources from CoCCoA survey presented in this work, respectively (the same as Fig. \ref{fig:O-COM ratios}). If the column densities of one species in the pair are upper limits, the ratios would be upper/lower limits, of which the data points are downward/upward triangles instead of circles. Data points in square without error bars mean that the column densities of both species in the pair are upper limits. Dashed lines in panels (a)-(d) indicate the ratio of 1. Shaded areas in gray slash/cross and green in panels (e)-(g) indicate the ranges of the experimental results in \cite{Chuang2017} and the simulation predictions by \cite{Simons2020}, respectively.}
      \label{fig:2COM_ratio}
\end{figure*}

It is also interesting to notice the relation between \ce{CH3CHO} and the two more hydrogenated O-COMs, GA and \ce{C2H5OH}. The ratios of \ce{CH3CHO}/GA and \ce{C2H5OH}/\ce{CH3CHO} shown in Fig.~\ref{fig:2COM_ratio}(b) and (c) behave differently; \ce{CH3CHO}/GA is $\sim$1 with a small scatter comparable to DMF/MF, while \ce{C2H5OH}/\ce{CH3CHO} exhibits a larger scatter. These results are counterintuitive since \ce{C2H5OH} is suggested to be a direct hydrogenation product of \ce{CH3CHO} \citep[e.g., experiments by][]{Fedoseev2022}, while GA and \ce{CH3CHO} are usually not present simultaneously in experiments. More investigation is needed to verify if GA and \ce{CH3CHO} are chemically linked. 

\subsubsection{\ce{CH2OHCHO} (GA) and \ce{(CH2OH)2} (EG)}\label{sec_GA_EG}
In contrast to MF, simulations and experiments usually produce more GA than observations \citep[e.g.,][]{Chuang2016, Chuang2017, Simons2020}, as do the G22 models. 
\cite{Fedoseev2015, Fedoseev2017} have confirmed experimentally that GA and EG can be formed in the solid phase through surface hydrogenation of CO molecules at $T$ = 13 K. They suggested that GA and EG are the outcome of successive hydrogenation of glyoxal (HC(O)CHO), which is formed from two HCO radicals. In the G22 models, EG is mainly formed through the addition of two CH$_2$OH radicals in the solid phase:
\begin{equation}
    \mathrm{CH_2OH + CH_2OH\rightarrow (CH_2OH)_2}.
\end{equation}
Most of the GA is formed through repetitive H abstraction from EG:
\begin{align}
    &\mathrm{H + (CH_2OH)_2 \rightarrow CH_2OH_2CH_2O + H_2}\label{EG2GA_1},\ \mathrm{and}\\
    &\mathrm{H + CH_2OHCH_2O \rightarrow CH_2OHCHO + H_2}\label{EG2GA_2}.
\end{align}
About 60\% of the conversion from EG to GA through reaction (\ref{EG2GA_1})-(\ref{EG2GA_2}) is finished in the cold collapse stage when $T\lesssim10$ K, while about 30\% occurs in the middle of the warm-up stage when T $\sim$100--200 K. 
A small portion of solid-phase GA is formed via the route proposed by \cite{Chuang2016}:
\begin{equation}
    \mathrm{CH_2OH + HCO\rightarrow CH_2OHCHO}.
\end{equation}

However, despite the large scatter in the observed abundance ratios of GA and EG, it is obvious that EG is overall more abundant than GA in observations, while the G22 models give the opposite results. This implies that the interconversion between EG and GA may not be well modeled in G22.

Another simulation work by \cite{Simons2020} computed their O-COM network with four H/CO input ratios (5-60\%) at six low temperatures (8--20 K). They summarized the flux distribution of the network for the fiducial model with $n$(CO) = 10.0 cm $^{-3}$, $n$(H) = 2.5 cm $^{-3}$ and $T = 10$ K (see Fig.~8 of Simons et al.). They found that the hydrogenation of glyoxal is more important to the GA formation than the H-abstraction from EG. Their results 
also show that the relative abundance of GA to EG is very sensitive to the $n$(H)/$n$(CO) ratio. The observed abundance ratios of GA over EG are $\sim$0.1--1 (Fig.~\ref{fig:2COM_ratio}g), which corresponds to $n_\mathrm{initial}$(H)/$n_\mathrm{initial}$(CO) > 0.25 in their models. \cite{Chuang2017} were able to reproduce a similar GA/EG ratio through the hydrogenation of ices with CO:\ce{CH3OH} = 4:1 at $T=14$ K, and also find it subject to the initial composition ratios. Besides pure hydrogenation, they tried introducing UV radiation, but the GA/EG ratio was not affected. 
The simulations and experiments mentioned above suggest that the formation of GA and EG is strongly regulated by the relative abundance of H atoms with respect to other ingredients such as CO. Parameters such as activation energy barriers and branching ratios of the related reactions of GA and EG formation may also attribute to the difference between simulations and observations. 

The abundance ratio of GA with respect to methanol predicted by the G22 models is nearly one order of magnitude higher than our observational results. A possible reason for the overproduction by chemical models is that GA has a higher desorption temperature than methanol \citep[see Fig. 2 in ][]{Fedoseev2015}. As a result, GA is expected to desorb and emit from a smaller region than methanol \citep[for quantification, see the toy model described in Appendix B of][]{Nazari2021}. If our spatial resolution is not high enough to resolve the actual emitting region of GA (suggested by the moment 0 map in Fig. \ref{fig:OCOM_mom0_G19.88}), there will be a beam dilution effect leading to an underestimated column density ratio of GA compared to the actual abundance ratio. 
An interesting fact is that EG has an even higher desorption temperature than GA \citep{Fedoseev2015} and is expected to suffer more from beam dilution, but it is not overproduced by the G22 models. This means that if beam dilution actually accounts for the underestimation of GA in observations, EG would have been underproduced by the G22 models, or there exist other reasons to explain the observed GA depletion.

\subsection{The influence of energetic processes}
As mentioned in Sect. \ref{sec_GA_EG}, \cite{Chuang2017} show that ratios among MF, GA, and EG can be altered by UV radiation. The ratios of MF with respect to GA and EG both increase by nearly two orders of magnitude when introducing UV radiation to the experiments, while the ratio between GA and EG is not much affected. The second row of Fig.~\ref{fig:2COM_ratio} shows the comparison among observations, simulations, and experiments \citep[this work,][respectively]{Simons2020, Chuang2017}. The UV intensity is not varied during the experiments, but the discrepancy between pure hydrogenation and UV-irradiation implies a positive correlation between the UV irradiation and the relative abundance of MF. However, the observed ratios of MF/GA and MF/EG are still overall higher than the values produced by simulations and experiments. 

In G22, one of the models tests the influence of cosmic-ray-induced ionization and UV-induced photodissociation, which shows an obvious enhancement of the O-COM abundances. The ``final'' model includes these energetic processes but with a fixed efficiency, as do the simulations in \cite{Simons2020}. The G22 models are able to reproduce enough MF by introducing a new set of non-diffusive reactions, but the abundances of GA and EG are not very consistent with our observations. Experiments by \cite{Oberg2009} suggest that the final product composition after irradiating \ce{CH3OH} ices with UV lamps depends more on the UV fluence and temperature, instead of the UV flux itself. 
There is a possible explanation that the short timescale offsets the high UV flux in high-mass sources, and the total UV fluence falls in the same order as low-mass sources.  
To figure out the influence of energetic processes on COM chemistry, especially in the solid phase, more experiments and simulations with varied parameters are needed. 
Observations of larger samples of protostellar objects with different masses and luminosities are also needed to provide a more reliable statistics of COMs ratios.

\subsection{O-COMs from clouds to comets}\label{sec_other_objects}
\begin{figure}[!t]
   \centering
   \includegraphics[width=\hsize]{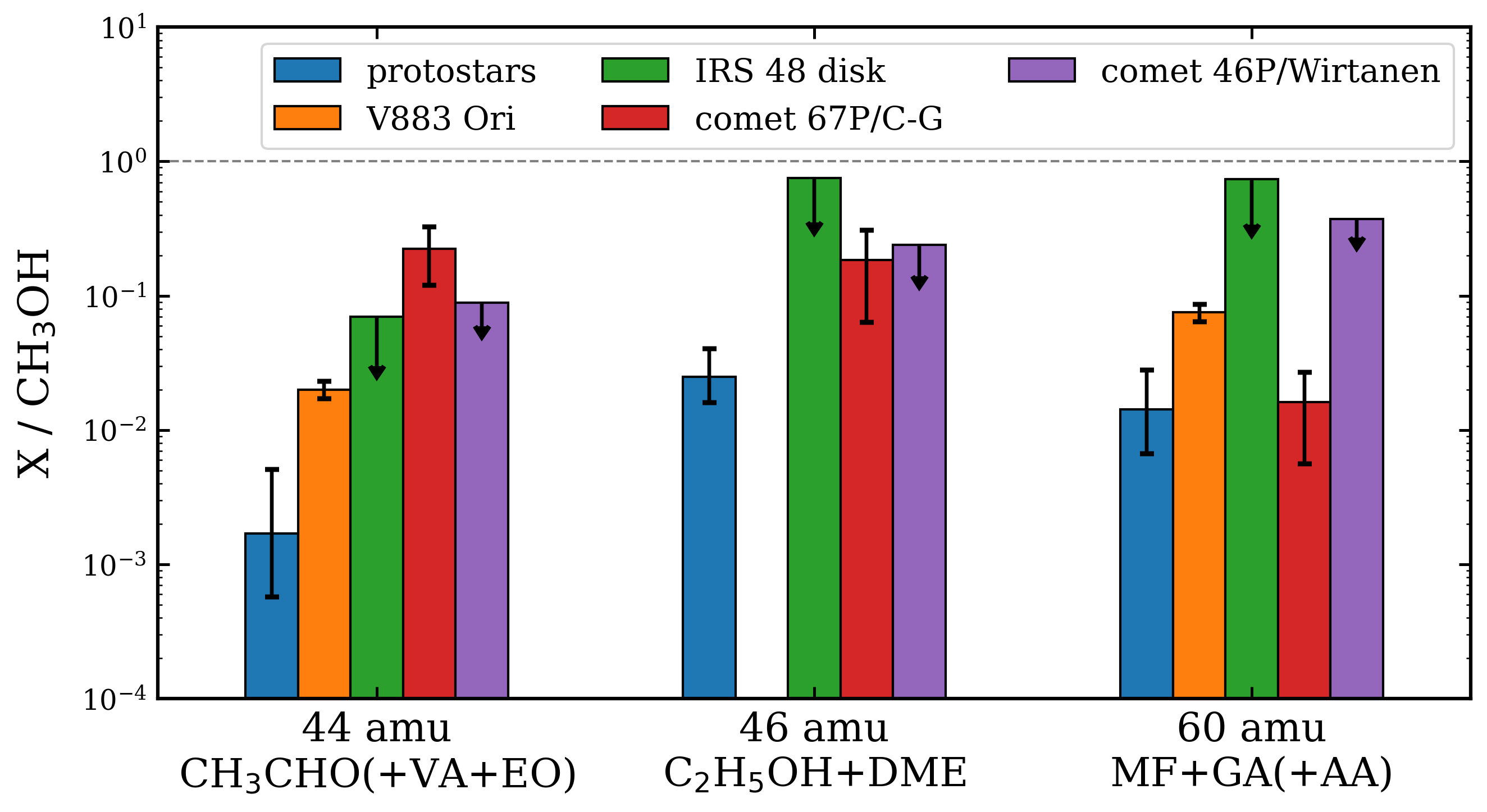}
      \caption{Column density ratios of selected O-COMs with respect to methanol in different astronomical objects (see references in Sect. \ref{sec_other_objects}). Species are divided into three groups with different atomic  mass unit (amu), and those in parentheses are only shown for the degenerate detections in the comet 67P/C-G by the ROSINA mass spectrometer; they have not been detected separately in other sources. The column densities of the species with the same amu are summed up. The data of IRS 48 are shown as upper limits due to the possible underestimation of the column density of methanol.}
      \label{fig:other_env}
\end{figure}

Figure~\ref{fig:other_env} summarizes the observed O-COM ratios with respect to methanol in different astronomical objects. The data of protostars are represented by the average ratios of the low-mass and high-mass sources discussed in this work, and the uncertainties correspond to the standard deviation. We compare them with the literature data of the outbursting protostar V883(FU) Ori \citep{Lee2019NatAs} and the protoplanetary disk around Oph IRS 48 \citep{Brunken2022}. Two comets, 67P/Churyumov–Gerasimenko (67P/C–G) \citep{Robin2019,Drozdovskaya2019} and 46P/Wirtanen \citep{Biver2021}, are also taken into account, which reflect the pristine chemical composition in our solar system. All the sources were observed by ALMA, except that the data of 67P/C-G were collected by the Rosetta Orbiter Spectrometer for Ion and Neutral Analysis (ROSINA), and the data of 46P/Wirtanen were taken by the Institut de Radio Astronomie Millim\'etrique (IRAM) 30-m telescope and the NOrthern Extended Millimeter Array (NOEMA). ROSINA is a mass spectrometer which cannot distinguish among isomers with the same mass (e.g., \ce{C2H5OH} and DME have the same atomic mass unit (amu) of 46). Considering the data availability, we chose three groups of O-COMs for comparison:
\begin{description}
    \item amu = 44: \ce{CH3CHO}, CH$_2$CHOH (vinyl alcohol, VA), c-C$_2$H$_4$O (ethylene oxide, EO)
    \item amu = 46: \ce{C2H5OH}, DME
    \item amu = 60: MF, GA, CH$_3$COOH (acetic acid, AA)
\end{description}
EO, VA, and AA are mentioned only for the potential degeneracy of the detection of 67P/C-G; they were not searched or detected in other sources except one low-mass protostellar object (IRAS 16293-2242 B) that we considered here. The O-COM ratios of IRS 48 may be overestimated since the column density of methanol was determined from lines that are likely to be optically thick (hence upper limit signs are used).

In general, protostellar sources have lower O-COM ratios than the other sources, especially for the group of amu = 44. 
For the other two groups, considering that the data of IRS 48 may be overestimated, the O-COM ratios are more comparable among the samples. However, the sample here is too small to draw any robust conclusion. 
A larger sample of sources at different evolutionary stages are needed to study the chemical inheritance throughout star formation.

\subsection{JWST ice observations}
With JWST successfully operating, it is now becoming more possible to detect COMs other than methanol in ices. The absorption features for the identification of O-COM ices are mainly located in the mid-infrared between 2 and 15 $\mu$m \citep{Boogert2015ARAA, Rocha2022}. \ce{CH3CHO} and \ce{C2H5OH} ices tentatively account for the spectroscopic features at 7.24 and 7.41 $\mu$m in ISO \citep{Schutte1999}, \textit{Spitzer}/IRS \citep{Oberg2011}, and the latest JWST/MIRI observations \citep{Yang2022, McClure2023}. Besides, the feature around 11.3 $\mu$m in the JWST spectra may also have some contributions from \ce{CH3CHO}, \ce{C2H5OH}, and MF \citep{TvS2018, TvS2021}. However, more detailed spectral modeling is needed to confirm the detection of these COMs. Once JWST observations with larger samples and higher sensitivities are available, we will be able to estimate the ratios of \ce{CH3CHO} and \ce{C2H5OH} with respect to methanol in ices, based on the absorption features at 7.2 and 7.4 $\mu$m. By comparing their ice ratios to the gas-phase ones in the same sources, we can get an idea whether and to what extent these species participate in gas-phase chemistry after they sublimate from the ice mantles of dust grains.

\section{Conclusions}
We analyzed the spectra of 14 high-mass protostellar objects in the CoCCoA survey. We focused on six selected O-COMs: \ce{CH3CHO}, \ce{C2H5OH}, DME, MF, GA, and EG, and derive their column densities and excitation temperatures for the 14 sources. We also performed various comparisons between the observed O-COM ratios with respect to methanol and the results from previous simulations and experiments. We summarize our conclusions below:
\begin{enumerate}
      \item The column density ratios of the six selected O-COMs with respect to methanol show no clear difference between the five low-mass studied previously and the 14 high-mass protostellar objects observed with ALMA, suggesting that these species are mainly formed under similar conditions. Current astrochemical simulations and experiments support the possibility of early formation of COM ices on dust grains in the pre-stellar stage, before the environments in low- and high-mass star-forming regions begin to diverge. However, the possibility exists that other gas-phase formation routes also play an important role in shaping the COM ratios, which needs more investigations to pin down.
      \item DME and MF show smaller scatter in their ratios with respect to methanol than \ce{CH3CHO}, \ce{C2H5OH}, GA, and EG. This may hint at some chemical links among DME, MF, and methanol, such as having the same precursor (e.g., \ce{CH3O}) in their formation routes.
      \item The ratios among pairs of O-COMs also show the same trends between the low-mass and the high-mass groups. In particular, the ratios of DME/MF and \ce{CH3CHO}/GA are quite consistently around 1, while others show larger scatter.
      \item Previous experiments show that the ratios of MF/GA and MF/EG can be significantly enhanced by UV irradiation, but the observed values are even higher than the laboratory ones. The ratio of GA/EG is not affected by UV in experiments and match well with our observations.
      \item Comparison with the state-of-the-art models shows consistency for some O-COMs such as DME and MF. The differences between models and observations may result from less constrained gas-phase chemistry (\ce{CH3CHO} and \ce{C2H5OH}) and the different emitting areas under limited spatial resolutions (GA and EG). Chemical simulations and laboratory experiments are important to testing and exploring possible explanations.
      \item 
      The comparison of O-COM ratios among sources at different evolutionary stages may probe the chemical inheritance during star formation processes. However, observations toward larger samples are needed to enable statistical analyses.
   \end{enumerate}

ALMA line surveys toward large samples of star-forming regions are shedding light on the origin of COMs and the chemical evolution in the early stages of star formation. We look forward to linking our results to more ALMA observations on gas-phase COMs and the upcoming JWST mid-infrared data on solid-phase COMs to probe their formation history.

\begin{acknowledgements}
      The authors are grateful to the entire CoCCoA team for their interest in this study.
      This paper makes use of the following ALMA data: ADS/JAO.ALMA\#2019.1.00246.S. ALMA is a partnership of ESO (representing its member states), NSF (USA) and NINS (Japan), together with NRC (Canada), MOST and ASIAA (Taiwan), and KASI (Republic of Korea), in cooperation with the Republic of Chile. The Joint ALMA Observatory is operated by ESO, AUI/NRAO, and NAOJ.
      Astrochemistry in Leiden is supported by the Netherlands Research School for Astronomy (NOVA), by funding from the European Research Council (ERC) under the European Union’s Horizon 2020 research and innovation programme (grant agreement No. 101019751 MOLDISK), by the Dutch Research Council (NWO) grants TOP-1 614.001.751 and 618.000.001, and by the Danish National Research Foundation through the Center of Excellence “InterCat” (Grant agreement no.: DNRF150).
      The National Radio Astronomy Observatory is a facility of the National Science Foundation operated under cooperative agreement by Associated Universities, Inc.
      Y.C. acknowledges Jiao He for useful discussion about astrochemical experiments.
      J.K.J. acknowledges support from the Independent Research Fund Denmark (grant number 0135-00123B).
      M.N.D. is supported by the Swiss National Science Foundation (SNSF) Ambizione grant 180079, the Center for Space and Habitability (CSH) Fellowship, and the IAU Gruber Foundation Fellowship.
      S.I. acknowledges the Danish National Research Foundation through the Center of Excellence “InterCat” (Grant agreement no.: DNRF150).
      B.M.K. is supported by the Swiss National Science Foundation (SNSF) Ambizione grant 180079.
      N.F.W.L. acknowledges funding by the Swiss National Science Foundation Ambizione grant 193453.
\end{acknowledgements}

\bibliographystyle{aa} 
\bibliography{references} 

\newpage
\onecolumn

\begin{appendix} 
\section{Additional figures}\label{sec_additional_fig}
\begin{figure*}[h!]
    \centering
    \includegraphics[width=0.95\textwidth]{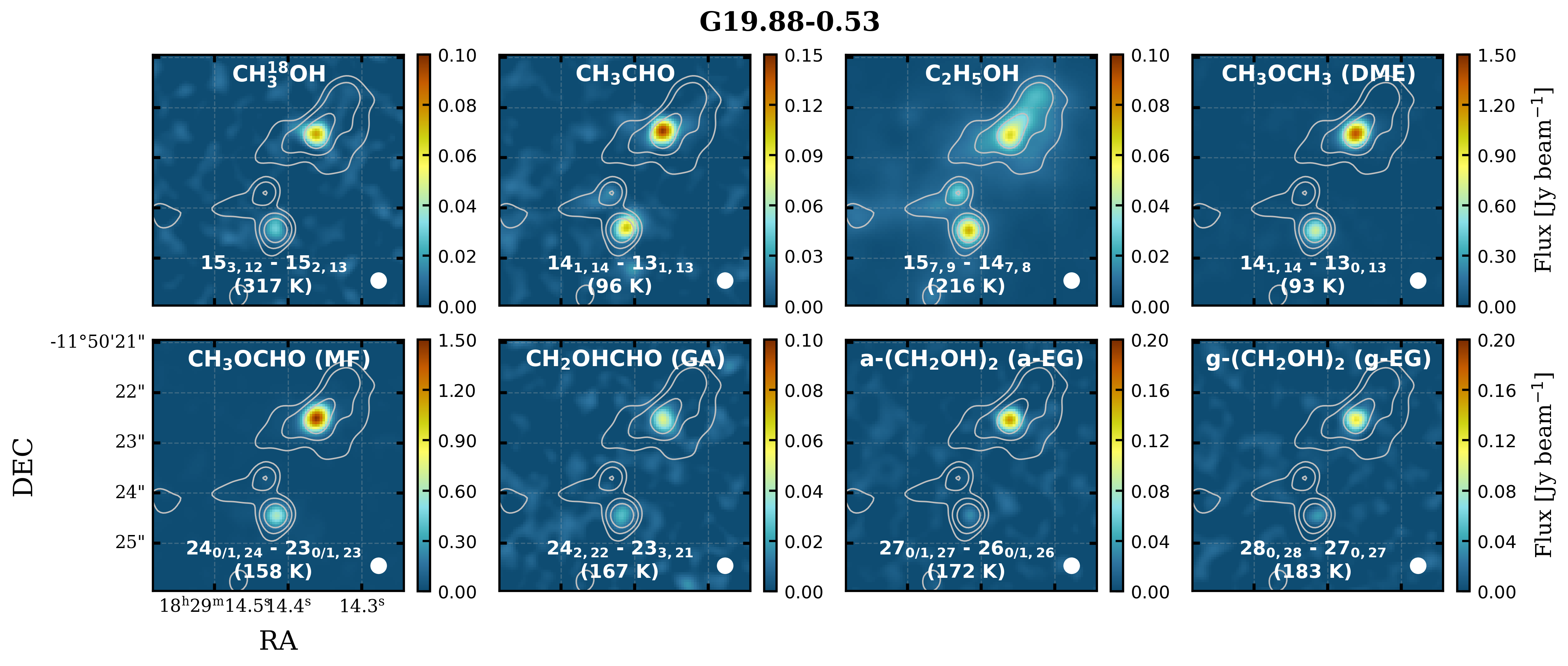}
    \caption{Moment 0 maps of selected O-COMs in the representative source G19.88-0.53. Contours and beams are the same as those in Fig. \ref{fig:maps}. We chose the strongest and unblended (or less blended) line in the upper tuning to make moment 0 map of each species, of which the quantum numbers and the upper energy level are indicated by white bold text at the bottom of each panel. The images are integrated over the FWHM with respect to the line center.}
    \label{fig:OCOM_mom0_G19.88}
\end{figure*}

\begin{figure*}[h!]
    \centering
    \begin{subfigure}{0.95\textwidth}
      \includegraphics[width=\hsize]{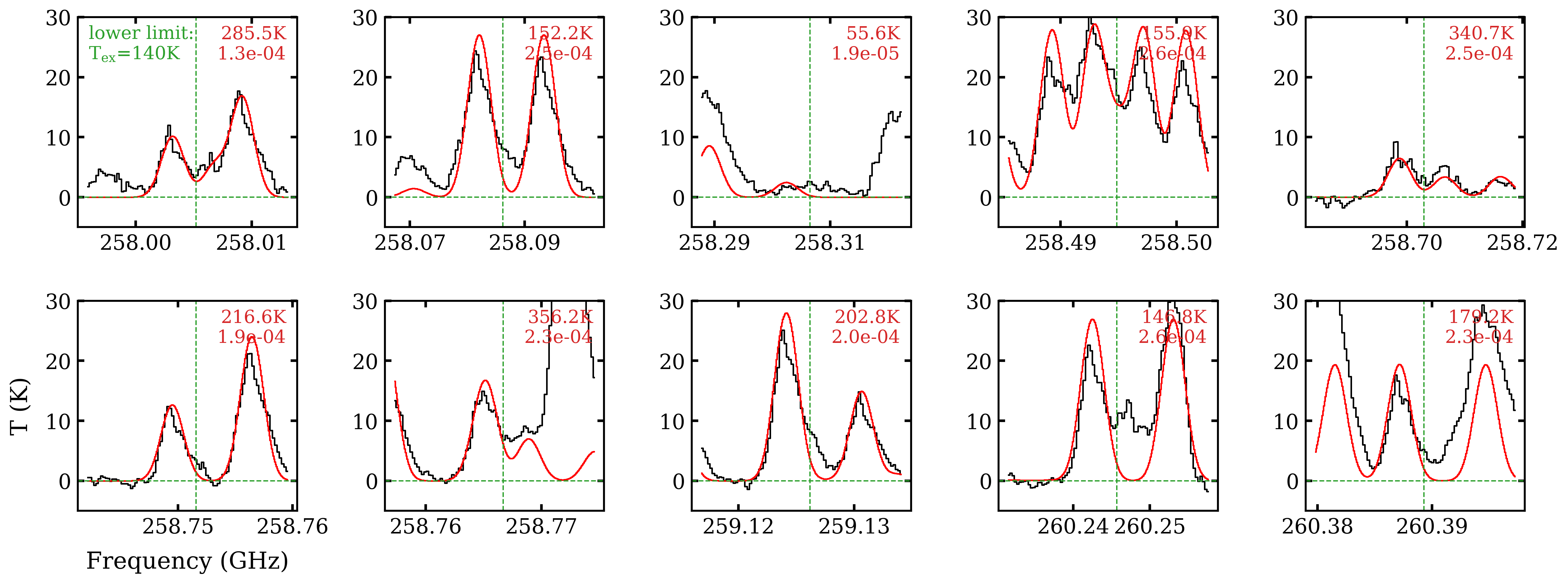}
      \caption{}
    \end{subfigure}
    \begin{subfigure}{0.95\textwidth}
      \includegraphics[width=\textwidth]{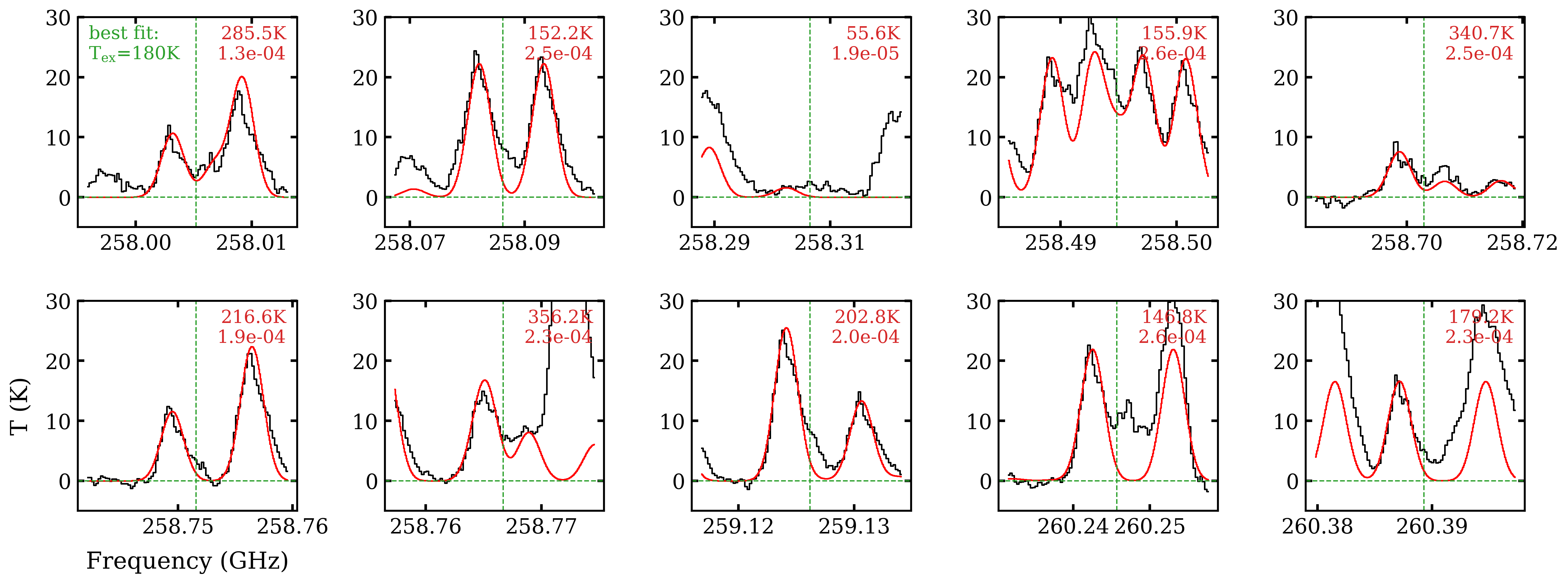}
      \caption{}
    \end{subfigure}
    \begin{subfigure}{0.95\textwidth}
      \includegraphics[width=\hsize]{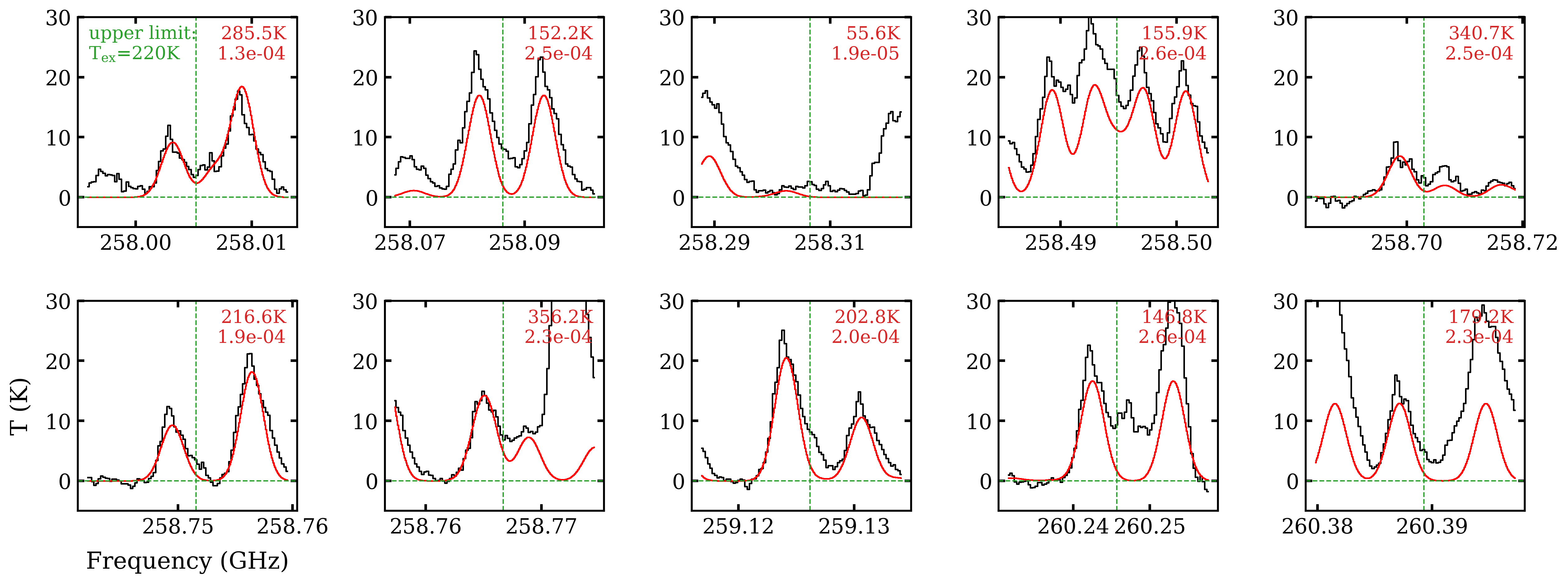}
      \caption{}
    \end{subfigure}
    \caption{An example of the best fit and the upper and lower limits of excitation temperatures ($T_\mathrm{ex}$) of MF (\ce{CH3OCHO}) in the spectrum of G19.88-0.53. Ten transitions with different upper energy $E_\mathrm{up}$ are chosen for comparison. Panel (b) shows the best fit while panel (a) and panel (c) show the lower and upper limits of $T_\mathrm{ex}$, respectively.}
    \label{fig:uncertainty}
\end{figure*}

\begin{figure*}[h!]
    \centering
    \includegraphics[width=\textwidth]{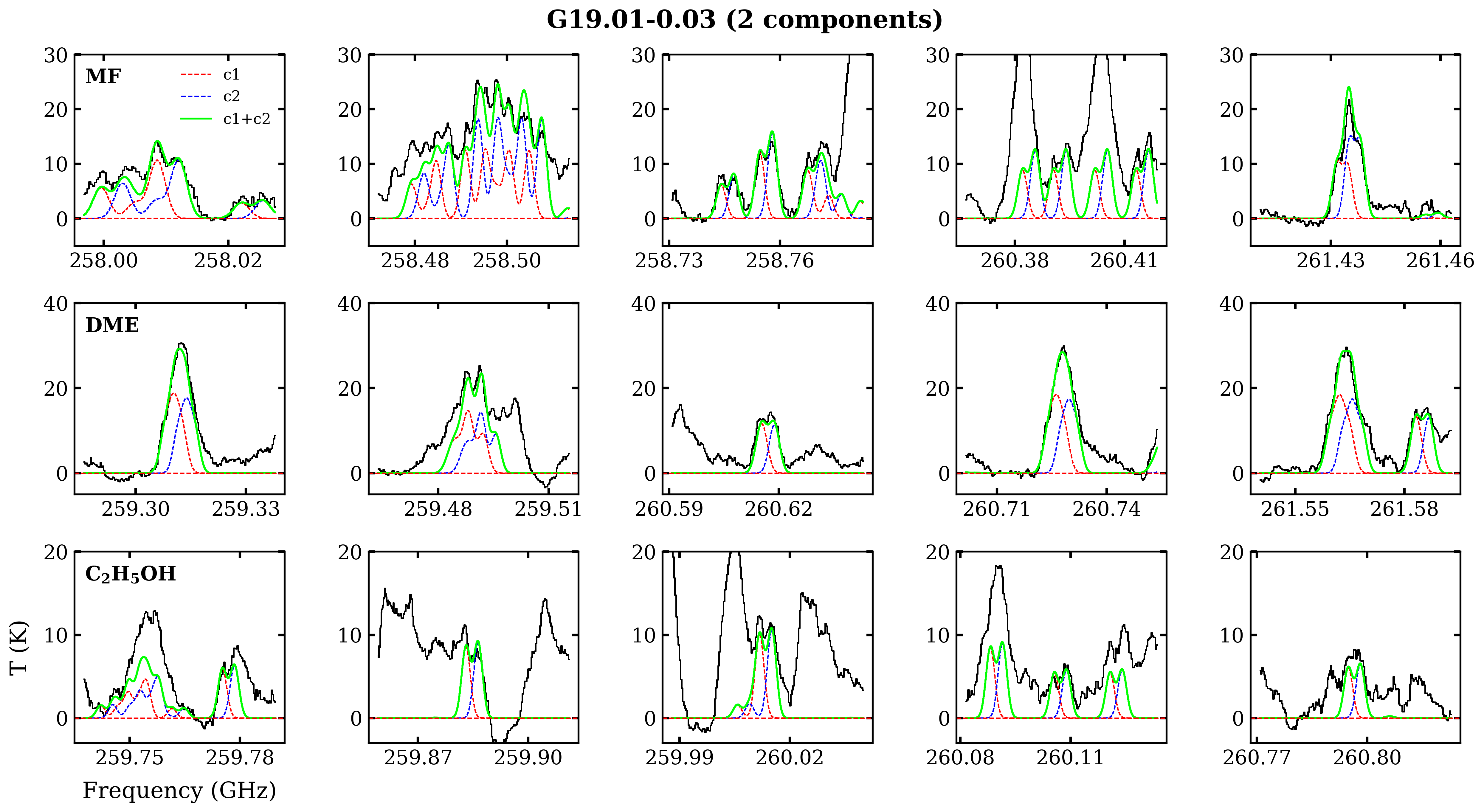}
    \caption{Example of spectral fitting that introduces two components, represented by three O-COMs of G19.01-0.03. The observed spectrum is shown in solid black. The two components are shown in dashed red and blue, respectively. The sum of the two components is shown in solid green. See Table \ref{table:results_1} for other sources and species that are fitted by two components.}
    \label{fig:fitting_example_2components}
\end{figure*}

\begin{figure*}[h!]
    \centering
    \includegraphics[width=0.95\textwidth]{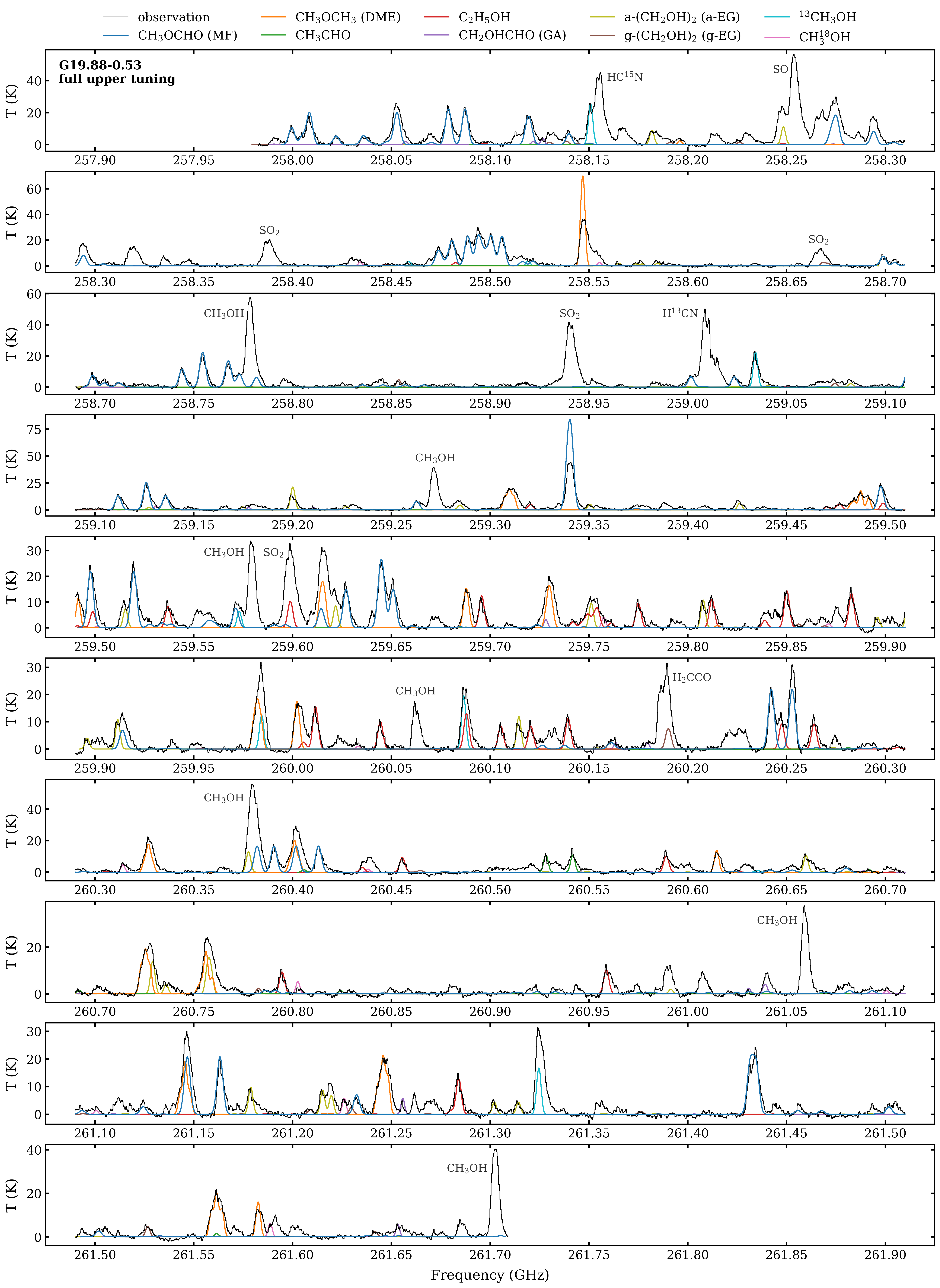}
    \caption{The same as Fig.~\ref{fig:fitting_example} but for the full upper tuning. Some identified lines from other species than the selected O-COMs are labeled in gray text.}
    \label{fig:fitting_example_full}
\end{figure*}

\begin{figure*}[h!]
    \centering
    \begin{subfigure}{0.95\textwidth}
      \includegraphics[width=\hsize]{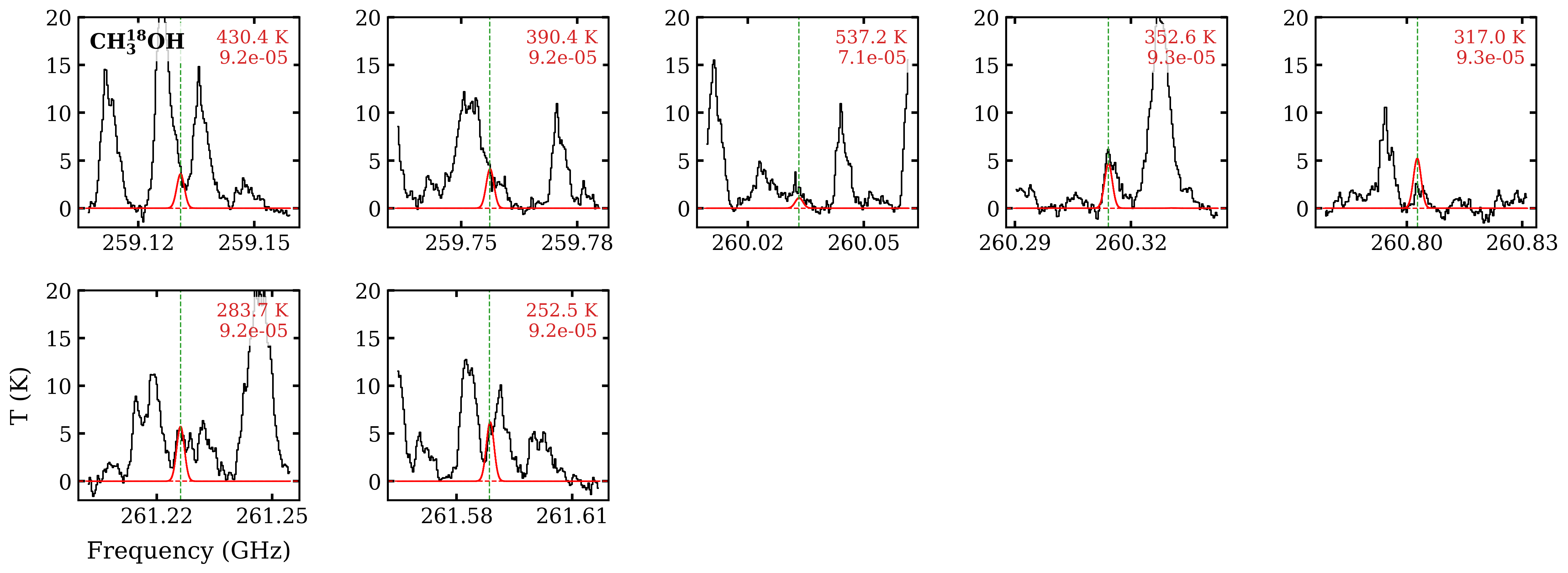}
      \caption{}
    \end{subfigure}
    \begin{subfigure}{0.95\textwidth}
      \includegraphics[width=\textwidth]{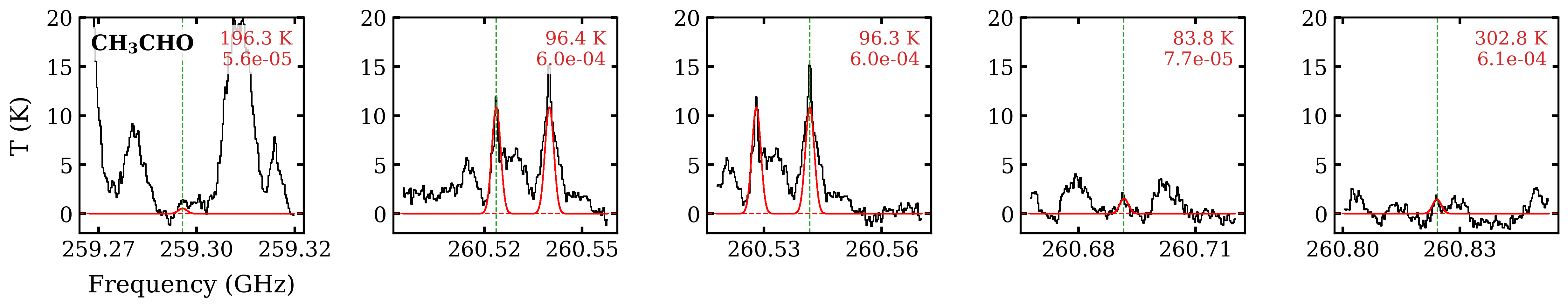}
      \caption{}
    \end{subfigure}
    \begin{subfigure}{0.95\textwidth}
      \includegraphics[width=\hsize]{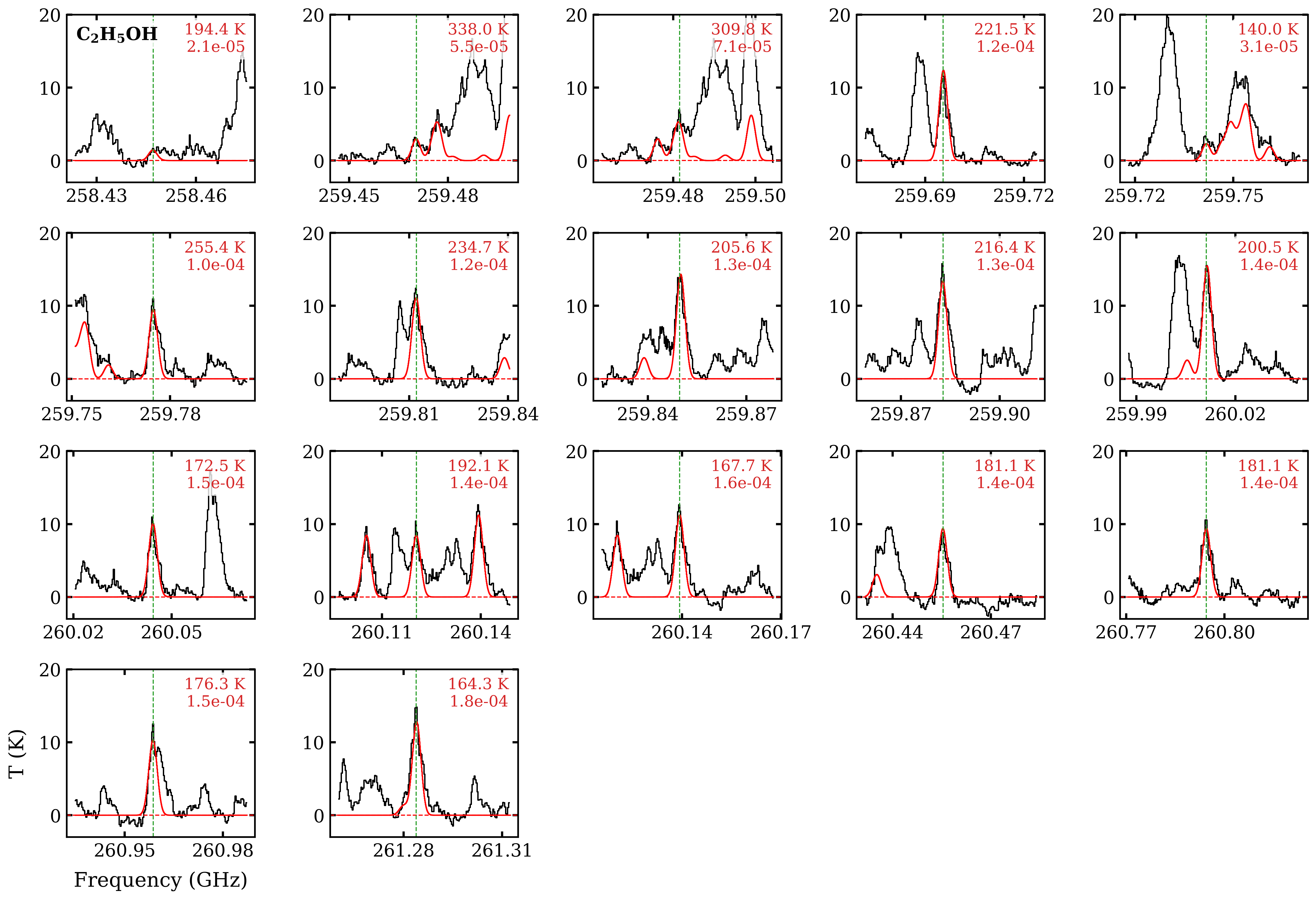}
      \caption{}
    \end{subfigure}
    \caption{The best-fit model for CH$_3^{18}$OH, \ce{CH3CHO} and \ce{C2H5OH}, taking G19.88-0.53 as an example. Only selected unblended lines are shown. In each panel, the centered transition is indicated by the vertical dashed green line; the upper energy level (in K) and the Einstein A coefficient (in s$^{-1}$) are listed in red text in the upper right.}
\label{fig:fitting_per_species_1}
\end{figure*}

\begin{figure*}[h!]
    \centering
    \begin{subfigure}{0.95\textwidth}
      \includegraphics[width=\hsize]{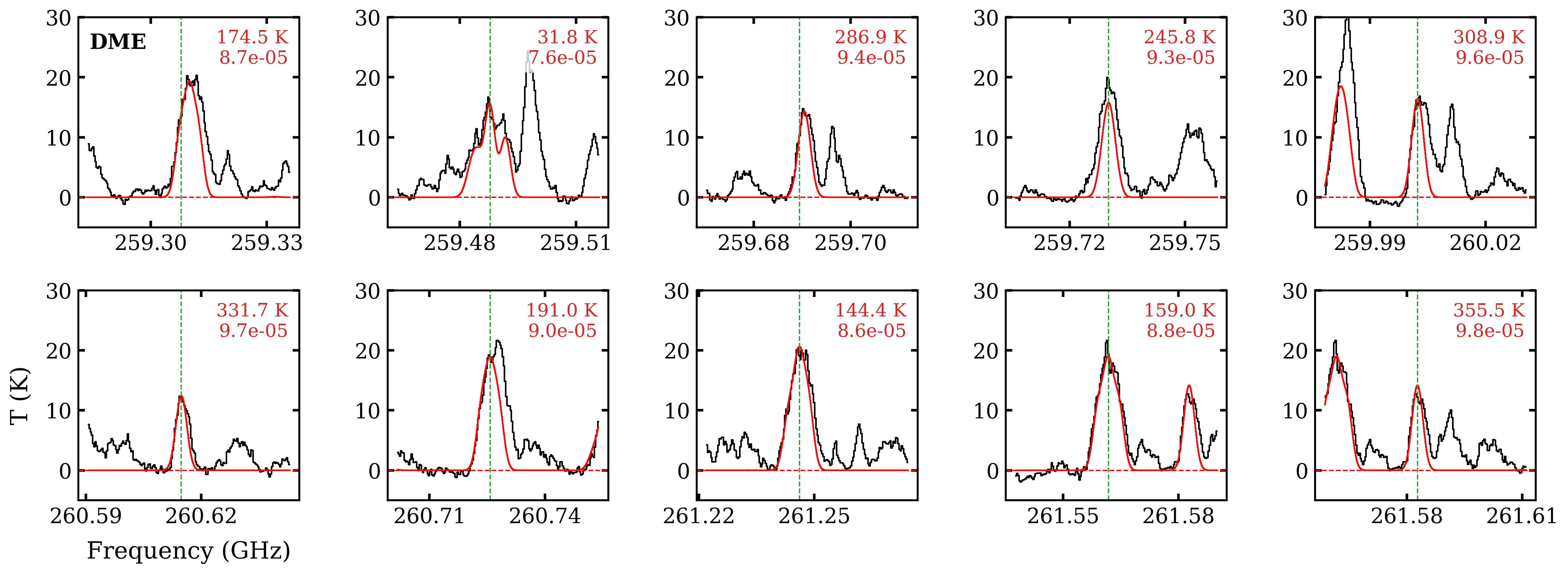}
      \caption{}
    \end{subfigure}
    \begin{subfigure}{0.95\textwidth}
      \includegraphics[width=\hsize]{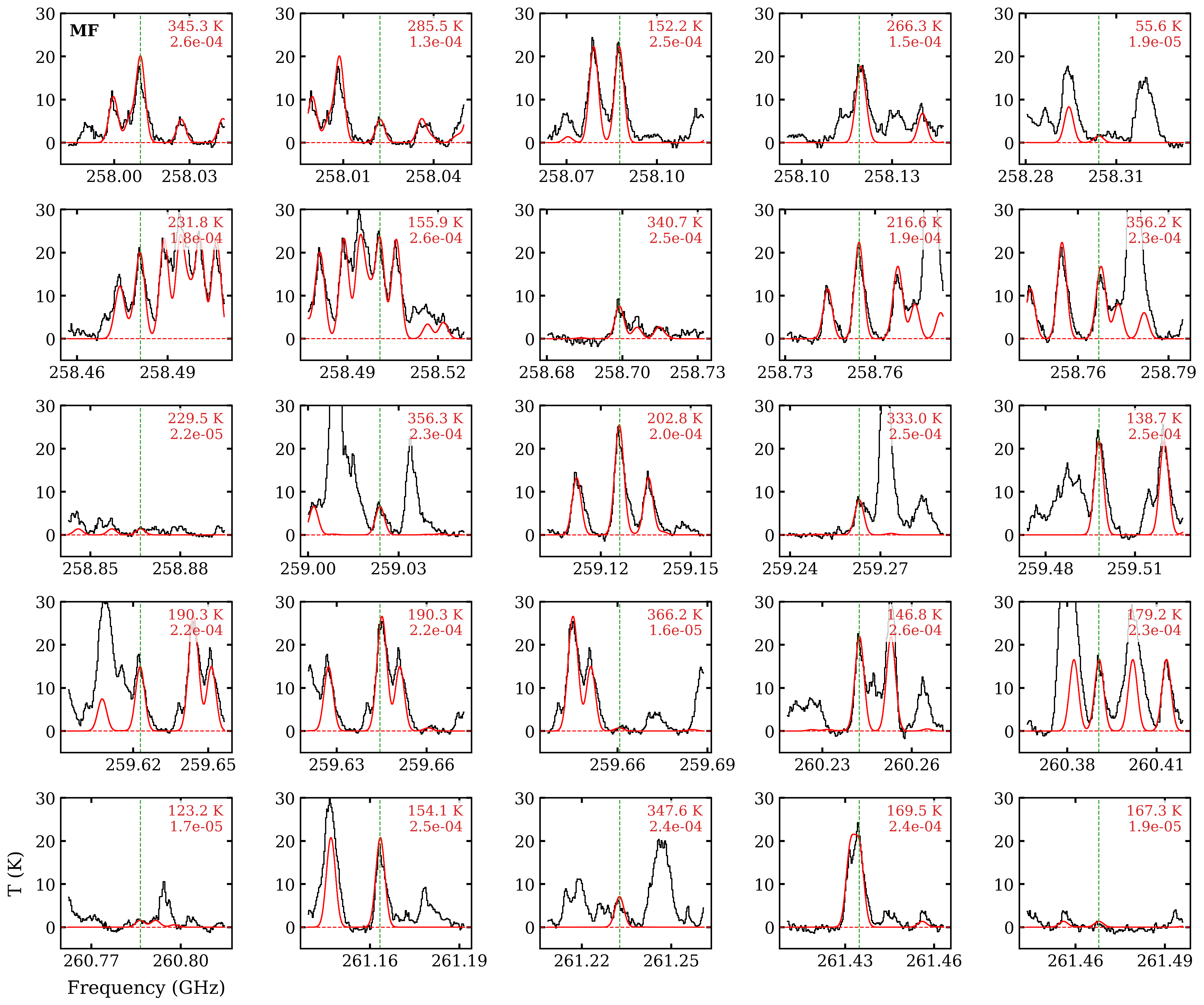}
      \caption{}
    \end{subfigure}
    \caption{The same as Fig. \ref{fig:fitting_per_species_1} but for \ce{CH3OCH3} (DME) and \ce{CH3OCHO} (MF).}
    \label{fig:fitting_per_species_2}
\end{figure*}

\begin{figure*}[h!]
    \centering
    \begin{subfigure}{0.95\textwidth}
      \includegraphics[width=\hsize]{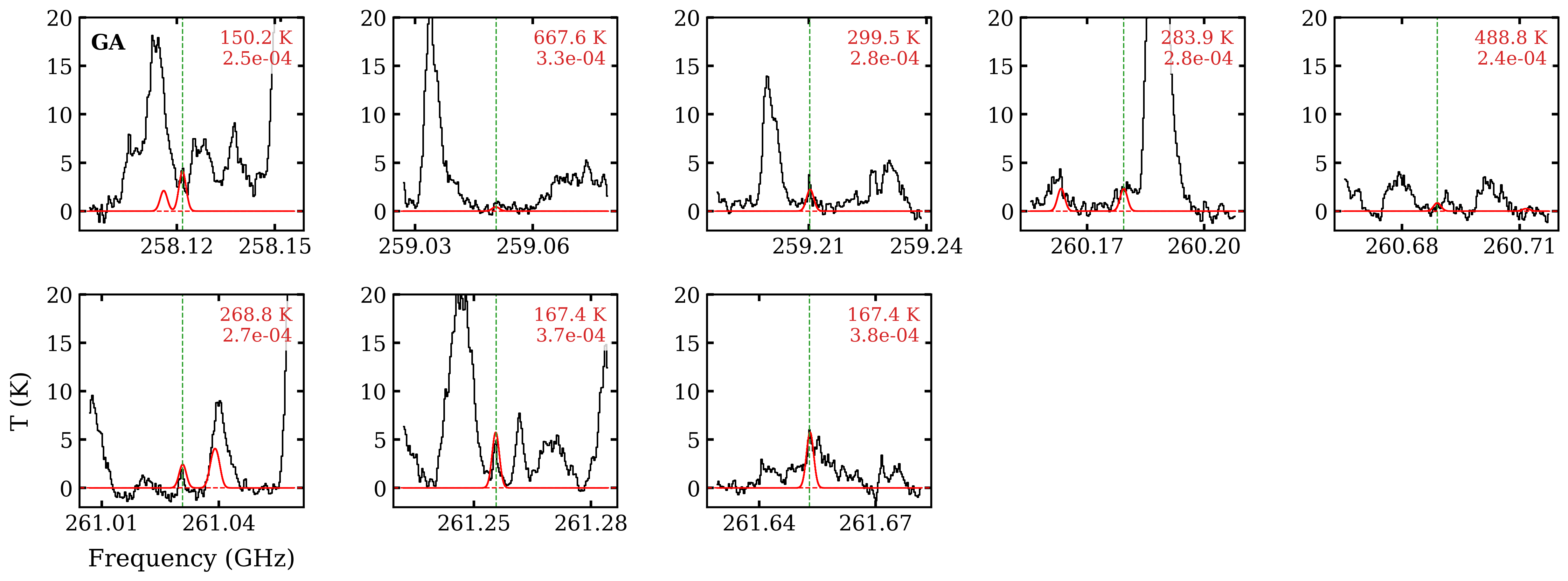}
      \caption{}
    \end{subfigure}
    \begin{subfigure}{0.95\textwidth}
      \includegraphics[width=\hsize]{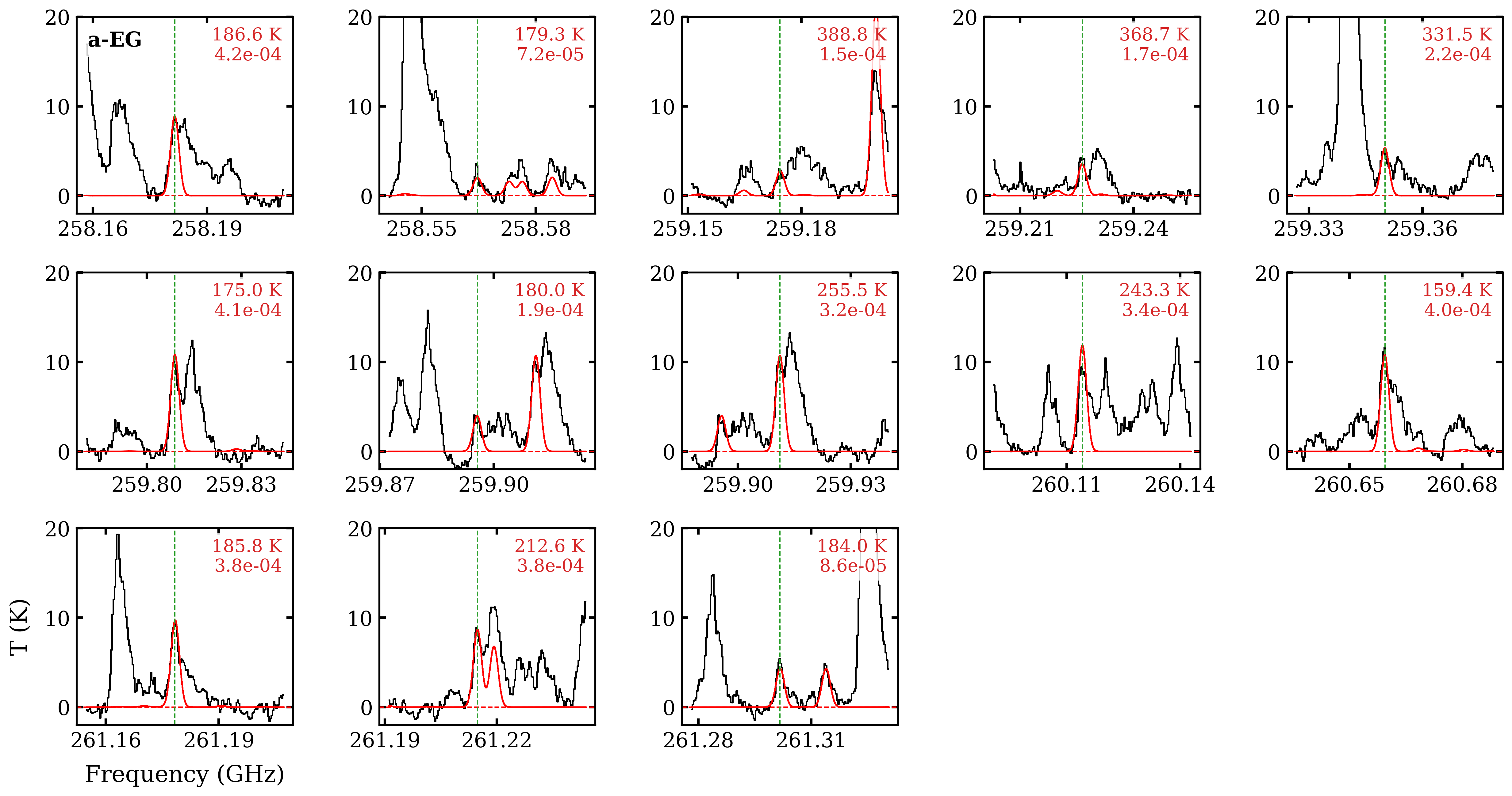}
      \caption{}
    \end{subfigure}
    \begin{subfigure}{0.95\textwidth}
      \includegraphics[width=\hsize]{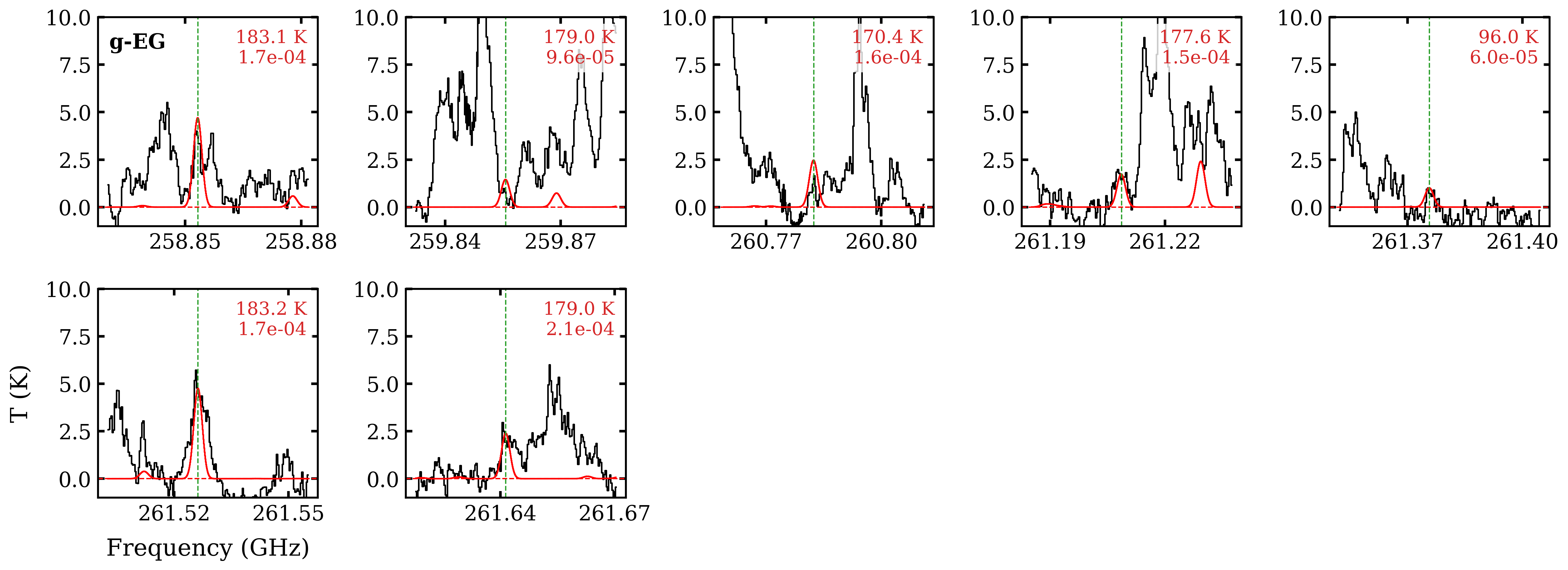}
      \caption{}
    \end{subfigure}
    \caption{The same as Fig. \ref{fig:fitting_per_species_1} but for \ce{CH2OHCHO} (GA), $a$- and $g$-\ce{(CH2OH)2} (EG).}
    \label{fig:fitting_per_species_3}
\end{figure*}

\begin{figure*}[h!]
    \centering
    \includegraphics[width=\textwidth]{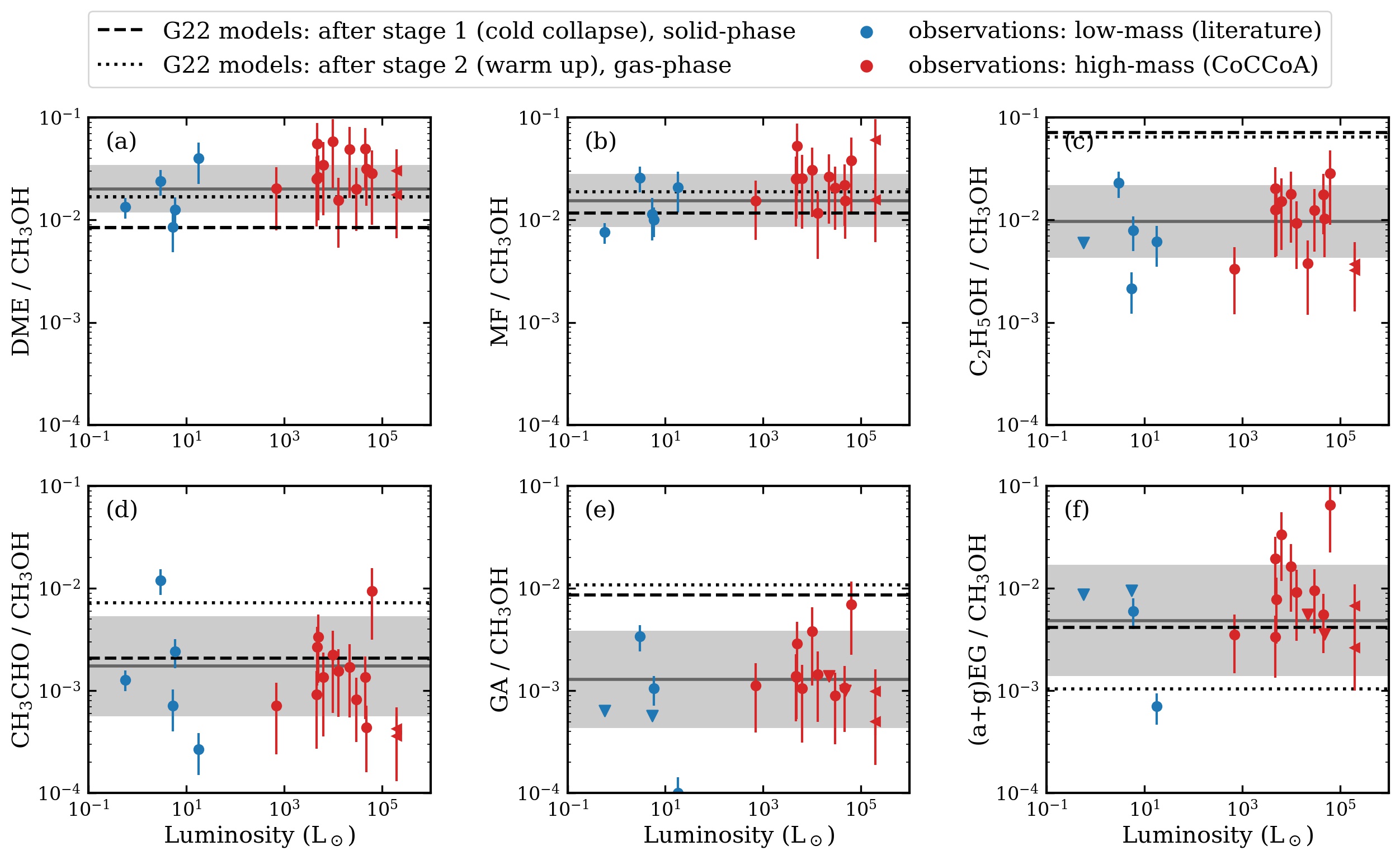}
    \caption{The same as Fig.~\ref{fig:O-COM ratios} but includes the intercept error in Eq.~(\ref{eqn:18O_ratio}).}
    \label{fig:O-COM ratio full error}
\end{figure*}

Figure \ref{fig:OCOM_mom0_G19.88} shows the moment 0 maps of selected O-COMs in G19.88-0.53. Figure \ref{fig:uncertainty} shows how the fitting looks like with the best fit and the upper and lower limits of $T_\mathrm{ex}$, taking MF and G19.88-0.53 as an example. Figure \ref{fig:fitting_example_2components} shows the spectral fitting with two components, taking G19.01-0.03 as an example. Figure \ref{fig:fitting_example_full} shows the full spectral fitting results of selected O-COMs for the representative source G19.88-0.53. Figures \ref{fig:fitting_per_species_1}-\ref{fig:fitting_per_species_3} show the zoom-in fitting of each O-COM. Figure \ref{fig:O-COM ratio full error} shows the ratios of the six selected O-COMs with respect to methanol, which is the same as Fig. \ref{fig:O-COM ratios} but with the intercept error in Eq.~(\ref{eqn:18O_ratio}) included.

\clearpage
\section{Additional tables}
Table \ref{table:transition} lists the identified transitions of two minor isotopologues of \ce{CH3OH} and the six selected O-COMs focused in this paper.
\begin{longtable}[t!]{rrclcll}
\caption{Identified transitions of O-COMs.\label{table:transition}}\\
\hline\hline
Species & \multicolumn{3}{c}{Transition} & Frequency (MHz) & $E_\mathrm{up}$ (K) & $A_\mathrm{ij}$ (s$^{-1}$) \\
Database & & & & & & $a(b)$ = $a\times10^b$\\
\hline
\endfirsthead
\caption{continued.}\\
\hline Species & \multicolumn{3}{c}{Transition} & Frequency (MHz) & $E_\mathrm{up}$ (K) & $A_\mathrm{ij}$ (s$^{-1}$) \\
Database & & & & & & $a(b)$ = $a\times10^b$\\
\hline
\endhead
\hline
\endfoot
\hline\hline
\endlastfoot

$^{13}$CH$_3$OH, vt=0,1 & 16 3 14 +0 & -- & 16 2 15 -0 & 258153.004 & 358.01 & 9.03(-5) \\
CDMS & 19 5 15 +0 & -- & 20 4 16 +0 & 258460.830 & 568.65 & 2.95(-5) \\
 & 19 5 14 -0 & -- & 20 4 17 -0 & 258521.376 & 568.65 & 2.95(-5) \\
 & 17 3 15 +0 & -- & 17 2 16 -0 & 259036.489 & 396.48 & 9.11(-5) \\
 & 12 -2 10 1 & -- & 13 -3 10 1 & 259575.198 & 540.96 & 7.21(-5) \\
 & 2 1 1 0 & -- & 1 0 1 0 & 259986.530 & 27.85 & 5.46(-5) \\
 & 18 3 16 +0 & -- & 18 2 17 -0 & 260088.839 & 437.21 & 9.21(-5) \\
 & 24 -3 21 1 & -- & 23 -2 21 1 & 260637.531 & 999.80 & 6.76(-5) \\
 & 19 3 17 +0 & -- & 19 2 18 -0 & 261326.838 & 480.20 & 9.31(-5) \\
\hline
CH$_3^{18}$OH, v=0-2 & 19 3 16 0 & -- & 19 2 17 0 & 258436.476 & 472.65 & 9.22(-5) \\
CDMS & 15 1 14 0 & -- & 14 2 13 0 & 258557.479 & 283.36 & 4.05(-5) \\
 & 18 3 15 0 & -- & 18 2 16 0 & 259133.097 & 430.41 & 9.24(-5) \\
 & 5 4 2 1 & -- & 6 3 4 1 & 259391.755 & 121.80 & 8.35(-6) \\
 & 17 3 14 0 & -- & 17 2 15 0 & 259759.660 & 390.39 & 9.25(-5) \\
 & 17 2 15 2 & -- & 16 3 13 2 & 259873.428 & 362.98 & 2.95(-5) \\
 & 12 2 10 4 & -- & 13 3 10 4 & 260035.462 & 537.16 & 7.08(-5) \\
 & 16 3 13 0 & -- & 16 2 14 0 & 260316.321 & 352.59 & 9.26(-5) \\
 & 15 1 15 1 & -- & 14 2 13 1 & 260440.344 & 272.16 & 2.76(-5) \\
 & 15 3 12 0 & -- & 15 2 13 0 & 260804.867 & 317.02 & 9.26(-5) \\
 & 15 4 11 2 & -- & 16 3 13 2 & 261102.783 & 363.04 & 3.00(-5) \\
 & 14 3 11 0 & -- & 14 2 12 0 & 261228.369 & 283.66 & 9.25(-5) \\
 & 13 3 10 0 & -- & 13 2 11 0 & 261590.879 & 252.53 & 9.24(-5) \\
\hline
\ce{CH3CHO} & 19 3 17 1 & -- & 19 2 18 1 & 259298.775 & 196.31 & 5.60(-5) \\
JPL & 13 1 12 6 & -- & 12 1 11 6 & 259375.784 & 459.31 & 5.70(-4) \\
 & 13 2 12 7 & -- & 12 2 11 7 & 259761.410 & 474.77 & 5.34(-4) \\
 & 19 3 17 0 & -- & 19 2 18 0 & 260283.405 & 196.35 & 5.69(-5) \\
 & 13 1 13 1 & -- & 12 0 12 2 & 260408.016 & 83.89 & 7.68(-5) \\
 & 14 1 14 1 & -- & 13 1 13 1 & 260530.403 & 96.39 & 6.02(-4) \\
 & 14 1 14 0 & -- & 13 1 13 0 & 260544.019 & 96.32 & 6.01(-4) \\
 & 13 1 13 0 & -- & 12 0 12 0 & 260694.002 & 83.82 & 7.73(-5) \\
 & 14 1 14 3 & -- & 13 1 13 3 & 260826.516 & 302.82 & 6.12(-4) \\
\hline
\ce{C2H5OH}, v=0 & 19 2 18 1 & -- & 18 3 16 0 & 258099.207 & 222.51 & 1.42(-5) \\
CDMS & 16 4 13 1 & -- & 16 3 13 0 & 258449.274 & 194.44 & 2.12(-5) \\
 & 25 1 24 2 & -- & 25 0 25 2 & 258484.664 & 272.70 & 3.90(-5) \\
 & 14 3 11 0 & -- & 13 2 11 1 & 259322.639 & 155.72 & 7.25(-5) \\
 & 15 12 3 1 & -- & 14 12 2 1 & 259472.683 & 337.99 & 5.52(-5) \\
 & 15 12 4 1 & -- & 14 12 3 1 & 259472.683 & 337.99 & 5.52(-5) \\
 & 15 13 2 1 & -- & 14 13 1 1 & 259477.660 & 368.56 & 3.81(-5) \\
 & 15 13 3 1 & -- & 14 13 2 1 & 259477.660 & 368.56 & 3.81(-5) \\
 & 15 11 4 1 & -- & 14 11 3 1 & 259479.315 & 309.83 & 7.09(-5) \\
 & 15 11 5 1 & -- & 14 11 4 1 & 259479.315 & 309.83 & 7.09(-5) \\
 & 15 10 5 1 & -- & 14 10 4 1 & 259501.052 & 284.11 & 8.53(-5) \\
 & 15 10 6 1 & -- & 14 10 5 1 & 259501.052 & 284.11 & 8.53(-5) \\
 & 15 9 6 1 & -- & 14 9 5 1 & 259539.131 & 260.82 & 9.82(-5) \\
 & 15 9 7 1 & -- & 14 9 6 1 & 259539.131 & 260.82 & 9.82(-5) \\
 & 15 8 8 1 & -- & 14 8 7 1 & 259601.059 & 239.96 & 1.10(-4) \\
 & 15 8 7 1 & -- & 14 8 6 1 & 259601.060 & 239.96 & 1.10(-4) \\
 & 15 7 9 1 & -- & 14 7 8 1 & 259697.897 & 221.55 & 1.20(-4) \\
 & 15 7 8 1 & -- & 14 7 7 1 & 259697.903 & 221.55 & 1.20(-4) \\
 & 13 1 12 1 & -- & 12 2 10 0 & 259744.133 & 140.01 & 3.11(-5) \\
 & 18 4 15 1 & -- & 18 3 15 0 & 259748.793 & 223.63 & 3.48(-5) \\
 & 15 11 4 0 & -- & 14 11 3 0 & 259751.561 & 304.31 & 7.53(-5) \\
 & 15 11 5 0 & -- & 14 11 4 0 & 259751.561 & 304.31 & 7.53(-5) \\
 & 15 12 3 0 & -- & 14 12 2 0 & 259754.443 & 332.42 & 5.88(-5) \\
 & 15 12 4 0 & -- & 14 12 3 0 & 259754.443 & 332.42 & 5.88(-5) \\
 & 15 10 5 0 & -- & 14 10 4 0 & 259756.535 & 278.65 & 9.03(-5) \\
 & 15 10 6 0 & -- & 14 10 5 0 & 259756.535 & 278.65 & 9.03(-5) \\
 & 15 13 2 0 & -- & 14 13 1 0 & 259763.493 & 362.98 & 4.08(-5) \\
 & 15 13 3 0 & -- & 14 13 2 0 & 259763.493 & 362.98 & 4.08(-5) \\
 & 15 9 6 0 & -- & 14 9 5 0 & 259777.143 & 255.43 & 1.04(-4) \\
 & 15 9 7 0 & -- & 14 9 6 0 & 259777.143 & 255.43 & 1.04(-4) \\
 & 15 14 1 0 & -- & 14 14 0 0 & 259778.002 & 395.98 & 2.13(-5) \\
 & 15 14 2 0 & -- & 14 14 1 0 & 259778.002 & 395.98 & 2.13(-5) \\
 & 15 8 8 0 & -- & 14 8 7 0 & 259814.437 & 234.67 & 1.16(-4) \\
 & 15 8 7 0 & -- & 14 8 6 0 & 259814.438 & 234.67 & 1.16(-4) \\
 & 32 5 28 2 & -- & 32 4 29 2 & 259841.221 & 477.18 & 1.17(-4) \\
 & 15 6 10 1 & -- & 14 6 9 1 & 259852.185 & 205.59 & 1.29(-4) \\
 & 15 6 9 1 & -- & 14 6 8 1 & 259852.573 & 205.59 & 1.29(-4) \\
 & 15 7 9 0 & -- & 14 7 8 0 & 259885.072 & 216.37 & 1.27(-4) \\
 & 15 7 8 0 & -- & 14 7 7 0 & 259885.079 & 216.37 & 1.27(-4) \\
 & 25 2 24 2 & -- & 25 1 25 2 & 260007.736 & 272.77 & 3.96(-5) \\
 & 15 6 10 0 & -- & 14 6 9 0 & 260013.579 & 200.53 & 1.36(-4) \\
 & 15 6 9 0 & -- & 14 6 8 0 & 260014.001 & 200.53 & 1.36(-4) \\
 & 15 3 13 1 & -- & 14 3 12 1 & 260046.644 & 172.46 & 1.46(-4) \\
 & 15 1 14 0 & -- & 14 1 13 0 & 260090.201 & 159.58 & 1.75(-4) \\
 & 15 5 11 1 & -- & 14 5 10 1 & 260107.609 & 192.10 & 1.37(-4) \\
 & 15 5 10 1 & -- & 14 5 9 1 & 260122.761 & 192.10 & 1.37(-4) \\
 & 15 3 13 0 & -- & 14 3 12 0 & 260141.678 & 167.72 & 1.59(-4) \\
 & 15 5 11 0 & -- & 14 5 10 0 & 260249.761 & 187.15 & 1.45(-4) \\
 & 15 5 10 0 & -- & 14 5 9 0 & 260266.127 & 187.15 & 1.45(-4) \\
 & 7 1 6 1 & -- & 6 0 6 0 & 260437.554 & 86.70 & 5.46(-5) \\
 & 15 4 12 1 & -- & 14 4 11 1 & 260457.726 & 181.10 & 1.43(-4) \\
 & 15 4 12 0 & -- & 14 4 11 0 & 260591.330 & 176.26 & 1.52(-4) \\
 & 31 4 28 2 & -- & 30 5 25 2 & 260627.301 & 437.93 & 1.86(-5) \\
 & 15 4 11 1 & -- & 14 4 10 1 & 260796.868 & 181.14 & 1.43(-4) \\
 & 15 4 11 0 & -- & 14 4 10 0 & 260960.989 & 176.30 & 1.53(-4) \\
 & 20 3 18 1 & -- & 19 4 16 0 & 261282.323 & 247.21 & 2.17(-5) \\
 & 15 1 14 1 & -- & 14 1 13 1 & 261286.306 & 164.30 & 1.79(-4) \\
\hline
\ce{CH3OCH3} (DME), v=0 & 30 5 26 3 & -- & 29 6 23 3 & 258197.879 & 459.88 & 2.47(-5) \\
CDMS & 30 5 26 5 & -- & 29 6 23 5 & 258197.880 & 459.88 & 2.47(-5) \\
 & 30 5 26 1 & -- & 29 6 23 1 & 258198.172 & 459.88 & 2.47(-5) \\
 & 30 5 26 0 & -- & 29 6 23 0 & 258198.464 & 459.88 & 2.47(-5) \\
 & 30 14 17 3 & -- & 31 15 17 3 & 258278.183 & 738.73 & 1.24(-5) \\
 & 14 1 14 3 & -- & 13 0 13 3 & 258548.819 & 93.33 & 1.31(-4) \\
 & 14 1 14 5 & -- & 13 0 13 5 & 258548.819 & 93.33 & 1.31(-4) \\
 & 14 1 14 1 & -- & 13 0 13 1 & 258549.063 & 93.33 & 1.31(-4) \\
 & 14 1 14 0 & -- & 13 0 13 0 & 258549.308 & 93.33 & 1.31(-4) \\
 & 17 5 12 5 & -- & 17 4 13 5 & 259309.472 & 174.54 & 8.77(-5) \\
 & 17 5 12 3 & -- & 17 4 13 3 & 259309.758 & 174.54 & 8.74(-5) \\
 & 17 5 12 1 & -- & 17 4 13 1 & 259311.947 & 174.54 & 8.76(-5) \\
 & 17 5 12 0 & -- & 17 4 13 0 & 259314.279 & 174.54 & 8.77(-5) \\
 & 6 3 4 3 & -- & 5 2 3 3 & 259484.856 & 31.77 & 7.40(-5) \\
 & 6 3 4 5 & -- & 5 2 3 5 & 259486.616 & 31.77 & 7.65(-5) \\
 & 6 3 4 1 & -- & 5 2 3 1 & 259489.733 & 31.77 & 7.58(-5) \\
 & 6 3 4 0 & -- & 5 2 3 0 & 259493.749 & 31.77 & 7.65(-5) \\
 & 22 5 18 3 & -- & 22 4 19 3 & 259615.889 & 265.87 & 9.34(-5) \\
 & 22 5 18 5 & -- & 22 4 19 5 & 259615.896 & 265.87 & 9.34(-5) \\
 & 22 5 18 1 & -- & 22 4 19 1 & 259617.339 & 265.87 & 9.34(-5) \\
 & 22 5 18 0 & -- & 22 4 19 0 & 259618.785 & 265.87 & 9.34(-5) \\
 & 23 5 19 3 & -- & 23 4 20 3 & 259688.831 & 286.90 & 9.44(-5) \\
 & 23 5 19 5 & -- & 23 4 20 5 & 259688.835 & 286.90 & 9.44(-5) \\
 & 23 5 19 1 & -- & 23 4 20 1 & 259690.065 & 286.90 & 9.44(-5) \\
 & 23 5 19 0 & -- & 23 4 20 0 & 259691.297 & 286.90 & 9.44(-5) \\
 & 21 5 17 3 & -- & 21 4 18 3 & 259730.487 & 245.76 & 9.26(-5) \\
 & 21 5 17 5 & -- & 21 4 18 5 & 259730.502 & 245.76 & 9.26(-5) \\
 & 21 5 17 1 & -- & 21 4 18 1 & 259732.149 & 245.76 & 9.26(-5) \\
 & 21 5 17 0 & -- & 21 4 18 0 & 259733.804 & 245.76 & 9.26(-5) \\
 & 33 6 28 3 & -- & 32 7 25 3 & 259816.557 & 562.74 & 2.54(-5) \\
 & 33 6 28 5 & -- & 32 7 25 5 & 259816.583 & 562.74 & 2.54(-5) \\
 & 33 6 28 1 & -- & 32 7 25 1 & 259817.273 & 562.74 & 2.54(-5) \\
 & 33 6 28 0 & -- & 32 7 25 0 & 259817.976 & 562.74 & 2.54(-5) \\
 & 20 5 16 3 & -- & 20 4 17 3 & 259982.535 & 226.58 & 9.18(-5) \\
 & 20 5 16 5 & -- & 20 4 17 5 & 259982.568 & 226.58 & 9.18(-5) \\
 & 20 5 16 1 & -- & 20 4 17 1 & 259984.408 & 226.58 & 9.18(-5) \\
 & 20 5 16 0 & -- & 20 4 17 0 & 259986.265 & 226.58 & 9.18(-5) \\
 & 24 5 20 3 & -- & 24 4 21 3 & 260003.375 & 308.85 & 9.55(-5) \\
 & 24 5 20 5 & -- & 24 4 21 5 & 260003.377 & 308.85 & 9.55(-5) \\
 & 24 5 20 1 & -- & 24 4 21 1 & 260004.389 & 308.85 & 9.55(-5) \\
 & 24 5 20 0 & -- & 24 4 21 0 & 260005.402 & 308.85 & 9.55(-5) \\
 & 19 5 15 3 & -- & 19 4 16 3 & 260327.133 & 208.31 & 9.10(-5) \\
 & 19 5 15 5 & -- & 19 4 16 5 & 260327.201 & 208.31 & 9.10(-5) \\
 & 19 5 15 1 & -- & 19 4 16 1 & 260329.221 & 208.31 & 9.10(-5) \\
 & 19 5 15 0 & -- & 19 4 16 0 & 260331.275 & 208.31 & 9.10(-5) \\
 & 16 5 11 5 & -- & 16 4 12 5 & 260400.539 & 159.03 & 8.73(-5) \\
 & 16 5 11 3 & -- & 16 4 12 3 & 260401.135 & 159.03 & 8.64(-5) \\
 & 16 5 11 1 & -- & 16 4 12 1 & 260403.244 & 159.03 & 8.71(-5) \\
 & 16 5 11 0 & -- & 16 4 12 0 & 260405.649 & 159.03 & 8.73(-5) \\
 & 25 5 21 3 & -- & 25 4 22 3 & 260616.059 & 331.72 & 9.68(-5) \\
 & 25 5 21 5 & -- & 25 4 22 5 & 260616.059 & 331.72 & 9.68(-5) \\
 & 25 5 21 1 & -- & 25 4 22 1 & 260616.851 & 331.72 & 9.68(-5) \\
 & 25 5 21 0 & -- & 25 4 22 0 & 260617.643 & 331.72 & 9.68(-5) \\
 & 18 5 14 3 & -- & 18 4 15 3 & 260725.448 & 190.97 & 9.02(-5) \\
 & 18 5 14 5 & -- & 18 4 15 5 & 260725.587 & 190.97 & 9.02(-5) \\
 & 18 5 14 1 & -- & 18 4 15 1 & 260727.767 & 190.97 & 9.02(-5) \\
 & 18 5 14 0 & -- & 18 4 15 0 & 260730.017 & 190.97 & 9.02(-5) \\
 & 6 3 3 5 & -- & 5 2 4 5 & 260754.380 & 31.77 & 7.71(-5) \\
 & 6 3 3 3 & -- & 5 2 4 3 & 260756.140 & 31.77 & 7.47(-5) \\
 & 6 3 3 1 & -- & 5 2 4 1 & 260758.402 & 31.77 & 7.65(-5) \\
 & 6 3 3 0 & -- & 5 2 4 0 & 260761.524 & 31.77 & 7.72(-5) \\
 & 17 5 13 3 & -- & 17 4 14 3 & 261145.206 & 174.54 & 8.91(-5) \\
 & 17 5 13 5 & -- & 17 4 14 5 & 261145.492 & 174.54 & 8.94(-5) \\
 & 17 5 13 1 & -- & 17 4 14 1 & 261147.803 & 174.54 & 8.93(-5) \\
 & 17 5 13 0 & -- & 17 4 14 0 & 261150.257 & 174.54 & 8.94(-5) \\
 & 15 5 10 5 & -- & 15 4 11 5 & 261245.097 & 144.44 & 8.66(-5) \\
 & 15 5 10 3 & -- & 15 4 11 3 & 261246.331 & 144.44 & 8.30(-5) \\
 & 15 5 10 1 & -- & 15 4 11 1 & 261248.113 & 144.44 & 8.56(-5) \\
 & 15 5 10 0 & -- & 15 4 11 0 & 261250.488 & 144.44 & 8.66(-5) \\
 & 16 5 12 3 & -- & 16 4 13 3 & 261560.797 & 159.03 & 8.74(-5) \\
 & 16 5 12 5 & -- & 16 4 13 5 & 261561.393 & 159.03 & 8.84(-5) \\
 & 16 5 12 1 & -- & 16 4 13 1 & 261563.781 & 159.03 & 8.81(-5) \\
 & 16 5 12 0 & -- & 16 4 13 0 & 261566.471 & 159.03 & 8.84(-5) \\
 & 26 5 22 3 & -- & 26 4 23 3 & 261584.210 & 355.51 & 9.84(-5) \\
 & 26 5 22 5 & -- & 26 4 23 5 & 261584.210 & 355.51 & 9.84(-5) \\
 & 26 5 22 1 & -- & 26 4 23 1 & 261584.781 & 355.51 & 9.84(-5) \\
 & 26 5 22 0 & -- & 26 4 23 0 & 261585.353 & 355.51 & 9.84(-5) \\
\hline
\ce{CH3OCHO (MF)} & 21 8 13 3 & -- & 20 8 12 3 & 257906.128 & 365.98 & 2.22(-4) \\
JPL & 21 16 5 0 & -- & 20 16 4 0 & 257910.566 & 306.01 & 1.09(-4) \\
 & 21 16 6 0 & -- & 20 16 5 0 & 257910.566 & 306.01 & 1.09(-4) \\
 & 21 16 5 2 & -- & 20 16 4 2 & 257919.890 & 306.01 & 1.09(-4) \\
 & 21 16 6 1 & -- & 20 16 5 1 & 257933.830 & 306.00 & 1.09(-4) \\
 & 21 15 6 0 & -- & 20 15 5 0 & 258001.757 & 285.47 & 1.28(-4) \\
 & 21 15 7 0 & -- & 20 15 6 0 & 258001.757 & 285.47 & 1.28(-4) \\
 & 21 15 6 2 & -- & 20 15 5 2 & 258007.150 & 285.47 & 1.28(-4) \\
 & 24 0 24 3 & -- & 23 1 23 3 & 258010.378 & 345.26 & 4.19(-5) \\
 & 24 1 24 3 & -- & 23 1 23 3 & 258010.754 & 345.26 & 2.59(-4) \\
 & 24 0 24 3 & -- & 23 0 23 3 & 258010.754 & 345.26 & 2.59(-4) \\
 & 24 1 24 3 & -- & 23 0 23 3 & 258011.019 & 345.26 & 4.19(-5) \\
 & 21 15 7 1 & -- & 20 15 6 1 & 258024.240 & 285.46 & 1.28(-4) \\
 & 21 9 13 4 & -- & 20 9 12 4 & 258037.974 & 376.94 & 2.13(-4) \\
 & 24 10 15 1 & -- & 24 9 16 1 & 258041.077 & 243.70 & 2.22(-5) \\
 & 24 10 14 2 & -- & 24 9 15 2 & 258049.562 & 243.71 & 2.22(-5) \\
 & 24 10 14 0 & -- & 24 9 15 0 & 258052.000 & 243.70 & 2.23(-5) \\
 & 24 0 24 5 & -- & 23 1 23 4 & 258054.742 & 344.52 & 4.10(-5) \\
 & 24 1 24 4 & -- & 23 1 23 4 & 258055.043 & 344.52 & 2.60(-4) \\
 & 24 0 24 5 & -- & 23 0 23 5 & 258055.043 & 344.52 & 2.60(-4) \\
 & 24 1 24 4 & -- & 23 0 23 5 & 258055.296 & 344.52 & 4.10(-5) \\
 & 24 10 15 0 & -- & 24 9 16 0 & 258072.452 & 243.70 & 2.23(-5) \\
 & 22 2 20 2 & -- & 21 2 19 2 & 258081.042 & 152.23 & 2.52(-4) \\
 & 22 2 20 0 & -- & 21 2 19 0 & 258089.491 & 152.22 & 2.52(-4) \\
 & 21 14 7 0 & -- & 20 14 6 0 & 258121.191 & 266.26 & 1.45(-4) \\
 & 21 14 8 0 & -- & 20 14 7 0 & 258121.191 & 266.26 & 1.45(-4) \\
 & 21 14 7 2 & -- & 20 14 6 2 & 258122.660 & 266.25 & 1.45(-4) \\
 & 21 14 8 1 & -- & 20 14 7 1 & 258142.090 & 266.25 & 1.45(-4) \\
 & 21 13 8 2 & -- & 20 13 7 2 & 258274.950 & 248.38 & 1.61(-4) \\
 & 21 13 8 0 & -- & 20 13 7 0 & 258277.434 & 248.38 & 1.61(-4) \\
 & 21 13 9 0 & -- & 20 13 8 0 & 258277.434 & 248.38 & 1.61(-4) \\
 & 21 13 9 1 & -- & 20 13 8 1 & 258296.300 & 248.36 & 1.61(-4) \\
 & 11 5 7 0 & -- & 10 4 6 0 & 258306.279 & 55.60 & 1.92(-5) \\
 & 27 10 17 3 & -- & 27 9 18 3 & 258380.810 & 476.33 & 2.37(-5) \\
 & 27 10 18 4 & -- & 27 9 19 4 & 258425.612 & 476.33 & 2.36(-5) \\
 & 18 4 15 4 & -- & 17 3 14 5 & 258450.614 & 298.62 & 1.49(-5) \\
 & 23 1 22 2 & -- & 22 2 21 1 & 258475.052 & 155.91 & 3.57(-5) \\
 & 21 12 9 2 & -- & 20 12 8 2 & 258476.450 & 231.83 & 1.76(-4) \\
 & 23 1 22 0 & -- & 22 2 21 0 & 258480.586 & 155.90 & 3.57(-5) \\
 & 21 12 9 0 & -- & 20 12 8 0 & 258482.981 & 231.83 & 1.76(-4) \\
 & 21 12 10 0 & -- & 20 12 9 0 & 258482.981 & 231.83 & 1.76(-4) \\
 & 23 2 22 1 & -- & 22 2 21 1 & 258490.870 & 155.91 & 2.57(-4) \\
 & 27 10 18 3 & -- & 27 9 19 3 & 258493.927 & 476.33 & 2.38(-5) \\
 & 23 2 22 0 & -- & 22 2 21 0 & 258496.242 & 155.90 & 2.57(-4) \\
 & 21 12 10 1 & -- & 20 12 9 1 & 258499.332 & 231.82 & 1.76(-4) \\
 & 23 1 22 2 & -- & 22 1 21 2 & 258502.735 & 155.91 & 2.57(-4) \\
 & 23 1 22 0 & -- & 22 1 21 0 & 258508.181 & 155.90 & 2.57(-4) \\
 & 23 2 22 1 & -- & 22 1 21 2 & 258518.554 & 155.91 & 3.57(-5) \\
 & 23 2 22 0 & -- & 22 1 21 0 & 258523.821 & 155.90 & 3.57(-5) \\
 & 41 8 34 0 & -- & 41 7 35 0 & 258541.136 & 557.04 & 2.34(-5) \\
 & 21 5 17 3 & -- & 20 5 16 3 & 258701.047 & 340.65 & 2.46(-4) \\
 & 22 3 20 1 & -- & 21 2 19 2 & 258706.904 & 152.26 & 2.85(-5) \\
 & 22 3 20 0 & -- & 21 2 19 0 & 258714.075 & 152.25 & 2.85(-5) \\
 & 21 11 10 2 & -- & 20 11 9 2 & 258746.248 & 216.63 & 1.91(-4) \\
 & 21 11 11 0 & -- & 20 11 10 0 & 258756.673 & 216.63 & 1.91(-4) \\
 & 21 11 10 0 & -- & 20 11 9 0 & 258756.673 & 216.63 & 1.91(-4) \\
 & 21 7 15 3 & -- & 20 7 14 3 & 258768.938 & 356.20 & 2.33(-4) \\
 & 21 11 11 1 & -- & 20 11 10 1 & 258769.974 & 216.62 & 1.91(-4) \\
 & 21 3 18 3 & -- & 20 3 17 3 & 258775.320 & 333.28 & 2.52(-4) \\
 & 21 8 14 4 & -- & 20 8 13 4 & 258783.896 & 365.70 & 2.25(-4) \\
 & 23 10 14 1 & -- & 23 9 15 1 & 258837.620 & 229.45 & 2.20(-5) \\
 & 23 10 13 2 & -- & 23 9 14 2 & 258847.920 & 229.46 & 2.20(-5) \\
 & 23 10 13 0 & -- & 23 9 14 0 & 258859.163 & 229.46 & 2.20(-5) \\
 & 23 10 14 0 & -- & 23 9 15 0 & 258868.739 & 229.46 & 2.20(-5) \\
 & 11 5 7 4 & -- & 10 4 7 4 & 258955.739 & 242.40 & 1.97(-5) \\
 & 21 7 14 3 & -- & 20 7 13 3 & 259003.875 & 356.22 & 2.34(-4) \\
 & 21 7 14 5 & -- & 20 7 13 5 & 259025.827 & 356.29 & 2.34(-4) \\
 & 21 10 11 2 & -- & 20 10 10 2 & 259113.950 & 202.78 & 2.04(-4) \\
 & 21 10 12 0 & -- & 20 10 11 0 & 259128.178 & 202.78 & 2.04(-4) \\
 & 21 10 11 0 & -- & 20 10 10 0 & 259128.178 & 202.78 & 2.04(-4) \\
 & 21 10 12 1 & -- & 20 10 11 1 & 259137.930 & 202.77 & 2.04(-4) \\
 & 21 3 18 5 & -- & 20 3 17 5 & 259264.990 & 333.04 & 2.54(-4) \\
 & 34 4 30 0 & -- & 34 3 31 0 & 259299.779 & 366.21 & 1.61(-5) \\
 & 24 0 24 2 & -- & 23 1 23 1 & 259341.865 & 158.23 & 4.24(-5) \\
 & 24 1 24 1 & -- & 23 1 23 1 & 259342.015 & 158.23 & 2.63(-4) \\
 & 24 0 24 2 & -- & 23 0 23 2 & 259342.143 & 158.23 & 2.63(-4) \\
 & 24 1 24 1 & -- & 23 0 23 2 & 259342.293 & 158.23 & 4.24(-5) \\
 & 24 0 24 0 & -- & 23 1 23 0 & 259342.727 & 158.22 & 4.24(-5) \\
 & 24 1 24 0 & -- & 23 1 23 0 & 259342.876 & 158.22 & 2.63(-4) \\
 & 24 0 24 0 & -- & 23 0 23 0 & 259343.004 & 158.22 & 2.63(-4) \\
 & 24 1 24 0 & -- & 23 0 23 0 & 259343.152 & 158.22 & 4.24(-5) \\
 & 11 5 6 0 & -- & 10 4 7 0 & 259376.253 & 55.61 & 1.94(-5) \\
 & 42 7 35 2 & -- & 42 6 36 2 & 259422.473 & 581.45 & 2.31(-5) \\
 & 11 5 6 2 & -- & 10 4 7 1 & 259445.549 & 55.62 & 1.08(-5) \\
 & 26 10 17 3 & -- & 26 9 18 3 & 259455.455 & 460.41 & 2.36(-5) \\
 & 42 7 35 0 & -- & 42 6 36 0 & 259463.004 & 581.46 & 2.31(-5) \\
 & 20 4 16 2 & -- & 19 4 15 2 & 259499.905 & 138.67 & 2.54(-4) \\
 & 20 4 16 0 & -- & 19 4 15 0 & 259521.812 & 138.67 & 2.54(-4) \\
 & 22 10 13 1 & -- & 22 9 14 1 & 259529.940 & 215.81 & 2.16(-5) \\
 & 22 3 19 2 & -- & 21 4 18 1 & 259536.089 & 159.72 & 2.05(-5) \\
 & 22 10 12 2 & -- & 22 9 13 2 & 259540.700 & 215.82 & 2.16(-5) \\
 & 22 10 12 0 & -- & 22 9 13 0 & 259558.582 & 215.81 & 2.16(-5) \\
 & 22 3 19 0 & -- & 21 4 18 0 & 259560.402 & 159.72 & 2.05(-5) \\
 & 22 10 13 0 & -- & 22 9 14 0 & 259562.851 & 215.81 & 2.16(-5) \\
 & 21 5 17 4 & -- & 20 5 16 4 & 259573.430 & 340.45 & 2.49(-4) \\
 & 34 5 30 1 & -- & 34 4 31 1 & 259592.949 & 366.22 & 1.62(-5) \\
 & 26 10 17 4 & -- & 26 9 18 4 & 259593.370 & 460.37 & 2.35(-5) \\
 & 21 6 16 3 & -- & 20 6 15 3 & 259616.603 & 347.84 & 2.43(-4) \\
 & 21 9 12 2 & -- & 20 9 11 2 & 259629.300 & 190.30 & 2.16(-4) \\
 & 21 9 13 0 & -- & 20 9 12 0 & 259646.531 & 190.29 & 2.17(-4) \\
 & 21 9 12 0 & -- & 20 9 11 0 & 259647.705 & 190.29 & 2.17(-4) \\
 & 21 9 13 1 & -- & 20 9 12 1 & 259653.078 & 190.28 & 2.17(-4) \\
 & 34 5 30 0 & -- & 34 4 31 0 & 259662.604 & 366.23 & 1.62(-5) \\
 & 21 7 15 4 & -- & 20 7 14 4 & 259916.051 & 355.90 & 2.37(-4) \\
 & 21 10 12 1 & -- & 21 9 13 1 & 260128.610 & 202.77 & 2.12(-5) \\
 & 21 10 11 2 & -- & 21 9 12 2 & 260139.772 & 202.78 & 2.12(-5) \\
 & 21 10 11 0 & -- & 21 9 12 0 & 260162.612 & 202.78 & 2.12(-5) \\
 & 21 10 12 0 & -- & 21 9 13 0 & 260164.454 & 202.78 & 2.12(-5) \\
 & 21 3 18 2 & -- & 20 3 17 2 & 260244.498 & 146.76 & 2.56(-4) \\
 & 21 3 18 0 & -- & 20 3 17 0 & 260255.080 & 146.75 & 2.56(-4) \\
 & 25 10 16 3 & -- & 25 9 17 3 & 260296.709 & 445.09 & 2.34(-5) \\
 & 21 8 13 2 & -- & 20 8 12 2 & 260384.268 & 179.20 & 2.28(-4) \\
 & 21 8 14 0 & -- & 20 8 13 0 & 260392.731 & 179.19 & 2.29(-4) \\
 & 21 8 14 1 & -- & 20 8 13 1 & 260404.026 & 179.19 & 2.28(-4) \\
 & 21 8 13 0 & -- & 20 8 12 0 & 260415.333 & 179.19 & 2.29(-4) \\
 & 20 10 11 1 & -- & 20 9 12 1 & 260643.800 & 190.33 & 2.07(-5) \\
 & 20 10 10 2 & -- & 20 9 11 2 & 260655.151 & 190.34 & 2.07(-5) \\
 & 20 10 10 0 & -- & 20 9 11 0 & 260682.101 & 190.34 & 2.07(-5) \\
 & 20 10 11 0 & -- & 20 9 12 0 & 260682.850 & 190.34 & 2.07(-5) \\
 & 19 4 16 0 & -- & 18 3 15 0 & 260788.406 & 123.25 & 1.70(-5) \\
 & 19 4 16 1 & -- & 18 3 15 2 & 260793.540 & 123.26 & 1.70(-5) \\
 & 11 5 6 5 & -- & 10 4 6 5 & 260799.899 & 242.97 & 1.90(-5) \\
 & 32 3 29 2 & -- & 32 2 30 2 & 260916.838 & 316.40 & 1.31(-5) \\
 & 32 3 29 0 & -- & 32 2 30 0 & 261005.113 & 316.40 & 1.31(-5) \\
 & 32 4 29 0 & -- & 32 3 30 0 & 261070.724 & 316.40 & 1.31(-5) \\
 & 19 10 10 1 & -- & 19 9 11 1 & 261084.120 & 178.50 & 2.01(-5) \\
 & 19 10 9 2 & -- & 19 9 10 2 & 261095.680 & 178.51 & 2.01(-5) \\
 & 19 10 9 0 & -- & 19 9 10 0 & 261126.435 & 178.50 & 2.01(-5) \\
 & 19 10 10 0 & -- & 19 9 11 0 & 261126.726 & 178.50 & 2.01(-5) \\
 & 21 5 17 1 & -- & 20 5 16 1 & 261148.904 & 154.10 & 2.54(-4) \\
 & 21 5 17 0 & -- & 20 5 16 0 & 261165.456 & 154.09 & 2.54(-4) \\
 & 21 6 16 4 & -- & 20 6 15 4 & 261234.608 & 347.62 & 2.43(-4) \\
 & 21 7 15 0 & -- & 20 7 14 0 & 261433.791 & 169.52 & 2.41(-4) \\
 & 21 7 15 1 & -- & 20 7 14 1 & 261436.771 & 169.52 & 2.36(-4) \\
 & 18 10 9 1 & -- & 18 9 10 1 & 261458.040 & 167.26 & 1.94(-5) \\
 & 18 10 8 2 & -- & 18 9 9 2 & 261469.927 & 167.28 & 1.94(-5) \\
 & 18 10 8 0 & -- & 18 9 9 0 & 261504.039 & 167.27 & 1.94(-5) \\
 & 18 10 9 0 & -- & 18 9 10 0 & 261504.039 & 167.27 & 1.94(-5) \\
 & 21 7 14 2 & -- & 20 7 13 2 & 261715.518 & 169.55 & 2.37(-4) \\
 & 21 6 15 5 & -- & 20 6 14 5 & 261727.149 & 348.04 & 2.45(-4) \\
\hline
\ce{CH2OHCHO} (GA), v=0 & 29 11 18 0 & -- & 29 10 19 0 & 258059.438 & 315.77 & 2.78(-4) \\
CDMS & 29 11 19 0 & -- & 29 10 20 0 & 258123.981 & 315.77 & 2.78(-4) \\
 & 22 4 19 0 & -- & 21 3 18 0 & 258128.769 & 150.22 & 2.53(-4) \\
 & 40 6 34 0 & -- & 40 5 35 0 & 259147.668 & 488.73 & 2.40(-4) \\
 & 28 11 17 0 & -- & 28 10 18 0 & 259179.986 & 299.53 & 2.77(-4) \\
 & 28 11 18 0 & -- & 28 10 19 0 & 259212.684 & 299.53 & 2.77(-4) \\
 & 27 11 16 0 & -- & 27 10 17 0 & 260165.716 & 283.86 & 2.75(-4) \\
 & 27 11 17 0 & -- & 27 10 18 0 & 260181.761 & 283.86 & 2.75(-4) \\
 & 40 7 34 0 & -- & 40 6 35 0 & 260691.347 & 488.81 & 2.44(-4) \\
 & 26 11 15 0 & -- & 26 10 16 0 & 261033.098 & 268.76 & 2.72(-4) \\
 & 26 11 16 0 & -- & 26 10 17 0 & 261040.709 & 268.76 & 2.72(-4) \\
 & 24 4 20 0 & -- & 23 5 19 0 & 261041.952 & 182.31 & 1.89(-4) \\
 & 24 2 22 0 & -- & 23 3 21 0 & 261257.959 & 167.42 & 3.73(-4) \\
 & 24 3 22 0 & -- & 23 2 21 0 & 261655.476 & 167.43 & 3.75(-4) \\
\hline
$a$-\ce{(CH2OH)2} ($a$-EG) & 27 3 25 0 & -- & 26 3 24 1 & 258184.073 & 186.63 & 4.19(-4) \\
CDMS & 27 2 25 0 & -- & 26 2 24 1 & 258250.590 & 186.62 & 4.10(-4) \\
 & 27 1 26 0 & -- & 26 2 25 0 & 258567.278 & 179.35 & 7.15(-5) \\
 & 27 1 26 1 & -- & 26 2 25 1 & 258575.594 & 179.69 & 7.16(-5) \\
 & 27 2 26 0 & -- & 26 1 25 0 & 258579.070 & 179.35 & 7.11(-5) \\
 & 27 2 26 1 & -- & 26 1 25 1 & 258586.940 & 179.69 & 7.21(-5) \\
 & 26 6 21 0 & -- & 25 6 19 0 & 259084.460 & 191.35 & 3.28(-5) \\
 & 26 23 3 0 & -- & 25 23 2 1 & 259084.899 & 431.87 & 9.23(-5) \\
 & 26 23 4 0 & -- & 25 23 3 1 & 259084.899 & 431.87 & 9.23(-5) \\
 & 26 21 5 0 & -- & 25 21 4 1 & 259177.002 & 388.80 & 1.48(-4) \\
 & 26 21 6 0 & -- & 25 21 5 1 & 259177.002 & 388.80 & 1.48(-4) \\
 & 27 1 27 1 & -- & 26 1 26 0 & 259202.265 & 171.76 & 4.28(-4) \\
 & 27 0 27 1 & -- & 26 0 26 0 & 259202.322 & 171.76 & 4.28(-4) \\
 & 11 8 3 0 & -- & 10 7 3 1 & 259222.484 & 63.87 & 1.55(-5) \\
 & 11 8 4 0 & -- & 10 7 4 1 & 259222.486 & 63.87 & 1.55(-5) \\
 & 26 20 6 0 & -- & 25 20 5 1 & 259229.051 & 368.72 & 1.74(-4) \\
 & 26 20 7 0 & -- & 25 20 6 1 & 259229.051 & 368.72 & 1.74(-4) \\
 & 26 19 7 0 & -- & 25 19 6 1 & 259286.996 & 349.62 & 1.98(-4) \\
 & 26 19 8 0 & -- & 25 19 7 1 & 259286.996 & 349.62 & 1.98(-4) \\
 & 26 18 8 0 & -- & 25 18 7 1 & 259352.696 & 331.48 & 2.22(-4) \\
 & 26 18 9 0 & -- & 25 18 8 1 & 259352.696 & 331.48 & 2.22(-4) \\
 & 26 17 9 0 & -- & 25 17 8 1 & 259428.525 & 314.32 & 2.44(-4) \\
 & 26 17 10 0 & -- & 25 17 9 1 & 259428.525 & 314.32 & 2.44(-4) \\
 & 26 16 10 0 & -- & 25 16 9 1 & 259517.576 & 298.14 & 2.65(-4) \\
 & 26 16 11 0 & -- & 25 16 10 1 & 259517.576 & 298.14 & 2.65(-4) \\
 & 26 15 11 0 & -- & 25 15 10 1 & 259623.977 & 282.95 & 2.85(-4) \\
 & 26 15 12 0 & -- & 25 15 11 1 & 259623.977 & 282.95 & 2.85(-4) \\
 & 26 14 12 0 & -- & 25 14 11 1 & 259753.385 & 268.74 & 3.03(-4) \\
 & 26 14 13 0 & -- & 25 14 12 1 & 259753.385 & 268.74 & 3.04(-4) \\
 & 25 5 20 0 & -- & 24 5 19 1 & 259810.009 & 174.98 & 4.14(-4) \\
 & 26 3 23 0 & -- & 25 4 22 0 & 259898.267 & 180.03 & 1.90(-4) \\
 & 26 13 13 0 & -- & 25 13 12 1 & 259913.769 & 255.51 & 3.21(-4) \\
 & 26 13 14 0 & -- & 25 13 13 1 & 259913.769 & 255.51 & 3.21(-4) \\
 & 26 12 15 0 & -- & 25 12 14 1 & 260116.738 & 243.28 & 3.38(-4) \\
 & 26 12 14 0 & -- & 25 12 13 1 & 260116.738 & 243.28 & 3.38(-4) \\
 & 26 11 16 0 & -- & 25 11 15 1 & 260379.902 & 232.05 & 3.53(-4) \\
 & 26 11 15 0 & -- & 25 11 14 1 & 260379.910 & 232.05 & 3.53(-4) \\
 & 24 4 20 1 & -- & 23 4 19 0 & 260662.054 & 159.41 & 4.03(-4) \\
 & 26 10 17 0 & -- & 25 10 16 1 & 260731.312 & 221.83 & 3.68(-4) \\
 & 26 10 16 0 & -- & 25 10 15 1 & 260731.522 & 221.83 & 3.68(-4) \\
 & 25 4 22 1 & -- & 24 4 21 0 & 260738.168 & 167.90 & 1.45(-4) \\
 & 28 2 27 0 & -- & 27 2 26 1 & 260759.249 & 192.20 & 4.32(-4) \\
 & 28 1 27 0 & -- & 27 1 26 1 & 260760.891 & 192.20 & 4.31(-4) \\
 & 27 4 23 1 & -- & 26 5 22 1 & 260993.665 & 198.65 & 9.51(-5) \\
 & 26 5 22 0 & -- & 25 5 21 1 & 261181.033 & 185.84 & 3.79(-4) \\
 & 26 9 18 0 & -- & 25 9 17 1 & 261217.419 & 212.64 & 3.82(-4) \\
 & 26 9 17 0 & -- & 25 9 16 1 & 261221.825 & 212.64 & 3.82(-4) \\
 & 28 0 28 0 & -- & 27 1 27 0 & 261304.289 & 183.96 & 8.62(-5) \\
 & 28 1 28 0 & -- & 27 0 27 0 & 261304.396 & 183.96 & 8.62(-5) \\
 & 28 0 28 1 & -- & 27 1 27 1 & 261316.502 & 184.30 & 8.64(-5) \\
 & 28 1 28 1 & -- & 27 0 27 1 & 261316.604 & 184.30 & 8.64(-5) \\
\hline
$g$-\ce{(CH2OH)2} ($g$-EG) & 19 3 16 0 & -- & 18 2 16 1 & 258132.256 & 99.64 & 8.77(-5) \\
CDMS & 25 6 20 1 & -- & 24 6 19 0 & 258140.799 & 177.14 & 1.49(-4) \\
 & 25 4 22 1 & -- & 24 3 21 1 & 258146.236 & 166.51 & 5.54(-5) \\
 & 25 7 18 1 & -- & 24 7 17 0 & 258193.293 & 183.22 & 1.40(-4) \\
 & 25 4 22 0 & -- & 24 3 21 0 & 258228.389 & 166.46 & 1.24(-4) \\
 & 27 4 23 0 & -- & 26 5 22 0 & 258328.455 & 196.89 & 5.30(-5) \\
 & 27 4 23 1 & -- & 26 5 22 1 & 258376.787 & 196.94 & 6.77(-5) \\
 & 27 2 26 1 & -- & 26 2 25 0 & 258670.537 & 178.47 & 1.62(-4) \\
 & 27 1 26 1 & -- & 26 1 25 0 & 258673.113 & 178.47 & 1.65(-4) \\
 & 28 1 28 0 & -- & 27 1 27 1 & 258855.869 & 183.15 & 1.66(-4) \\
 & 28 0 28 0 & -- & 27 0 27 1 & 258855.890 & 183.15 & 1.66(-4) \\
 & 17 2 16 1 & -- & 16 1 16 0 & 258880.434 & 74.76 & 3.25(-5) \\
 & 25 6 19 0 & -- & 24 6 18 1 & 259076.676 & 177.52 & 1.52(-4) \\
 & 26 4 23 0 & -- & 25 4 22 1 & 259858.291 & 178.98 & 9.58(-5) \\
 & 28 0 28 1 & -- & 27 1 27 1 & 260191.593 & 183.21 & 1.70(-4) \\
 & 28 1 28 1 & -- & 27 0 27 1 & 260191.662 & 183.21 & 1.70(-4) \\
 & 28 0 28 0 & -- & 27 1 27 0 & 260193.105 & 183.15 & 1.70(-4) \\
 & 28 1 28 0 & -- & 27 0 27 0 & 260193.175 & 183.15 & 1.70(-4) \\
 & 19 3 16 1 & -- & 18 2 16 0 & 260466.788 & 99.70 & 9.11(-5) \\
 & 25 4 21 0 & -- & 24 4 20 1 & 260784.969 & 170.44 & 1.63(-4) \\
 & 25 6 19 1 & -- & 24 6 18 0 & 261211.160 & 177.57 & 1.54(-4) \\
 & 26 3 23 0 & -- & 25 3 22 1 & 261231.929 & 178.81 & 2.07(-4) \\
 & 18 5 14 1 & -- & 17 4 13 1 & 261293.071 & 96.03 & 4.57(-5) \\
 & 18 5 14 0 & -- & 17 4 13 0 & 261378.330 & 95.98 & 6.03(-5) \\
 & 28 1 28 1 & -- & 27 1 27 0 & 261528.877 & 183.21 & 1.71(-4) \\
 & 28 0 28 1 & -- & 27 0 27 0 & 261528.899 & 183.21 & 1.72(-4) \\
 & 26 4 23 1 & -- & 25 4 22 0 & 261644.038 & 179.02 & 2.06(-4) \\
\end{longtable}

\end{appendix}

\end{document}